\documentclass[useAMS,usenatbib, usegraphicx]{emulateapj}
\usepackage{aas_macros}
\usepackage{graphicx,epsfig}

\received{}
\revised{}
\accepted{Accepted for publication in ApJ}
\shorttitle{A tale of a rich cluster at z$\sim$0.8}
\shortauthors{A. Ferr\'e-Mateu et al.}

\begin{document}

\title{A tale of a rich cluster at z$\sim$0.8 as seen by the Star Formation Histories of its ETGs}

\author{Anna Ferr\'e-Mateu$^{1}$, Patricia S\'anchez-Bl\'azquez$^{2}$, Alexandre Vazdekis$^{3,4}$ and Ignacio G. de la Rosa$^{3,4}$} 
\affil{$^{1}$Subaru Telescope, Hilo, HI 96720, USA}
\affil{$^{2}$ Universidad Aut\'onoma de Madrid, Departamento de F\'isica Te\'orica, E28049 Cantoblanco, Madrid, Spain }
\affil{$^{3}$Instituto de Astrof\'{\i}sica de Canarias,  E38205 La Laguna,Tenerife, Spain}
\affil{$^{4}$Universidad de La Laguna, Departamento de Astrof\'isica, E38206 La Laguna,Tenerife, Spain}
\email{aferre@naoj.org (AFM)}   

\begin{abstract}
We present a detailed stellar population analysis for a sample of 24 early-type galaxies belonging to the rich cluster RX\,J0152.7-1357 at z\,=\,0.83. We have derived the age, metallicity, abundance pattern and star formation history for each galaxy \textit{individually}, to further characterize this intermediate-z reference cluster. We then study how these stellar population parameters depend on local environment. This provides a better understanding on the formation timescales and subsequent evolution of the substructures in this cluster. We have also explored the evolutionary link between z$\sim$0.8 ETGs and those in the local Universe by comparing the trends that the stellar population parameters follow with galaxy velocity dispersion at each epoch. We find that the ETGs in Coma are consistent with being the (passively-evolving) descendants of the ETG population in RXJ10152.7-1357. Furthermore, our results favor a downsizing picture, where the subclumps centers were formed first. This central parts contain the most massive galaxies, which formed the bulk of their stars in a short, burst-like event at high-z. On the contrary, the cluster outskirts are populated with less massive, smaller galaxies, which show a wider variety of Star Formation Histories. In general, they present extended star formation episodes over cosmic time, which seems to be related to their posterior incorporation into the cluster, around 4\,Gyr later after the initial event of formation.
\end{abstract}


\section{Introduction}
One of the main goals in modern astrophysics is to understand the formation and evolution of galaxies, particularly that of Early-Type Galaxies (ETGs), as they contain most of the luminous matter in the Universe. A striking case is that of the most massive ones, as these galaxies experience a strong size and morphological evolution, whereas their masses remain almost unvaried over cosmic time (e.g. \citealt{Daddi2005}; \citealt{Trujillo2006}; \citealt{Buitrago2008}; \citealt{vanDokkum2008}). In order to explain this puzzle, a two-phase formation mechanism has been proposed. First, a fast, monolithic-like phase occurs at high redshift (z$\sim$\,2-\,3), which creates the central massive galaxy. This phase is dominated by a dissipational in-situ star formation, fed by cold flows (e.g. \citealt{Kerevs2005}; \citealt{Dekel2009}, \citealt{Oser2010}) and/or gas rich mergers (e.g. \citealt{Ricciardelli2010}; \citealt{Wuyts2010}). Then, a late-time accretion phase takes place, with a gradually build up of the galaxy outskirts via the accretion of gas-poor satellites (e.g. \citealt{Naab2009}; \citealt{Oser2010}; \citealt{Hilz2013}). This minor merger mechanism leaves the properties of the central massive galaxy almost untouched while it grows in size (e.g. \citealt{Bezanson2009}; \citealt{Trujillo2011}; \citealt{Lopez-Sanjuan2012}). However, recent studies have pointed out to a dependence between this mass-size relation and the age of the stellar populations of the galaxies, with the stellar ages increasing for more massive and more compact galaxies (e.g. \citealt{Saracco2009}; \citealt{Williams2010}; \citealt{Poggianti2013}; but see \citealt{Trujillo2011} and \citealt{Andreon2013}). If this dependence is true, it implies that the selected galaxies will
be the most compact ones, posing a strong selection bias when comparing to a local population.  If this is taken into account, the size evolution that massive galaxies experience from high to low redshift is found to be milder \citep{Poggianti2013}. \\
But the story is even more complex, as environment also plays a crucial role in determining galaxy evolution. It is well known from the morphology-density relation of \citet{Dressler1980} that ETGs mainly dominate high-density regions, such as clusters. It is also assumed that a large fraction of the massive (and passive) galaxies from the early Universe, above depicted, will evolve into present day cluster compact galaxies. Therefore, clusters of galaxies are an excellent laboratory to study the physical mechanisms by which a massive galaxy can transform both its morphology and global properties. Most studiess during the last decade have focused on studying the differences on the stellar populations of galaxies in high-density (cluster) and low-density (field) environment (e.g. \citealt{Dressler1980}; \citealt{Trager2000}; \citealt{Kuntschner2002}; \citealt{Caldwell2003}; \citealt{Thomas2005}; \citealt{Sanchez-Blazquez2006b}; \citealt{Gobat2008}; \citealt{Rettura2010}). The overall picture is that massive ETGs in low-density environments seem to be, on average, $\sim$2\,Gyr younger and slightly more metal rich than their analogues in higher-density environment. Furthermore, the observed physical properties of the galaxies maintain a tight relation with the local environment (e.g. \citealt{Postman2005}; \citealt{Tanaka2005}; \citealt{Holden2007}; \citealt{Hilton2009}; \citealt{Vulcani2012}). It is assumed that the cluster ETGs are largely passively evolving since at least z$\sim$\,1.2 (e.g. \citealt{Andreon2008}; \citealt{dePropris2013}), with its red-sequence being already in place at even higher redshifts (e.g. \citealt{Bower1992},  \citealt{Kodama1997}; \citealt{Blakeslee2003}; \citealt{DeLucia2007}; \citealt{Faber2007}) and showing no significant evolution on the Fundamental Plane (e.g. \citealt{vanDokkum2003}; \citealt{Holden2005}), on the luminosity function (e.g. \citealt{Moustakas1997}; \citealt{Kodama2004}; \citealt{Rudnick2008}), or on the mass function (e.g. \citealt{Pozzetti2003}; \citealt{Perez-Gonzalez2008}; \citealt{Mortlock2011}). Consequently, a detailed study of the stellar content of ETGs in clusters at different redshifts provides new means to test the different formation and evolutionary models while unraveling the influence of the environment. \\
Going back to approximately the era when star formation was quenched presents several advantages, in particular when studying the stellar populations, as these become younger and their integrated spectrum becomes more sensitive to the age. However, such studies at high redshift are challenging, as obtaining data with enough S/N to accurately measure relevant absorption lines is difficult and time-consuming. So far, most studies have been performed by stacking the available spectra to achieve the required quality, mostly up to intermediate redshifts (z$\sim$\,1, approximately half the age of the Universe; e.g. \citealt{Kelson2001}; \citealt{Barr2005}; \citealt{Kelson2006}; \citealt{Schiavon2006}; \citealt{Tran2007}; \citealt{Sanchez-Blazquez2009}, PSB09 hereafter). We present here a revised analysis of the stellar populations of a sample of ETGs belonging to RX\,J0152.7-1357, a rich galaxy cluster at moderate redshift (z\,=\,0.83). The main novelty of this work is that we perform this study based on \textit{each galaxy individually}. This detailed treatment of the stellar populations allows us to constrain the downsizing scenario through the analysis of the individual Star Formation Histories (SFHs) and their relationship with the local environment. Note that by ``local environment" we will always refer here to the location of the galaxies within the cluster, not to the local galaxy density. This allows to trace the formation timescales of the different substructures, revealing the formation history of this cluster. We apply new state-of-the-art stellar population models and methods to show that the stellar populations of RX\,J0152.7-1357 already resemble those seen in nearby clusters. Section 2 comprises a summary of the available information on this cluster and the new reduction process. Section 3 presents the derived kinematics for the individual ETGs. In Section 4 we present the analysis of the stellar populations of each individual galaxy, both the single-stellar parameters (age, metallicity, abundance patterns) and the derived SFHs. In Section 5 we study the observed properties as a function of the location of the galaxies within the cluster. In Section 6 we study how all the previous properties relate to galaxy velocity dispersion, and discuss a passive evolution scenario by comparing our findings to Coma cluster galaxies. Section 7 describes the emerging picture of the formation history for this cluster and its implications on galaxy evolution theories. Finally,  a summary of the main results can be found in Section 8. We have adopted a concordance cosmology with $\Omega_{m}$=\,0.3, $\Omega_{\Lambda}$=\,0.7 and H$_{0}$=\,70\,km\,s$^{-1}$ Mpc$^{-1}$ throughout the paper.

\section{RX J0152.7-1357: a rich cluster at intermediate redshift}
\subsection{What we know so far}
The ETGs analyzed in this study belong to RX\,J0152.7-1357, a luminous X-ray galaxy cluster at z=\,0.83. It has been the target of many observing programs such as the \textit{ROSAT Deep Cluster Survey}, the \textit{Wide Angle ROSAT Pointed Survey} \citep{Ebeling2000}, the \textit{Bright Serendipitous High-redshift Archival Cluster survey} \citep{Nichol1999}, the \textit{BeppoSAX} \citep{DellaCeca2004}, the \textit{XMM-Newton} and \textit{Chandra} (\citealt{Jones2004}; \citealt{Maughan2003}). In the optical range, it has been observed for the \textit{ACS Intermediate Redshift Cluster Survey} \citep{Blakeslee2006} that allowed a morphological classification of its members \citep{Postman2005}. Moreover, RX\,J0152.7-1357 is one of the nineteen clusters observed for the \textit{Gemini/HST Galaxy Cluster Project} (\citealt{Jorgensen2005}; J05 from now onwards), which was devoted to study galaxy evolution until approximately half the age of the Universe.\\
Previous X-ray studies of RX\,J0152.7-1357 have shown that this is a complex substructured cluster in a merging state, similar to Coma in the local Universe. The total mass of the cluster is also similar to Coma (\citealt{Maughan2003}; see Table 1). Two big subclumps, the northern at z=\,0.838 (N-SubCl) and the southern at z=\,0.830 (S-SubCl) form the main structure. A third small eastern group at z=\,0.845 (E-group) and a diffuse group of galaxies off to the west at z=\,0.866 are also part of the cluster (\citealt{Demarco2005}; \citealt{Girardi2005}, see also Figure 1). A dynamical analysis is presented in \citet{Girardi2005}, showing that this cluster is not yet dynamically relaxed, with velocity gradients and substructure. The northern subclump is a more evolved system, also confirmed by its higher lensing mass \citep{Jee2005}.\\
Many studies have focused on understanding the properties of its galaxies. Details about RX\,J0152.7-1357 structure, colors and physical properties can be found in \citet{Blakeslee2006} and \citet{Nantais2013}. In \citet{Patel2009}, the red-sequence galaxies are analyzed out to large galacto-centric distances to study the impact of the environment. In \citet{Homeier2005}, the authors focused on the study of its few star forming galaxies in order to describe their transformation into the red sequence. These studies find that the morphological evolution is especially active in this cluster, with the center being almost entirely populated by ETGs. This cluster has been also used in several studies to address the evolution of cluster properties with redshift, such as the role of mass and environment \citep{diSeregoAlighieri2006}, the evolution on the K-band luminosity function \citep{Ellis2004}, the morphology-density relation \citep{Postman2005}, and the evolution of the Fundamental Plane (\citealt{Holden2005}; \citealt{Jorgensen2006}; \citealt{Chiboucas2009} and \citealt{Jorgensen2013}).\\
Most of the work presented here is compared to the one published by J05, which we use as a reference. In J05 they discussed possible evolutionary scenarios by comparing the averaged stellar population properties from RX\,J0152.7-1357 with a sample of galaxies in clusters of the local Universe. However, the emphasis of these authors was given to the general trends on the cluster rather than looking at the properties of each galaxy individually. Despite the exhaustive study performed in J05, the authors could not totally confirm or rule out the passive evolution scenario, as the results strongly depended on the different indicators they were using. They already pointed out that those inconsistencies could be related to the not fully developed methodology. But now, with new tools and updated models, we are in the position to fully characterize this cluster, as it can be considered as a reference one at intermediate redshifts.  As will be discussed in section 4, we deepen our understanding of the formation and evolution of the ETGs in this cluster by introducing a new methodology to study the stellar populations. This methodology combines the power of the full-spectral-fitting, a more novel technique to derive the SFHs of galaxies with better tolerance to data quality than the spectral indices \citep{CidFernandes2010} and an hybrid approach to analyze the abundance patterns, not carried before for clusters at such redshift.

\begin{figure*}
\centering
\includegraphics[scale=0.58]{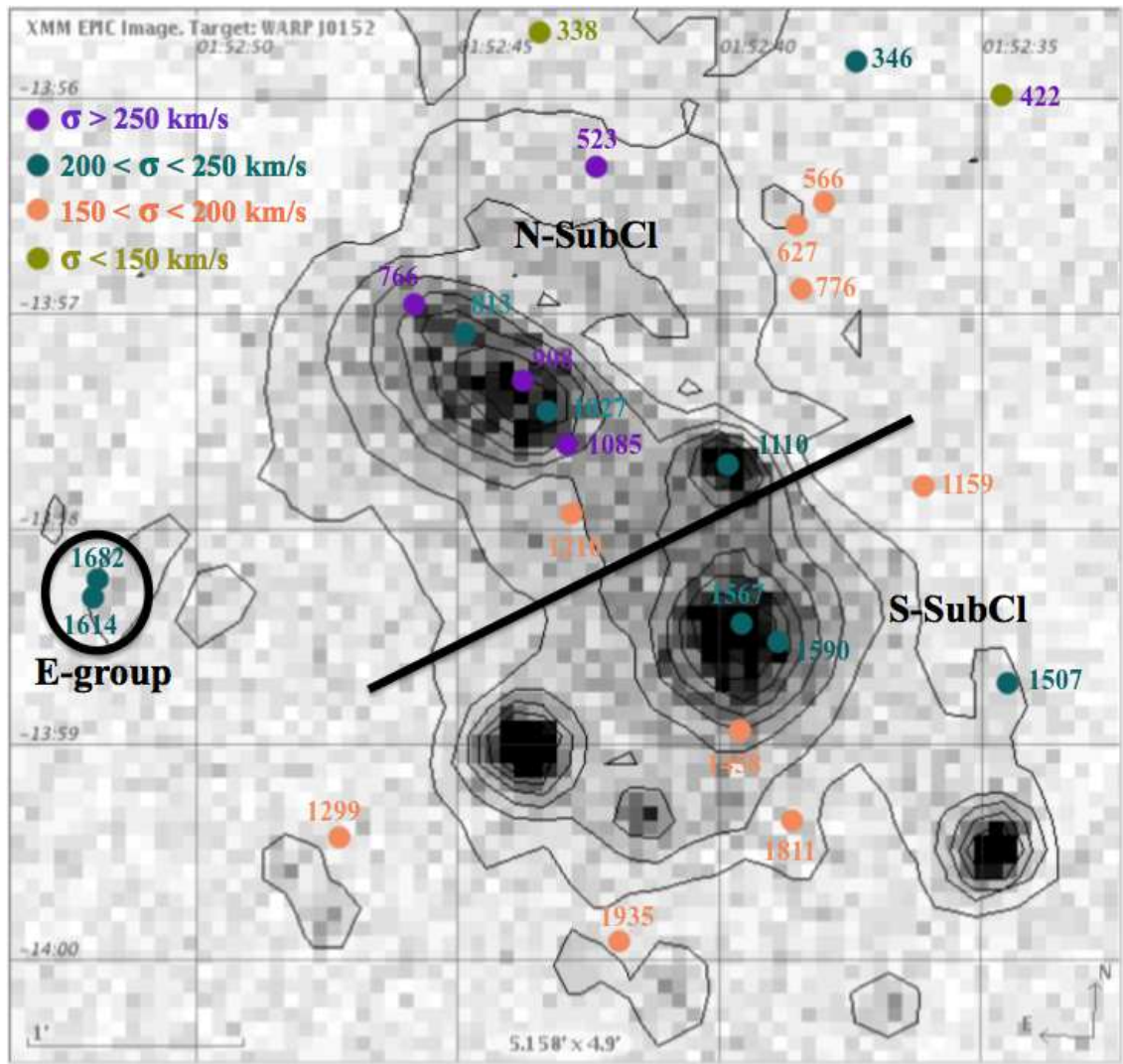}
\label{figure:1}
\caption{XMM-Newton image with the contours of the X-ray data superposed. The image covers approximately 5'$\times$\,5'. ETG members considered in this study are color-coded by their velocity dispersion derived with {\tt PPxF} in Section 3: green for $\sigma\,<\,$150\,$\rm{km\,s^{-1}}$, orange for 150\,$<\,\sigma\,\leq\,$200\,$\rm{km\,s^{-1}}$, dark green for 200\,$<\,\sigma\,\leq\,$250\,$\rm{km\,s^{-1}}$ and purple for galaxies with $\sigma\,>\,$250\,$\rm{km\,s^{-1}}$. The black line separates the main two regions here considered (north and south subclumps), following Girardi et al. (2005) structure. The east group is also shown, located at the left of the figure.}
\end{figure*} 

\subsection{Data and data reduction}
The spectroscopic data for this cluster were obtained with GMOS-N at Gemini in semester 2002B under the \textit{Gemini/HST Galaxy Cluster Project} (program GN-2002B-Q-29, see J05 for more details). A single mask was employed, with slits of 1$"$ width and the R400 grating. This provided an instrumental resolution of $\sigma$=\,116\,$\rm{km\,s^{-1}}$ at 4300\,$\rm\AA{}$ (rest frame). 25 individual exposures of the mask resulted in a total observing time of 21.7\,hours. This delivered 41 galaxy spectra covering an approximated rest-frame wavelength range of  $\lambda$\,3200-5200\,$\rm\AA{}$. 29 of these 41 spectra were ETGs cluster members. We performed a typical reduction process with bias subtraction and flat-field correction, cosmic ray removal, sky subtraction, wavelength calibration, telluric lines correction and relative flux calibration. Before adding all the frames corresponding to each galaxy, we extracted a 1.15'' aperture to match J05 data for comparison purposes. Our final sample contains 24 ETGs, as 5 objects were discarded due to a lower S/N ($<$\,10) than the required to perform the analysis proposed here.\\
One of the reasons beyond this new data reduction is that J05 results were all index-based and, despite being high quality data for such redshift (S/N$\sim$\,20), the S/N of the individual spectra was not enough to derive robust estimates (see e.g. \citealt{Conroy2013} and references therein). This could explain some of the inconsistent results on various of the indices. To overcome this problem, one possibility is to perform a very accurate and detailed data reduction, with particular attention to the errors that are propagated during this process. We used {\tt REDUCEME} \citep{Cardiel1999}, an optimized reduction package for slit spectroscopy, whose principal advantage is to give the error that is propagated in parallel to the reduction process. Furthermore, carrying out this new data reduction step-by-step for each single frame, without making use of any pipeline, allowed us to track the critical steps in the process, such as the wavelength calibration or the sky subtraction. The main differences between the reduction process performed by both works can be summarized as follows: \textit{(i)} J05 used 6th-order polynomials to fit the dispersion function on the wavelength calibration, while we used 2nd-order polynomials that equally give an accurate calibration while being more restrictive; \textit{(ii)} we were also more restraining on the sky subtraction, using linear fits to the regions selected as sky  (2nd-order polynomials were only used when the first clearly failed at removing the lines, which occurred only for a few frames). Note, however, that the reddest part of the spectra could not be completely cleaned, presenting still strong sky residuals. 

\subsubsection{Local sample: Coma cluster}
In Section 6, we compare the scaling relations of our cluster with those in a local cluster in order to see if the stellar populations of the galaxies in the intermediate redshift cluster could become the ones we see today assuming a passive evolution scenario. We have chosen Coma as the control local sample, as both clusters have similar masses, luminosities and densities, hence both are considered high-density, rich clusters (see Table 1). Coma has been used also in previous studies of RX\,J0152.7-1357 as its low-z reference cluster (e.g. \citealt{Homeier2005}, J05 and \citealt{Jorgensen2013}) and in particular, it has been considered as a plausible descendant of JKCS\,041, a cluster at z\,=\,1.8 \citep{Andreon2014}. Despite Coma not being considered a typical nearby cluster (see e.g. \citealt{Pimbblet2014}), our evolutionary hypothesis is supported by the fact that the majority of its properties and its degree of substructure highly resemble those in our intermediate redshift cluster (as seen in Table 1).\\
Before establishing a possible evolutionary link between the galaxies in the two depicted clusters, it is important to understand the expected evolution of the mass and substructure of the cluster. Simulations from \citet{Knebe2002} show that $\sim$\,30$\%$ of local clusters should have a high degree of substructure due to the inter-cluster merger and infall activity. However, as shown by e.g. \citet{White1995} and \citet{Roettiger1997}, individual galaxies can survive cluster mergers relatively unaffected, i.e. clusters can traverse each other without affecting their individual galaxies. For the cluster mass evolution, it is expected that it will progressively grow with time as new field and group galaxies are accreted into it (e.g. \citealt{Perez-Gonzalez2008}; \citealt{Marchesini2014} and references therein). However, the amount of mass gained from z=\,0.8 until today is almost negligible (a factor of $\sim$1.5; e.g. \citealt{Wechsler2002}; \citealt{Fakhouri2010}). Summarizing, the galaxy strength against mergers and the mild evolution seen for the stellar mass, both further support the viability of our study on the cluster evolution from the perspective of the stellar populations of their individual galaxies.\\
The spectra for Coma used here are those of \citet{Sanchez-Blazquez2006b} (PSB06, hereafter), obtained with the ISIS double spectrograph at the WHT (Roque de los Muchachos, La Palma). The data were reduced in a similar way than ours and the final spectra have a spectral resolution of 6.56\,$\rm\AA{}$. We have considered an aperture comparable to the one employed for our cluster at intermediate-z. The set of galaxies in Coma were selected to cover the same velocity dispersion range and a similar number of objects (22 galaxies in Coma \textit{vs} 24 in RX\,J0152.7-1357) to our cluster. The reader is referred to PSB06 for a more extended description on the Coma data.

\begin{table*}
\label{table:1}                      
\centering
\caption{Cluster's properties}    
\begin{tabular}{c|c c c c}   
\hline\hline      
Cluster & z &  Total Mass & Luminosity & $\sigma_{v}$ \\  
            &    &($\times$10$^{15}$M$_\odot$)&($\times$10$^{44}$ergs\,s$^{-1}$)& ($km\,s^{-1}$) \\
\hline  
Coma                    & 0.023 & 1.8\,-\,2.5\,$^{a}$ & 7.7 (0.1\,-\,2.4 keV)\,$^{b}$& 1015\,$^{c}$\\  
RX\,J0152.7-1357& 0.833 & 1.2\,-\,2.2\,$^{d}$ & 6.5 (0.5\,-\,2.0 keV)\,$^{e}$& 1322\,$^{d}$\\ 
\hline                                  
\end{tabular}

{(1) Clusters name; (2) redshift; (3) Total Mass, a range is shown as this estimate depends on the radii used to compute it; (4) X-ray luminosity; (5) Line-of-sight velocity dispersion from dynamical studies.\\
$^{a}$ \citealt{Kubo2007}; $^{b}$ \citealt{Reiprich2002}; $^{c}$ \citealt{Colless2000}; $^{d}$ \citealt{Girardi2005}; $^{e}$ \citealt{Vikhlinin2009}}
\end{table*}

\section{Stellar Kinematics}
To extract the information about the stellar kinematics we used {\tt pPXF} (Penalized Pixel-Fitting;  \citet{Cappellari2004}). We used as templates the MILES library of Single Stellar Population (SSP) models (\citealt{Vazdekis2010}, V10 hereafter) and the errors were derived from 1000 Monte-Carlo simulations. Our measured velocity dispersions are in general slightly larger than the ones of J05, as seen in Figure 2 and Table 2. These differences could be attributed to a different choice of spectral regions for the fit. We fitted the region  $\lambda$\,4100-4900\,$\rm\AA{}$, avoiding the D4000-break. Our spectra reach 5200\,$\rm\AA{}$, but we cannot rely on this reddest region as strong sky residuals are present. {\tt GANDALF} \citep{Sarzi2006} was used to clean the spectra from emission lines. Only three galaxies showed emission-line features in [OII] and [OIII], which are marked with an asterisk on Table 2. These galaxies are located in the outskirts of the main subclumps, which is in agreement with \citet{Jaffe2014}, who found that emission-line ETGs at $0.3\,<\,z\,<$\,0.9 are typically seen in the field and in the infall regions of clusters. We will not use these galaxies in the stellar population analysis based on the measurements of the indices but they will be considered in the full-spectrum-fitting approach, as it allows for masking them. \\
Figure 1 shows the position of the 24 ETGs within the cluster. Their measured velocity dispersions are color-coded: soft green for galaxies with $\sigma\,<\,$150\,$\rm{km\,s^{-1}}$, orange for 150\,$<\,\sigma\,\leq\,$200\,$\rm{km\,s^{-1}}$, dark green for 200\,$<\,\sigma\,\leq\,$250\,$\rm{km\,s^{-1}}$ and purple for galaxies with $\sigma\,>\,$250\,$\rm{km\,s^{-1}}$. We can see that galaxies with different velocity dispersions populate different regions of the cluster. The center of the N-SubCl contains the galaxies with the highest velocity dispersions ($\sim$\,260\,$\rm{km\,s^{-1}}$). The center of the S-SubCl is populated by galaxies with slightly lower velocity dispersions, though still being considered massive ($\sim$\,230\,$\rm{km\,s^{-1}}$), like in the small E-group. On the contrary, it is clear that the outskirts of the subclumps are mainly populated by galaxies with the lowest velocity dispersions ($\sigma\,\le\,$180\,$\rm{km\,s^{-1}}$). This is in agreement with other studies that show that the central regions are preferably populated by the most massive galaxies (e.g. \citealt{Rosati2009}; \citealt{Mei2012}; \citealt{Muzzin2012}; \citealt{Strazzullo2013}).

\begin{figure}
\centering
\includegraphics[scale=0.5]{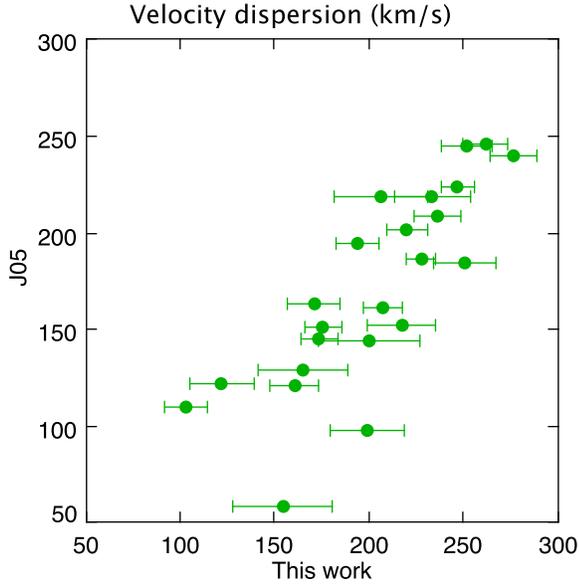}
\label{figure:2}
\caption{Comparison of the measured velocity dispersion for the cluster members in this study with those of J05. It is clear that for some galaxies we tend to recover higher values, which can be attributed to the selection of different regions for the fit.}
\end{figure}
 
\begin{table}
\label{table:2}                      
\centering
\caption{The sample}    
\begin{tabular}{c|c c c c c}   
\hline\hline      
ID & z &  subclump & S/N & $\sigma$ &  $\sigma_{J05}$\\  
    &    &                   &        & ($km\,s^{-1}$) & ($km\,s^{-1}$)\\
\hline  
338   & 0.8193 & N-SubCl & 14.32 &  122$\pm$\,17  & 121 \\      
346   & 0.8367 & N-SubCl & 17.63 &   217$\pm$\,18 & 151 \\
422   & 0.8342 & N-SubCl & 17.73 &  103$\pm$\,11  & 108 \\
523   & 0.8206 & N-SubCl & 25.25 &   277$\pm$\,12 & 239 \\
566*  & 0.8369 & N-SubCl & 17.63 &  170$\pm$\,13 & 162\\
627   & 0.8324 & N-SubCl & 13.79 &  194$\pm$\,11  & 193 \\
766   & 0.8346 & N-SubCl & 17.44 &  262$\pm$\,11  & 244 \\
776   & 0.8325 & N-SubCl & 20.56 &  160$\pm$\,13  & 119 \\
813   & 0.8351 & N-SubCl & 23.41 &  233$\pm$\,20  & 218 \\
908   & 0.8393 & N-SubCl & 23.12 &  250$\pm$\,16  & 183 \\
1027  & 0.8357 & N-SubCl & 14.44 &  236$\pm$\,12 & 207 \\
1085  & 0.8325 & N-SubCl & 22.91 &  252$\pm$\,13 & 244 \\ 
1110  & 0.8322 & N-SubCl & 16.84 &  220$\pm$\,11  & 200 \\
1159* & 0.8357 & N-SubCl & 18.71 &  175$\pm$\,9   & 150 \\
1210  & 0.8372 & N-SubCl & 13.54 &  165$\pm$\,23 & 127 \\
1299* & 0.8374 & S-SubCl & 15.17 &  200$\pm$\,26 & 143 \\
1458  & 0.8324 & S-SubCl & 12.18 &  173$\pm$\,10 & 144 \\
1507  & 0.8289 & S-SubCl & 10.76 &  206$\pm$\,24 & 217\\ 
1567  & 0.8291 & S-SubCl & 14.38 &  211$\pm$\,15 & 307 \\
1590  & 0.8317 & S-SubCl & 17.76 &  199$\pm$\,19 &  96 \\
1614  & 0.8433 & E-group & 19.66 &  247$\pm$\,9   & 222 \\
1682  & 0.8463 & E-group & 20.89 &  227$\pm$\,8   & 185 \\
1811  & 0.8351 & N-SubCl & 13.83 &  154$\pm$\,26 &  57 \\
1935  & 0.8252 & S-SubCl & 18.38 &  207$\pm$\,10 & 159 \\
\hline                                  
\end{tabular}

{Main properties of the galaxies selected from the J05 mask. (1) ID; (2) redshift; (3) subclump membership--N-SubCl (north subclump)--S-SubCl (south subclump)--E-group (east group), see Figure\,1; (4) S/N per $\rm\AA{}$ for the fitted range; (5) velocity dispersion from this work obtained with {\tt pPXF}; (6)velocity dispersion from table 12 in J05. Asterisks mark the galaxies with emission-lines.}
\end{table}
      
\section{Stellar populations}
\subsection{Line index analysis}
We have performed a detailed stellar population analysis galaxy by galaxy using the V10 SSP models. These models cover a wide range of both ages (from 0.1 to 17.78\,Gyr) and metallicities (from [Z/H]\,=\,$-$\,1.71 to [Z/H]\,=\,$+$\,0.22) for different initial mass functions. We have adopted the newly-defined LIS system of indices (V10), which is characterized for having a constant resolution and a flux-calibrated response curve. The advantage over the classical Lick/IDS system is that there is no need to smooth the data to the Lick/IDS wavelength dependent resolution \citep{Worthey1997}. Instead, we broaden our spectra to the LIS resolution that best matches our data, LIS-8.4$\rm\AA{}$ in this case. \\
We have further characterized the cluster members by deriving their individual ages, metallicities and abundance ratios with {\tt RMODEL} \citep{Cardiel2003}, a program that interpolates in the SSP model grids. The spectra wavelength range not affected by sky residuals, $\lambda$3600-4800\,$\rm\AA{}$ (rest frame), encompasses most of the commonly used indices for the stellar population analysis. The models of V10, which start at 3540$\rm\AA{}$, are thus appropriate for studying the bluest indices. Note that no such models were available when J05 published their analysis. However, the rest-frame wavelength range does not allow us to measure the most commonly used indices at low redshift studies, as H$\beta$, Mgb or [MgFe]. Therefore, we have created the pairs of index-index models grids using H$\gamma$F as our main age indicator, with CN$_{2}$, Fe4383 and C$_{2}$4668 as metallicity indices (e.g. \citealt{Thomas2003}; \citealt{Thomas2004}, PSB09). We also measured the index CN3883 to derive safer CN abundance ratios. From one side, CN$_{2}$ in massive galaxies might not be well fitted with scaled-solar stellar population models (\citealt{Sanchez-Blazquez2003}; \citealt{Carretero2004}, see also the fits in Section 4.2). From the other side, there was a telluric line covering the spectral range of the CN$_{2}$ index definition. \\
Appendix A presents the stellar population properties derived for each galaxy individually. However, it is easily seen in Figure A2 that the non-orthogonality of the model grids makes it difficult to infer accurate estimates, in particular for the abundance patterns. With this current method we find that galaxies are, on average $\sim$\,3.5\,Gyr old. This is slightly younger than the one stated in J05 ($\sim$\,5\,Gyr) but in better agreement with the value obtained on the basis of the colors \citep{Blakeslee2006} and compatible with studies at slightly higher redshift (e.g. at z\,=\,1.2, \citealt{Rettura2010}). We will further extent on the individual single-stellar population parameters in Section 4.2, where a more powerful methodology is described. 

\subsection{Star Formation Histories}
Spectra at high redshift do not usually have enough S/N and the sky subtraction is a difficult task, thus any feature that remains from the reduction process can give misleading line strength values. To pose further constrains on the stellar populations, we have also employed the full-spectrum-fitting approach, a more novel technique that is less sensitive to the these effects, as the affected regions can be masked.\\
We used {\tt STARLIGHT} \citep{CidFernandes2005} with a set of SSP SED models from V10. As templates, we only use those models younger than 8\,Gyr. The reason is that the age of the Universe at the cluster redshift is $\sim$\,7\,Gyr, but we have allowed an extra 1\,Gyr in order not to bias the errors for the oldest galaxies. Removing approximately half the age of the Universe has the advantage of avoiding the models that are more affected by the age-metallicity degeneracy. The highest metallicity of the models is [Z/H]=$+$\,0.22. When the output from the code is this exact value, we consider the derived metallicity a lower limit, implying that the galaxy could be younger and more metal rich (marked with a diamond symbol in Table 3).\\
Recent studies have highlighted the importance of the Initial Mass Function (IMF). There seems to be a dependence with galaxy mass or velocity dispersion, in the sense that more massive galaxies demand steeper IMF slopes (e.g. \citealt{Cenarro2003}; \citealt{Falcon-Barroso2003}; \citealt{Cappellari2012}; \citealt{Conroy2012a}; \citealt{Ferreras2013}; \citealt{LaBarbera2013}). Moreover, in \citet{Ferre-Mateu2013} we quantified the impact that such variations have on the derived SFHs from the full-spectrum-fitting approach. For these reasons, we will use hereafter the IMF slope that corresponds to the velocity dispersion of each galaxy, according to the relation from \citet{LaBarbera2013} for the unimodal shape case, i.e. considering a single-power law slope throughout the whole stellar mass range. A unimodal slope of 1.3 represents the Salpeter case in these models . The slope employed in each case is as specified in Table 3. For robustness, we have studied how the estimated parameters would change if assuming a universal IMF. We find a small impact on the derived ages of the galaxies at intermediate redshift (less than 1$\%$, see Appendix B), due to the fact that we are avoiding the old models that are more dependent on age variations \citep{Ferre-Mateu2013}.\\
The derived SFHs for all our ETGs in RX\,J0152.7-1357 are presented in Appendix B. Three types are clearly distinguishable, as illustrated in Figure 3: \textit{(i)} galaxies with a single burst-like episode of star formation with completely old ages ($\sim$\,6\,-7\,Gyr, Pop$\_$O hereafter); \textit{(ii)} galaxies with a single burst-like episode of star formation with intermediate ages ($\sim$\,2\,-4\,Gyr, Pop$\_$M); \textit{(iii)} and galaxies showing an extended star formation episode or with a non-negligible contribution from a recent burst of star formation on top of an old/intermediate burst (Pop$\_$E). Table 3 lists the mean light- and mass-weighted ages and the total metallicities derived from this approach as well as the SFH type for each galaxy.

\begin{table*}
\centering 
\caption{Mean luminosity- and mass-weighted ages and total metallicities from the full-spectral-fitting}        
\label{table:3}     
\centering                      
\begin{tabular}{c| c c c c c c}   
\hline\hline      
ID & IMF slope & Age$\_L$ & [Z/H]$\_L$ & Age$\_M$ & [Z/H]$\_M$ &SFH$\_$type \\  
    & (unimodal) & (Gyr)       & dex            &    (Gyr)       &  dex           &                       \\
\hline  
338  & 0.8 & 2.1$\pm$\,0.4 & 0.14$\pm$\,0.08 & 2.9$\pm$\,0.5 & 0.13$\pm$\,0.08 & Pop$\_$E  \\   
346  & 1.3 & 4.8$\pm$\,0.7 &-0.04$\pm$\,0.07 & 5.4$\pm$\,0.8 &-0.08$\pm$\,0.07 & Pop$\_$E  \\   
422  & 0.8 & 2.2$\pm$\,0.3 & 0.17$\pm$\,0.07 & 2.5$\pm$\,0.4 & 0.20$\pm$\,0.07 & Pop$\_$E \\   
523  & 1.8 & 3.7$\pm$\,0.3 & 0.16$\pm$\,0.05 & 3.6$\pm$\,0.3 & 0.16$\pm$\,0.05 & Pop$\_$M \\   
566  & 1.0 & 2.1$\pm$\,0.3 & 0.13$\pm$\,0.07 & 2.4$\pm$\,0.3 & 0.12$\pm$\,0.07 & Pop$\_$E  \\   
627  & 1.3 & 6.7$\pm$\,1.2 &-0.12$\pm$\,0.08 & 6.8$\pm$\,1.2 &-0.13$\pm$\,0.08 & Pop$\_$E  \\   
766  & 1.8 & 5.6$\pm$\,0.8 & 0.08$\pm$\,0.07 & 5.4$\pm$\,0.8 & 0.12$\pm$\,0.07 & Pop$\_$M  \\   
776  & 1.0 & 3.5$\pm$\,0.4 & 0.06$\pm$\,0.06 & 4.7$\pm$\,0.6 &-0.02$\pm$\,0.06 & Pop$\_$E  \\   
813$^{\diamond}$  & 1.8 & 2.9$\pm$\,0.3 & 0.19$\pm$\,0.06 & 3.3$\pm$\,0.3 & 0.22$\pm$\,0.06 & Pop$\_$M \\   
908  & 1.8 & 7.9$\pm$\,0.8 &-0.58$\pm$\,0.06 & 7.9$\pm$\,0.8 &-0.56$\pm$\,0.06 & Pop$\_$O  \\   
1027 & 1.8 & 7.0$\pm$\,1.2 & 0.00$\pm$\,0.08 & 7.0$\pm$\,1.2 & 0.06$\pm$\,0.08 & Pop$\_$O  \\   
1085$^{\diamond}$ & 1.8 & 4.7$\pm$\,0.5 & 0.20$\pm$\,0.06 & 5.9$\pm$\,0.6 & 0.22$\pm$\,0.06 & Pop$\_$E  \\   
1110$^{\diamond}$ & 1.3 & 3.2$\pm$\,0.5 & 0.20$\pm$\,0.07 & 3.3$\pm$\,0.5 & 0.22$\pm$\,0.07 & Pop$\_$M \\   
1159 & 1.0 & 3.8$\pm$\,0.5 &-0.00$\pm$\,0.07 & 4.7$\pm$\,0.6 &-0.09$\pm$\,0.07 & Pop$\_$E  \\   
1210 & 1.0 & 2.8$\pm$\,0.5 &-0.20$\pm$\,0.08 & 2.9$\pm$\,0.5 &-0.20$\pm$\,0.08 & Pop$\_$E  \\   
1299 & 1.3 & 2.7$\pm$\,0.4 & 0.17$\pm$\,0.08 & 3.8$\pm$\,0.6 & 0.20$\pm$\,0.08 & Pop$\_$E \\   
1458$^{\diamond}$ & 1.0 & 2.1$\pm$\,0.4 & 0.20$\pm$\,0.09 & 2.2$\pm$\,0.4 & 0.22$\pm$\,0.09 & Pop$\_$M \\   
1507 & 1.3 & 1.8$\pm$\,0.3 & 0.17$\pm$\,0.09 & 1.9$\pm$\,0.4 & 0.18$\pm$\,0.09 & Pop$\_$M \\   
1567 & 1.3 & 3.2$\pm$\,0.6 &-0.21$\pm$\,0.08 & 3.3$\pm$\,0.6 &-0.20$\pm$\,0.08 & Pop$\_$M \\   
1590$^{\diamond}$ & 1.3 & 3.2$\pm$\,0.5 & 0.19$\pm$\,0.07 & 3.6$\pm$\,0.5 & 0.22 $\pm$\,0.07 & Pop$\_$M \\   
1614 & 1.8 & 3.0$\pm$\,0.4 & 0.19$\pm$\,0.07 & 3.1$\pm$\,0.4 & 0.19$\pm$\,0.07 & Pop$\_$M \\   
1682 & 1.3 & 3.1$\pm$\,0.4 & 0.12$\pm$\,0.06 & 3.1$\pm$\,0.4 & 0.12$\pm$\,0.06 & Pop$\_$E \\   
1811 & 0.8 & 7.9$\pm$\,1.7 & 0.14$\pm$\,0.09 & 7.9$\pm$\,1.7 & 0.18$\pm$\,0.09 & Pop$\_$O  \\   
1935 & 1.3 & 3.4$\pm$\,0.5 & 0.15$\pm$\,0.07 & 3.5$\pm$\,0.5 & 0.15$\pm$\,0.07 & Pop$\_$E  \\   
\hline                                                            
\end{tabular}\\
{Galaxy ID (column 1) and the IMF slope considered for each of them (column 2). Columns 3 to 6 show the mean luminosity- and mass-weighted ages and total metallicities derived from the full-spectrum fitting code {\tt STARLIGHT}. The type of SFH is specified in column 7 as Pop$\_$O (old burst), Pop$\_$M (intermediate-age burst) or Pop$\_$E (extended or residual SFH). Galaxies labeled with $^{\diamond}$ are those that saturate on metallicity and the employed IMF slope is stated on column 2.}
\end{table*}   

\begin{figure}
\centering
\label{figure:3}
        \includegraphics[scale=0.49]{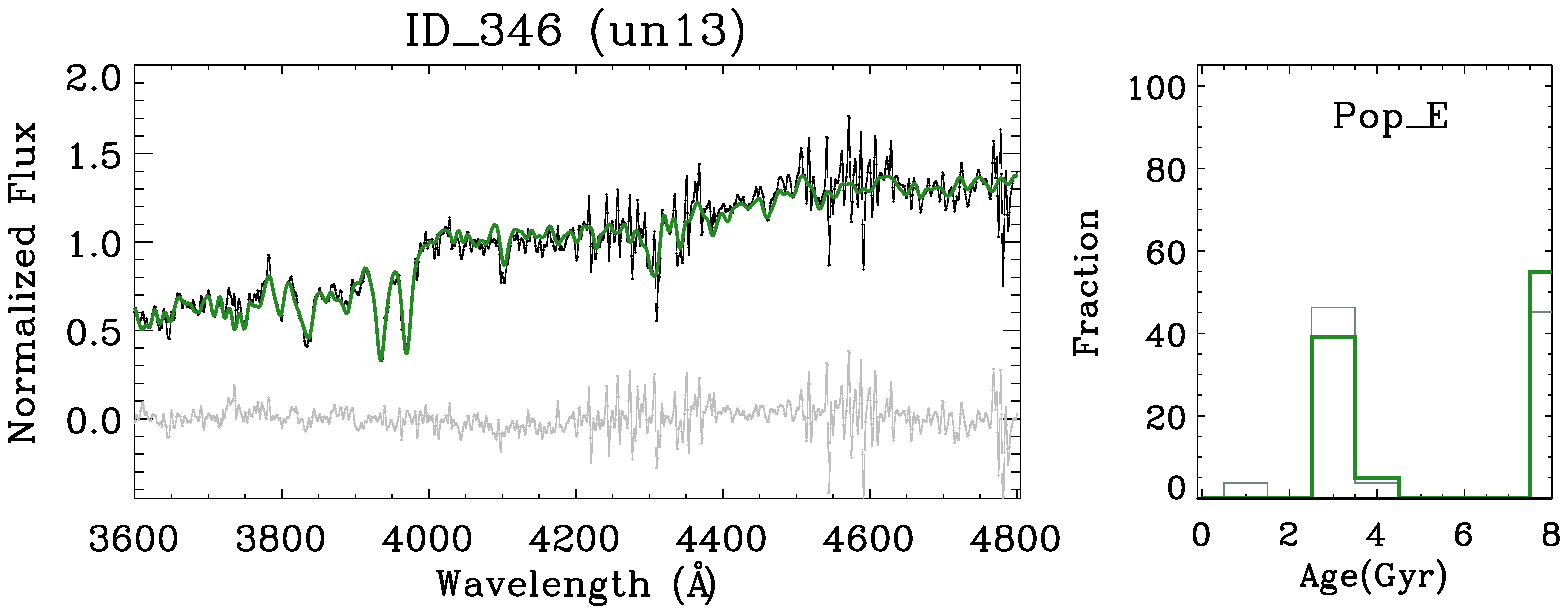}  
        \includegraphics[scale=0.49]{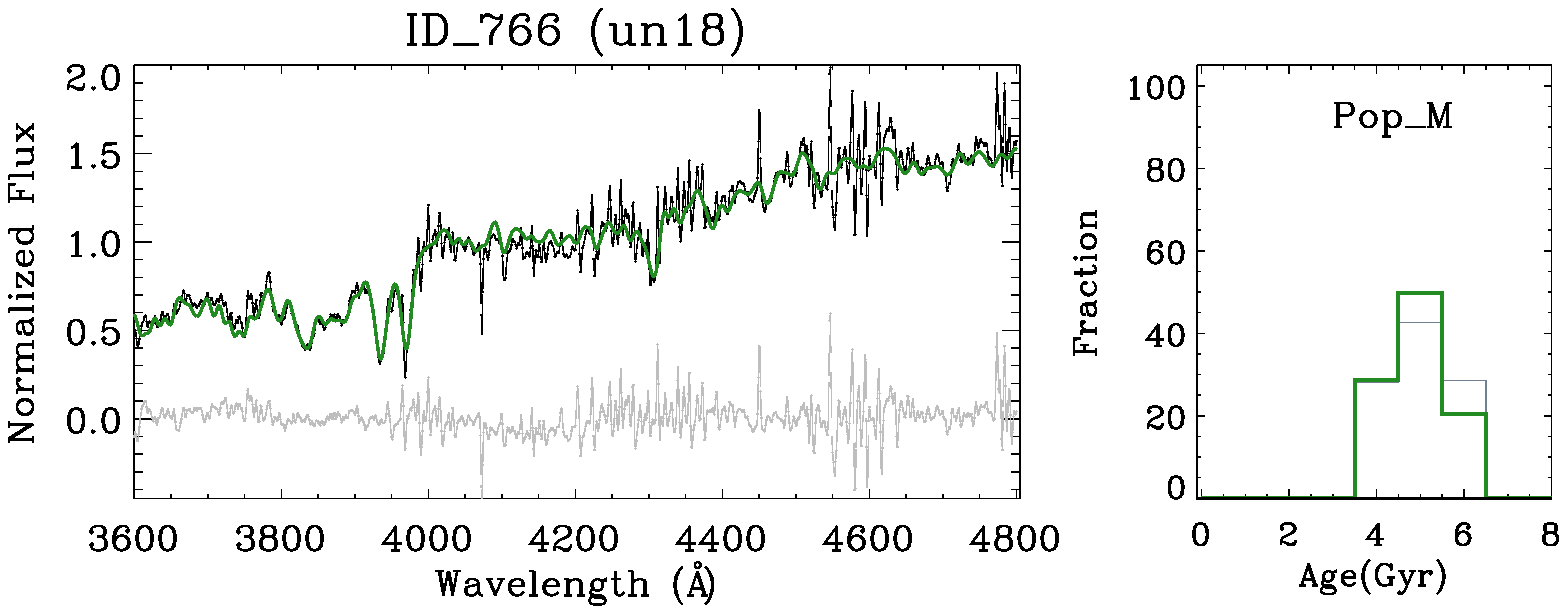} 
        \includegraphics[scale=0.49]{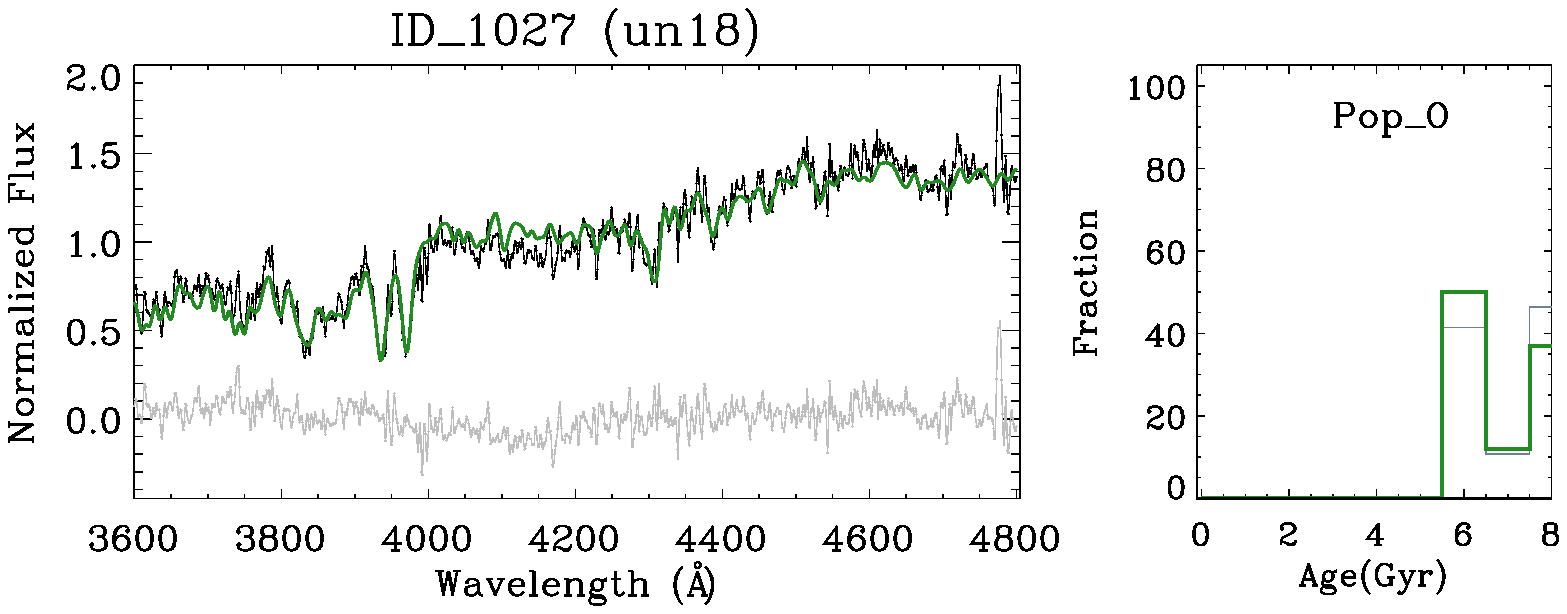}  
\caption{Three representative galaxy spectra (black) are plotted on the left panels, together with the best fit from {\tt STARLIGHT} (green) and their residuals (gray). The IMF slope adopted is stated on top of each panel. We plot in green the histograms corresponding to the mass-weighted SFHs whereas in gray we plot the luminosity-weighted SFHs (right panels). On each panel, we state the three different types of SFHs, classified according to the age of the dominant burst. The rest of galaxies can be found in Appendix B1.} 
\end{figure} 
                
\subsection{Combined method}
The various index-index diagrams presented on Appendix A2 show that these pairs of indices do not provide orthogonal grids. Moreover, galaxies with non-solar abundance patterns fall outside the model grids. This all makes difficult to extrapolate the model prediction grids to obtain reliable estimates. We present here a new approach where we remove to a great extent the dependence on the age. This new hybrid approach \citep{LaBarbera2013} combines the luminosity-weighted age that is derived from the full-spectrum-fitting technique (on the y-axis) with the model predictions from the metallicity indicators (on the x-axis). The majority of our galaxies fall now inside this nearly orthogonal hybrid grids, as seen in Figure 4. This allows for better inter/extrapolations that result in more reliable metallicities (see Table 4). The robustness of this method was tested by comparing these estimates with those few obtained from the classical index-index diagrams  (Appendix A, see also La Barbera et al. (2013) for a more extensive description).\\
Because the elements are produced in the stars on different timescales, they encode crucial information about the SFH of each galaxy (e.g. \citealt{Peletier1989}; \citealt{Worthey1992}; \citealt{Vazdekis2001b}; PSB06). The difference between the metallicity from a metallic-element \textit{A} with respect to the metallicity derived from a Fe-sensitive one, such as Fe4383, [Z$_{A}$/Z$_{Fe}$], has been shown to be a good proxy for these abundance ratios (\citealt{Carretero2004}; \citealt{Yamada2006}, V10, \citealt{LaBarbera2013}). With this new approach, we can also obtain more robust abundance ratios from the derived metallicities. We thus derive [C/Fe]$\sim$\,0.37\,dex and [CN/Fe]$\sim$\,0.3\,dex (Table 4) by averaging the derived values for galaxies in the velocity dispersion range 150\,-\,250\,$\rm{km\,s^{-1}}$. This range is selected as these abundance patters are known to depend on the velocity dispersion \citep{Carretero2004}. Despite the bad fitting in the CN$_{2}$ region, we obtain similar ratios for the CN using both CN3883 and CN$_{2}$ metallic indices.

\begin{figure*}[H!]
\centering
\includegraphics[scale=0.77]{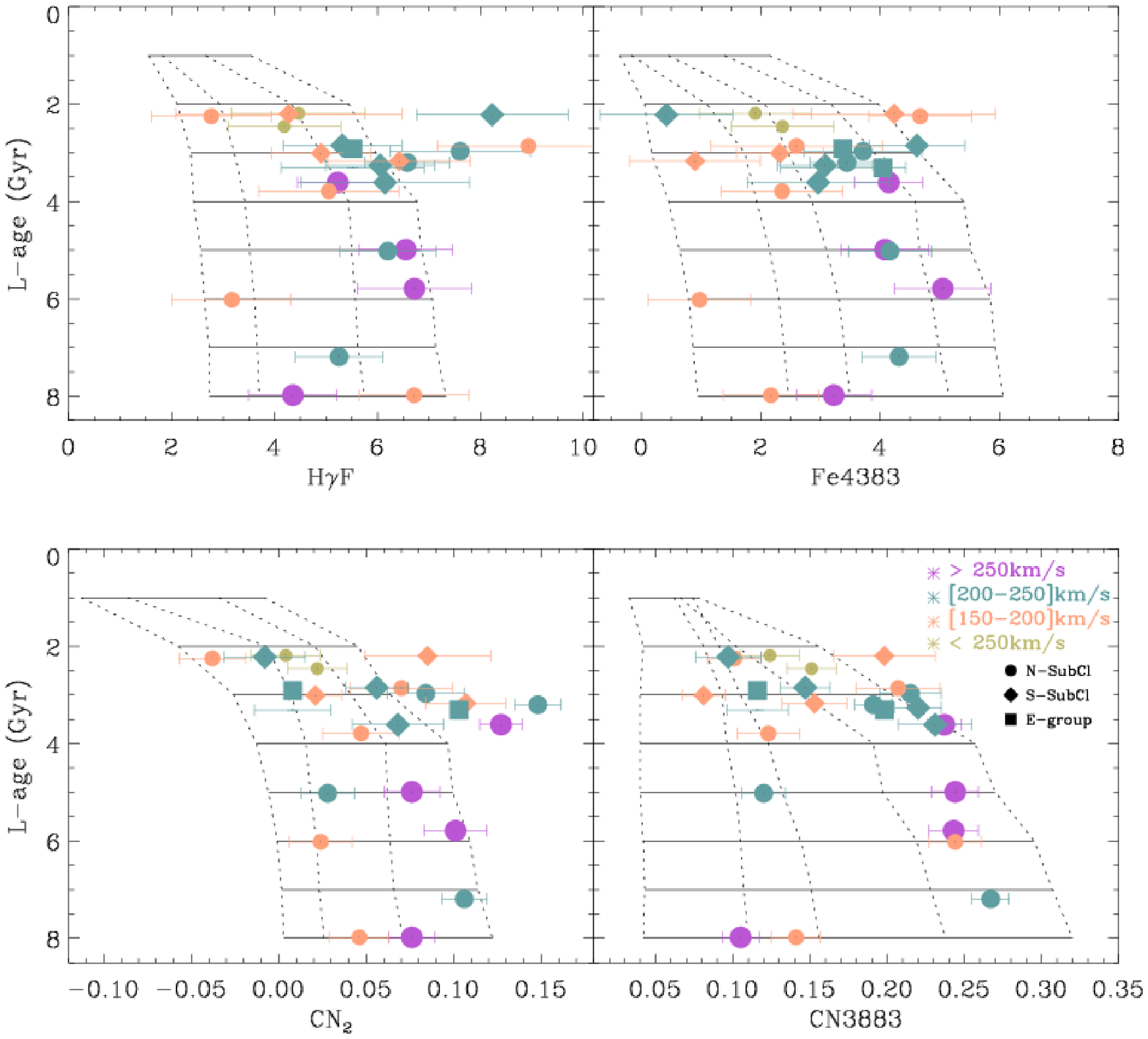}
\label{Figure:4}
\caption{Diagnostic diagrams where the derived mean luminosity-weighted ages from the full-spectrum-fitting are plotted against various metallicity indicators. The SSP model grids of V10 are plotted, with age (in Gyr) increasing from top to bottom as indicated in the labels, and metallicity from left to right ([Z/H]= -0.71, -0.40, 0.00, +0.22). Note that the resulting model grids are virtually orthogonal compared to the classical index-index grids of Fig. A2, which permits to extrapolate the metallicities and derive abundance patterns. The velocity dispersion of the galaxies is color-coded (as in Fig.1) and emphasized with sizes (larger symbols for larger velocity dispersions), while the symbols represent to which substructure the galaxy belongs to: circles for the N-SubCl, diamonds for the S-SubCl and squares for the E-group.}
\end{figure*} 

\begin{table*}
\centering    
\caption{Metallicities and abundance ratios from the hybrid method}
\label{table:4}                     
\begin{tabular}{r| r r r r r r r}   
\hline\hline      
ID &  [Z/H] & [Z/H]&  [Z/H] & [Z/H] &  [C/Fe] & [CN/Fe] & [CN/Fe]\\
 & C$_{2}$4668-$L\_age$ & Fe4383-$L\_age$ & CN$_{2}$-$L\_age$ & CN3883-$L\_age$ & from C$_{2}$4668 & from CN$_{2}$ & from CN3883\\
\hline             
338   & 0.00\,$^{+0.30}_{-0.38}$&-0.45\,$^{+0.35}_{-0.35}$&-0.35\,$^{+0.35}_{-0.25}$& 0.10\,$^{+0.05}_{-0.20}$& 0.45\,$^{+0.46}_{-0.51}$ & 0.10\,$^{+0.49}_{-0.43}$ & 0.55\,$^{+0.35}_{-0.40}$ \\ 
346   & 0.10\,$^{+0.20}_{-0.20}$&-0.15\,$^{+0.25}_{-0.20}$&-0.35\,$^{+0.20}_{-0.15}$& 0.55\,$^{+0.17}_{-0.10}$& 0.25\,$^{+0.32}_{-0.28}$ &-0.20\,$^{+0.32}_{-0.25}$ &-0.40\,$^{+0.30}_{-0.20}$ \\
422   &-0.10\,$^{+0.28}_{-0.30}$&-0.40\,$^{+0.31}_{-0.27}$&-0.10\,$^{+0.25}_{-0.25}$& 0.10\,$^{+0.10}_{-0.05}$& 0.30\,$^{+0.41}_{-0.40}$ & 0.30\,$^{+0.39}_{-0.36}$ & 0.51\,$^{+0.35}_{-0.27}$ \\
523   & 0.00\,$^{+0.18}_{-0.15}$& 0.00\,$^{+0.15}_{-0.20}$&   -                     & 0.22\,$^{+0.17}_{-0.17}$& 0.00\,$^{+0.23}_{-0.25}$ &  -                       & 0.22\,$^{+0.22}_{-0.26}$ \\
627   &-0.35\,$^{+0.35}_{-0.45}$&-1.30\,$^{+0.55}_{-0.70}$&-0.40\,$^{+0.25}_{-0.28}$& 0.10\,$^{+0.05}_{-0.10}$& 0.75\,$^{+0.65}_{-0.83}$ & 0.90\,$^{+0.60}_{-0.75}$ & 0.31\,$^{+0.55}_{-0.70}$ \\
766   & 0.18\,$^{+0.22}_{-0.18}$& 0.05\,$^{+0.20}_{-0.20}$& 0.20\,$^{+0.10}_{-0.05}$& 0.10\,$^{+0.08}_{-0.05}$& 0.13\,$^{+0.29}_{-0.26}$ & 0.15\,$^{+0.22}_{-0.20}$ & 0.05\,$^{+0.21}_{-0.20}$ \\
776   &-0.10\,$^{+0.26}_{-0.14}$&-0.55\,$^{+0.25}_{-0.20}$&-0.10\,$^{+0.15}_{-0.10}$&-0.40\,$^{+0.10}_{-0.15}$& 0.45\,$^{+0.36}_{-0.24}$ & 0.45\,$^{+0.29}_{-0.22}$ & 0.15\,$^{+0.26}_{-0.25}$ \\
813   &  -                      & 0.00\,$^{+0.15}_{-0.20}$& 0.40\,$^{+0.10}_{-0.10}$& 0.30\,$^{+0.05}_{-0.08}$&  -                       & 0.40\,$^{+0.18}_{-0.22}$ & 0.30\,$^{+0.15}_{-0.21}$ \\
908   &-0.30\,$^{+0.20}_{-0.15}$&-0.45\,$^{+0.15}_{-0.30}$& 0.02\,$^{+0.08}_{-0.07}$&-0.75\,$^{+0.15}_{-0.05}$& 0.15\,$^{+0.25}_{-0.33}$ & 0.43\,$^{+0.17}_{-0.30}$ &-0.30\,$^{+0.21}_{-0.30}$ \\
1027  &-0.10\,$^{+0.28}_{-0.25}$&-0.20\,$^{+0.25}_{-0.25}$& 0.18\,$^{+0.12}_{-0.03}$& 0.10\,$^{+0.08}_{-0.05}$& 0.10\,$^{+0.37}_{-0.35}$ & 0.38\,$^{+0.27}_{-0.25}$ & 0.30\,$^{+0.26}_{-0.25}$ \\
1085  & 0.15\,$^{+0.17}_{-0.10}$&-0.15\,$^{+0.15}_{-0.15}$& 0.10\,$^{+0.08}_{-0.10}$& 0.18\,$^{+0.10}_{-0.06}$& 0.30\,$^{+0.22}_{-0.18}$ & 0.25\,$^{+0.17}_{-0.18}$ & 0.33\,$^{+0.15}_{-0.16}$ \\ 
1110  & 0.30\,$^{+0.20}_{-0.25}$&-0.15\,$^{+0.25}_{-0.25}$&   -                     & 0.18\,$^{+0.04}_{-0.08}$& 0.45\,$^{+0.32}_{-0.35}$ &   -                      & 0.33\,$^{+0.25}_{-0.26}$ \\
1299  &  -                      &-0.40\,$^{+0.20}_{-1.20}$& 0.30\,$^{+0.10}_{-0.20}$& 0.30\,$^{+0.10}_{-0.10}$&   -                      & 0.70\,$^{+0.29}_{-0.29}$ & 0.70\,$^{+0.29}_{-0.24}$ \\
1458  & 0.00\,$^{+0.30}_{-0.50}$& 0.22\,$^{+0.20}_{-0.20}$&  -                      &  -                      &-0.22\,$^{+0.42}_{-0.35}$ &   -                      &   -                      \\
1507  &   -                     &-1.20\,$^{+0.20}_{-1.20}$&-0.40\,$^{+0.40}_{-0.25}$&-0.30\,$^{+0.35}_{-0.60}$&   -                      & 0.80\,$^{+0.44}_{-1.22}$ & 0.90\,$^{+0.40}_{-1.34}$ \\ 
1567  & 0.20\,$^{+0.30}_{-0.30}$&-0.30\,$^{+0.30}_{-0.30}$&   -                     & 0.22\,$^{+0.08}_{-0.04}$& 0.50\,$^{+0.42}_{-0.42}$ &   -                      & 0.55\,$^{+0.31}_{-0.30}$ \\
1590  & 0.30\,$^{+0.20}_{-0.20}$&-1.00\,$^{+0.30}_{-0.40}$&   -                     & 0.00\,$^{+0.10}_{-0.10}$& 1.30\,$^{+0.36}_{-0.44}$ &   -                      & 1.00\,$^{+0.31}_{-0.41}$ \\
1614  &   -                     & 0.00\,$^{+0.20}_{-0.20}$&   -                     & 0.18\,$^{+0.02}_{-0.08}$&   -                      &   -                      & 0.18\,$^{+0.20}_{-0.21}$ \\
1682  & 0.12\,$^{+0.18}_{-0.22}$&-0.10\,$^{+0.20}_{-0.20}$&-0.40\,$^{+0.20}_{-0.10}$&-0.35\,$^{+0.20}_{-0.10}$& 0.22\,$^{+0.26}_{-0.29}$ &-0.30\,$^{+0.28}_{-0.22}$ &-0.25\,$^{+0.28}_{-0.22}$ \\
1811  & 0.15\,$^{+0.55}_{-0.45}$&-0.80\,$^{+0.45}_{-0.70}$&-0.20\,$^{+0.25}_{-0.35}$&-0.50\,$^{+0.20}_{-0.20}$& 0.95\,$^{+0.71}_{-0.83}$ & 0.60\,$^{+0.51}_{-0.78}$ & 0.30\,$^{+0.49}_{-0.72}$ \\
1935  & 0.20\,$^{+0.30}_{-0.20}$&-0.35\,$^{+0.20}_{-0.25}$& 0.10\,$^{+0.12}_{-0.10}$& 0.22\,$^{+0.03}_{-0.04}$& 0.55\,$^{+0.36}_{-0.32}$ & 0.45\,$^{+0.23}_{-0.26}$ & 0.57\,$^{+0.20}_{-0.25}$ \\
\hline                                                                                                              
\end{tabular}\\

{Metallicities obtained with the new hybrid method that combines a metallic-index indicator with the luminosity-weighted age from the full-spectral-fitting. Different different metallicity indices are used (columns 2-5) and their abundance ratios are derived from the proxy (columns 6-8).}               
\end{table*}

\section{Distribution of the stellar populations within the cluster}
We then study the distribution of the previous stellar population properties within the cluster to find any existing relation as a function of local environment. The top-left panel of Figure 5 shows that our galaxies exhibit, in general, ages of $\sim\,$3\,-4\,Gyr, which is in agreement with the color-based estimates from \citet{Blakeslee2006}, although the center of the N-SubCl contains the galaxies with the oldest ages. The small range in metallicities derived for our galaxies (top-right panel) only shows that except for three galaxies (out of 24) being metal poor, the rest of galaxies have already reached the high metallicites observed in low-z clusters.\\
The bottom panel is the most relevant, showing the distribution of the galaxies according to the three different types of SFH described in Sect. 4.2. It highlights that galaxies with different SFHs clearly populate different regions of the cluster. This is in agreement with the general trends of \citet{Demarco2010}. These authors studied the SFHs of a set of stacked spectra (combined by different properties) related to the projected angular distribution. However, as mentioned above, in this paper we are studying the cluster galaxies individually for the first time, hence we can now trace back the history of each of them and relate it to the local environment. Our results show that the N-SubCl is generally populated by galaxies with SFhs classified as Pop$\_$O. The ones classified as Pop$\_$M mostly populate the S-SubCl, and those showing extended SFHs, Pop$\_$E galaxies, are preferably located at the outskirts of both subclumps. If we assume that the star formation is truncated as galaxies fall into the cluster, we can roughly estimate the cluster substructures formation timescales from their last episode of star formation. The N-SubCl was formed first and stopped forming stars in an early epoch. Approximately 3\,Gyr later, the S-SubCl was formed following similar timescales. The outskirts of the cluster were populated by the later accretion of new galaxies, $\sim$\,4\,Gyr after the initial star formation event.\\
In addition, there seems to be a second dependence of the SFH type, this time with the velocity dispersion of the galaxies: the most massive ones tend to be classified as Pop$\_$O type, intermediate-mass galaxies tend to be Pop$\_$M and the galaxies with the smaller velocity dispersions are those with Pop$\_$E SFHs. We will further extend on this issue in Section 6.

\begin{figure*}
\centering
\includegraphics[scale=0.36]{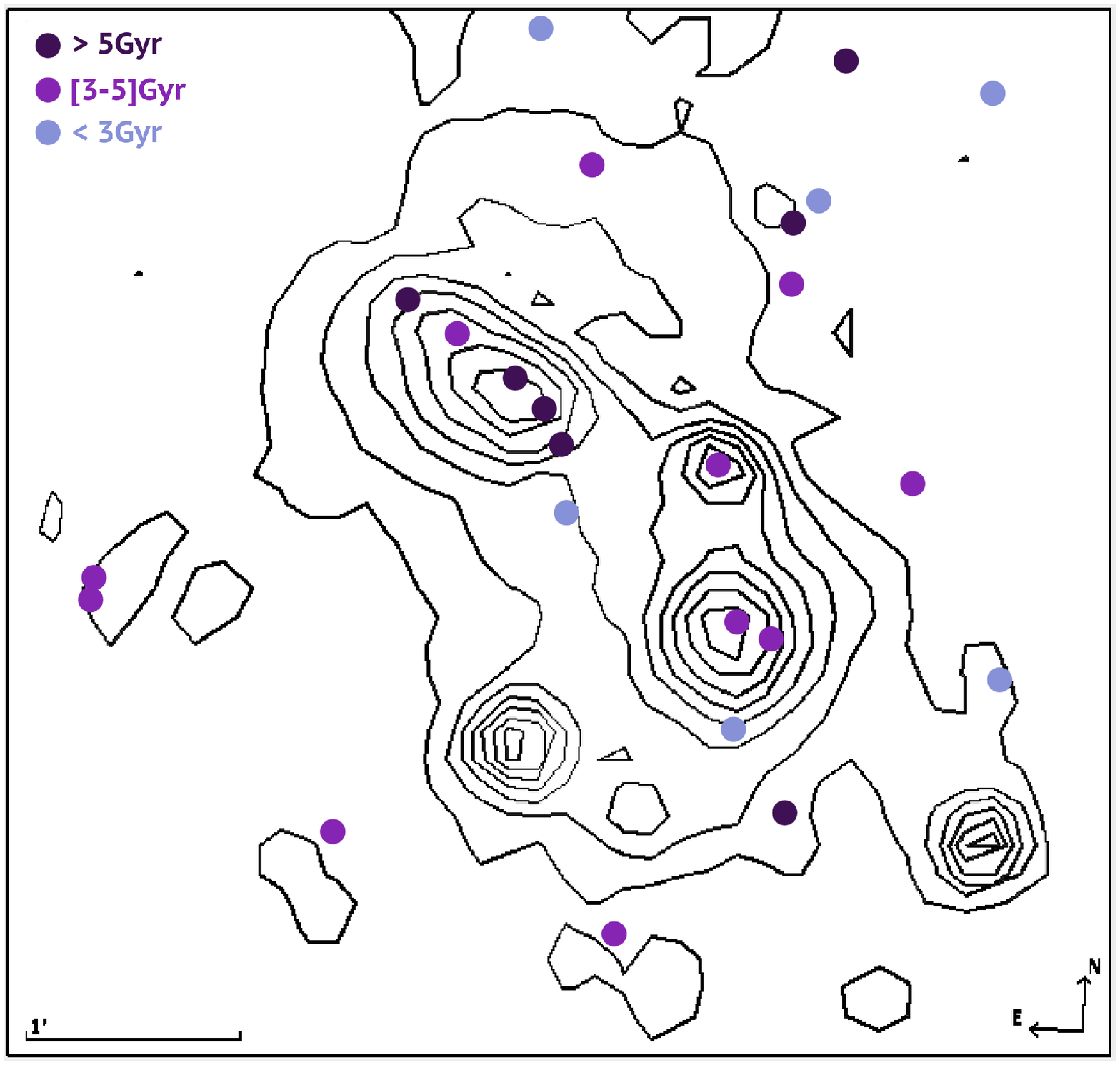}
\includegraphics[scale=0.36]{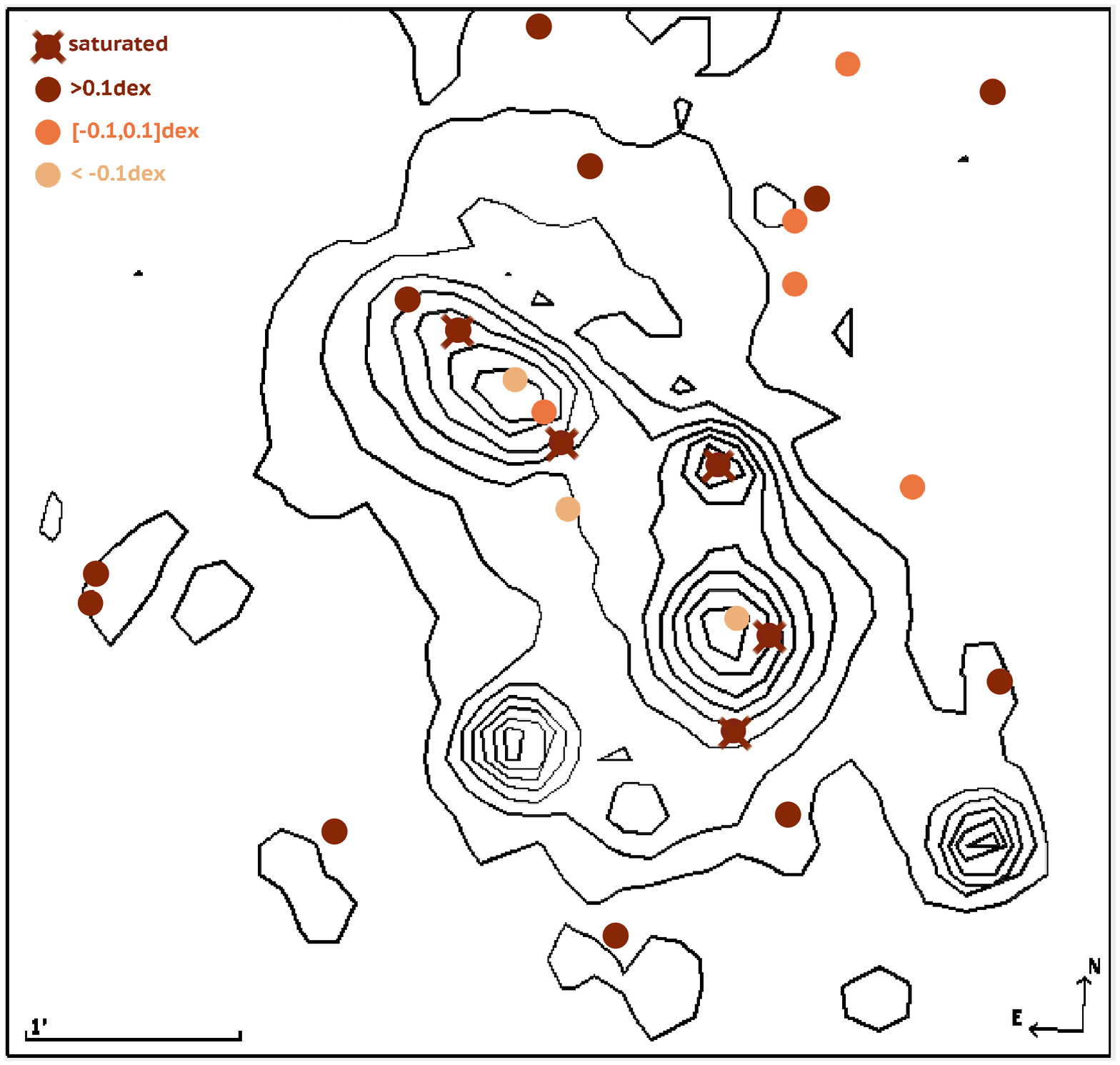}\\
\includegraphics[scale=0.53]{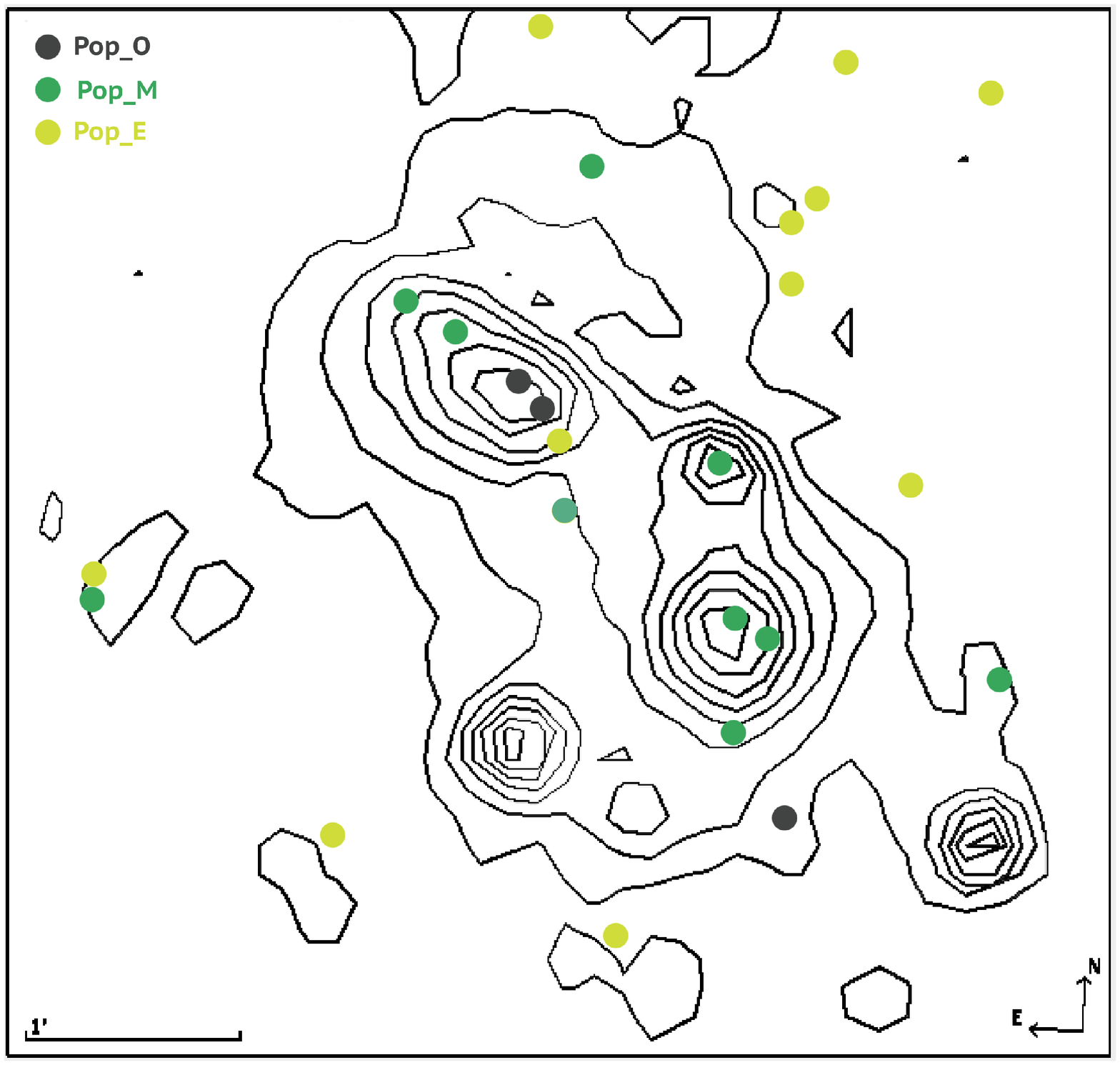}
\label{figure:5}
\caption{Galaxy stellar populations and their location within the cluster, color-coded as specified in each panel. Contours represent the X-ray data like in Figure 1. \textit{Top-left panel}: mean-mass weighted ages from {\tt STARLIGHT}; \textit{top-right panel}: mass-weighted metallicities including the galaxies which saturate in metallicity (crossed circles, marked with a diamond on Table\,3); \textit{bottom panel}: different SFH types, color-coded depending on their main burst of formation: grey for the Pop$\_$O, green for the Pop$\_$M and light green for the Pop$\_$E.}
\end{figure*} 

\section{Evolution through cosmic time}
To pose further constrains on galaxy evolution theories, we investigate in this section how the relations between the velocity dispersion and the stellar population properties derived in the previous sections vary over cosmic time. We want to see if the stellar populations of the ETGs in our intermediate-z cluster could evolve into the ones we see in todays cluster ETGs. In other words, we want to see if the stellar populations of the ETGs in RX\,J0152.7-1357 could evolve into those we see in the ETGs of analog local clusters, such as Coma. For this purpose, we use data from the Coma cluster (see Sec. 2.2.1). In particular, we artificially evolve back in time the observed local relations to the redshift of RX\,J0152.7-1357 and see if they match our observations at this redshift. We assume a passive evolution scenario, where all stars were formed in a single-burst at a redshift of formation of \textit{z$_{f}$}=\,3 (to be consistent with previous studies such as e.g. \citealt{Demarco2010}). Therefore, we stress here that when we discuss the passive evolution scenario this will always refer to the expected behavior \textit{with respect to} Coma. 

\subsection{Index-$\sigma$ relations}
Index-$\sigma$ relations help to disentangle the age-metallicity degeneracy, as each index has a different sensitivity to it (e.g. J05, PSB06, PSB09, \citealt{Harrison2011}). In fact, indices sensitive to the metallicity are positively correlated, while those related to the age anticorrelate (e.g. \citealt{Bender1993}; \citealt{Jorgensen1999}; \citealt{Kuntschner2000}; \citealt{Bernardi2003}; \citealt{Caldwell2003}, PSB06). Figure 6 shows the index-$\sigma$ relations for some relevant line indices of our cluster ETGs at z$\sim$0.83 (circles, color coded as in Fig.\,5 by their SFH type) and for the ones in Coma (blue triangles). The solid line shows the linear fit to the data, while the dashed line shows the expected relation for Coma at z\,=\,0.83 (assuming passive evolution and a \textit{z$_{f}$}=\,3). The indices D4000, CN$_{2}$ and C$_{2}$4668 all show relations compatible with a passive evolution of the stellar populations. In fact, we find a positive relation for the D4000 break (e.g. \citealt{Barbaro1997}, PSB09), in contrast to J05, where no correlation was found. However, we find that in general the remaining indices are compatible with a passive evolution \textit{only} for the most massive galaxies. We see a large dispersion for CN3883, H$_{\delta}$F or Fe3883. These dispersions, in particular the extremely low values seen for Fe4383, have been previously reported to be a real effect and not a consequence of a large data scatter (e.g. J05, PSB09). The latter authors hypothesized that less massive galaxies would need a more extended SFH. We can now confirm this trend, showing that all those galaxies deviating from the expected relation are those presenting extended or residual SFHs (Pop$\_$E). This is all compatible with a downsizing scenario (\citealt{Cowie1996}; \citealt{Treu2005}; \citealt{Peng2010}), where the properties of most massive galaxies would be settled in the very early Universe (z$>$1) whereas less massive galaxies would still be evolving down to the current time.

\begin{figure*}
\centering
\includegraphics[scale=0.8]{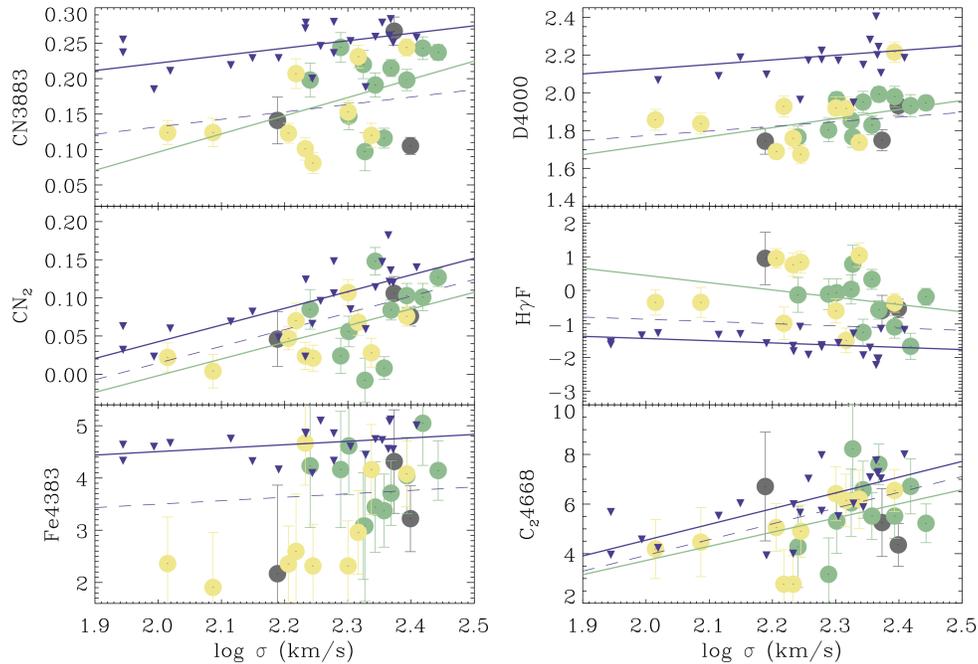}
\label{figure:6}
\caption{Relation between the measured line-strength indices and the velocity dispersion for the galaxies in Coma (blue triangles), and RX\,J0152.7-1357 (circles, color-coded by their SFH type as in Fig.\,5). The solid lines correspond to the linear fits to the data. The dashed line represents the expected relation from evolving the z=\,0 linear fit to the redshift of RX\,J0152.7-1357 assuming a \textit{z$_{f}$}=\,3. Note that the most massive galaxies are always compatible with passive evolution.}
\end{figure*} 

\subsection{Mean ages and metallicities} 
Figure 7 shows the dependence of the age and the metallicity (from the full-spectral-fitting approach, both luminosity- and mass-weighted) with the velocity dispersion. Galaxies from the intermediate-z cluster are again color-coded based on their SFH type. The age-$\sigma$ relation for the intermediate-z cluster is virtually flat, although a weak positive trend is seen with a large scatter (as also seen in e.g. \citealt{Proctor2004}; PSB09, \citealt{Rettura2010}; \citealt{Harrison2011}). Note however, that the ages for Coma seem to decrease with velocity dispersion. This is because we are missing the most massive galaxies in PSB06s original sample (see fig.\,2 in \citealt{Sanchez-Blazquez2006c}) as we selected them to cover the same $\sigma$ range as in our cluster.  PSB06 reported a mild positive correlation between age and $\sigma$ (also seen in \citealt{Trager2000}), as we do for RX\,J0152.7-1357. To guide the eye, the dashed lines correspond to the mean value obtained for each parameter, whereas the solid line indicates the age difference of the Universe between the redshifts of the two depicted clusters. The difference in the mean ages ($\sim$\,4\,Gyr \textit{vs} $\sim$\,12\,Gyr) is compatible with the lookback time corresponding to the redshift of the cluster. From the lower panels in Figure 7, it is seen that the total metallicity does not evolve within this redshift interval. We also find a positive relation with velocity dispersion, in the sense that more massive galaxies show a higher total metallicity (e.g. \citealt{Greggio1997}; \citealt{Thomas2005}, PSB09, \citealt{Harrison2011}).\\
Altogheter, this is compatible with a scenario where the most massive galaxies, located at the center of the cluster, have undergone passive evolution since their early formation, while the low-mass galaxies at the outskirts have suffered a more extended SFH, most likely related to their posterior infall into the cluster. This is in agreement with other studies that suggest that not all cluster galaxies were fully in place at z$\sim$\,1 (e.g. \citealt{DeLucia2004}; \citealt{DeLucia2007}; \citealt{Kodama2004}).

\begin{figure*}
\centering
\includegraphics[scale=1.0]{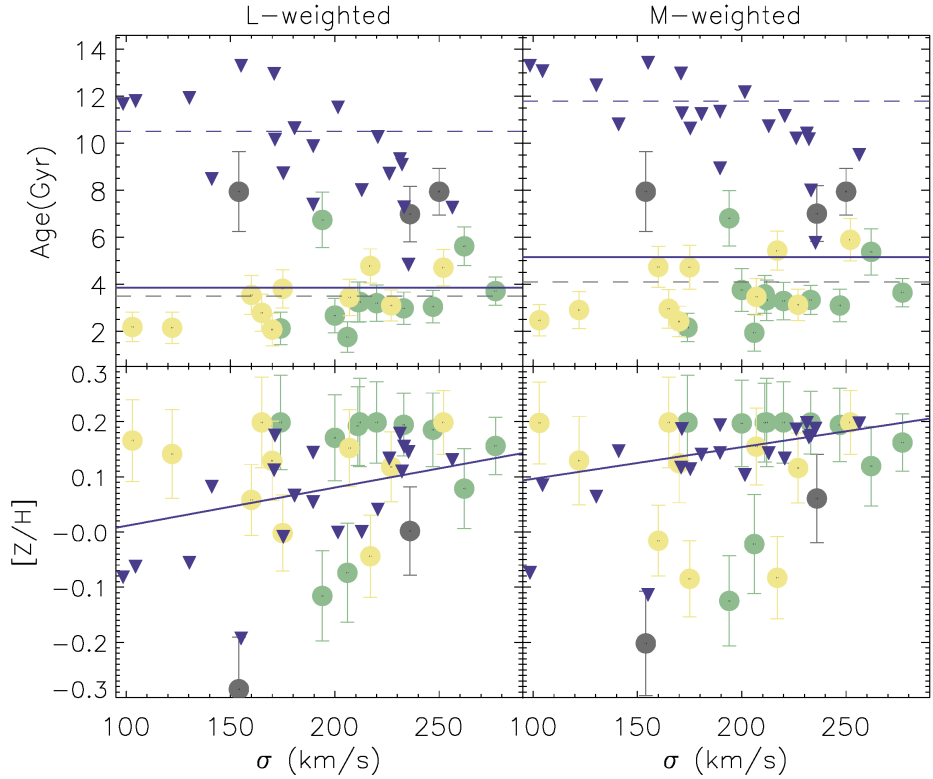}
\label{figure:7}
\caption{Relation between the mean ages and metallicities derived from the full-spectrum-fitting approach and the velocity dispersion.  Coma galaxies are marked as blue triangles, while the galaxies in RX\,J0152.7-1357 are the circles color-coded by their SFH type as in Fig.\,5. On the upper panels, the dashed lines correspond to the mean age for each cluster, whereas the solid line indicates the expected evolution for Coma. The difference between both clusters mean ages is compatible with the expected evolution over cosmic time assuming a passive evolution scenario. Note in the lower panels that the total metallicity is similar for the two clusters, showing a mild trend with the velocity dispersion, as indicated by the solid line.}
\end{figure*}

\subsection{Abundance patterns}
The abundances of C and CN have been reported to be related to the environment and are interpreted as different SFHs and formation timescales for each element (e.g. \citealt{Sanchez-Blazquez2003}; \citealt{Carretero2004}; \citealt{Sanchez-Blazquez2006c}). We have studied how these abundance patterns relate to velocity dispersion (instead of L$_{X}$) for both clusters, as shown in Figure 8. The pattern in the intermediate-z cluster shows a larger scatter compared to Coma due to the lower S/N of our spectra, but within the errors, both clusters show similar mean values (marked by the dashed lines). This has two implications. On one hand, this result suggests that the ETGs in both clusters settled on rather similar timescales because the abundance patterns are thought to be chemical clocks for clusters of similar densities \citep{Carretero2004}.  On the other hand, this also shows that the abundance pattern of the CN and the C are already at place at z$\sim$0.8 and therefore no significant evolution of these properties is expected for the galaxies in this cluster. 

\begin{figure*}
\centering
\includegraphics[scale=0.78]{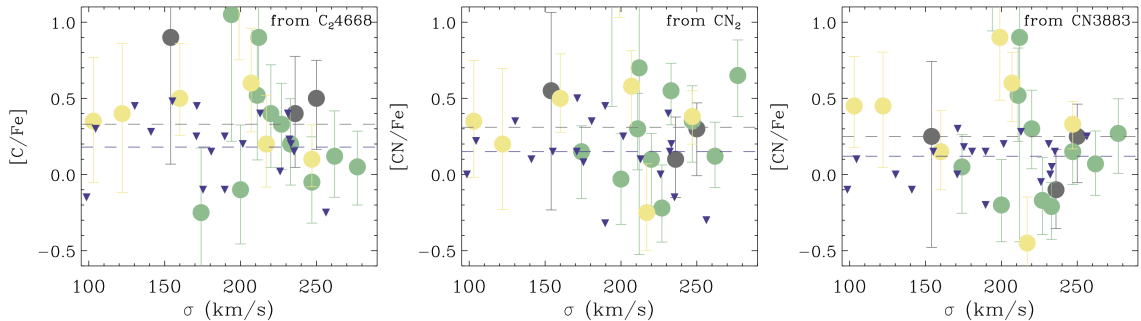}
\label{figure:8}
\caption{The abundance pattern for both clusters is plotted against the velocity dispersion, parametrized by SFH type as in the previous figure. Dashed lines show the mean abundance ratios for each cluster. It is seen that, like the total metallicity, the abundance pattern is already settled at the redshift of RX\,J0152.7-1357, suggesting that ETGs in both clusters where formed on similar timescales.}
\end{figure*}

\section{Discussion: the emerging picture}
This study has been devoted to find out if cluster ETGs at intermediate redshift could become those we see in analog clusters in the local Universe, investigating the evolutionary link between their stellar populations. The differences in the index-$\sigma$ and the stellar population parameters-$\sigma$ relations between RX\,J0152.7-1357 and Coma can be explained with an scenario where massive ETGs formed all their stars at z$\ge$3 and evolved passively since then. However, the low-mass galaxies in the cluster show a larger scatter for many of these relations, as it happens in the nearby Universe. These differences can be better understood through the SFHs of the galaxies. The low-mass galaxies tend to show more extended SFHs, probably due to their posterior incorporation into the cluster. The fact that our proxy for the abundance ratio provides very similar values for the two clusters studied here, indicates that the ETGs in both clusters were settled following similar timescales. Altogether, this supports as plausible the hypothesis of the stellar populations of the ETGs in RX\,J0152.7-1357 evolving into ETGs like the ones seen in Coma.\\
We find that the SFHs and local environment clearly narrate the story of the formation and subsequent evolution of this cluster substructures. The last episode of star formation, as seen from the SFHs, allows us to estimate the formation timescales of the cluster substructures. The center of the N-SubCl was formed first, as it contains the most massive ($\sigma\sim$\,260\,$\rm{km\,s^{-1}}$) and old galaxies ($\sim$\,6\,-7\,Gyr), which were formed in a single burst-like episode at high redshift. In a similar event but around 3\,Gyr later, the S-SubCl was formed, populated by massive ($\sigma\sim$\,230\,$\rm{km\,s^{-1}}$) and intermediate-age galaxies ($\sim$3\,-4\,Gyr). It is known that the N-SubCl is slowly moving towards the S-SubCl. Although a study of the dynamical evolution of the cluster is out of the scope of this paper, we estimate a merger time for these substructures of $\sim$2.3\,Gyr from Equation 5 in \citet{Boylan-Kolchin2008}. In the meantime, new low-mass galaxies were added at the outskirts of both subclumps, showing a more extended SFH that may be related to a posterior epoch of accretion into the cluster. Their last burst of formation occurred at $\sim$\,2\,Gyr, almost 4\,Gyr after the main epoch of star formation for the most massive galaxies at the centers of the subclumps.\\
However, we have to be careful when interpreting these results. First, several studies have found that massive galaxies evolve in size by a factor of $\sim$\,4 since z$\sim$\,2.5 (e.g. \citealt{Daddi2005}; \citealt{Trujillo2007}; \citealt{vanderWel2011}; \citealt{Toft2012}). They evolve over cosmic time by enlarging their sizes and changing their morphologies but without a substantial increase on their stellar masses. There is only a mild increase in the mass function from z\,$\sim$\,0.8 to z\,$\sim$\,0 (by a factor of $\sim$\,1.5, see e.g. \citealt{Wechsler2002}; \citealt{Cimatti2006}; \citealt{Marchesini2014}) that can be explained by the accretion of smaller satellites at the periphery of the massive galaxy (e.g. \citealt{Wuyts2010}, \citealt{Lopez-Sanjuan2012}; \citealt{Quilis2012}). However, if the reported trend of size with age at a fixed stellar mass is real (e.g. \citealt{Saracco2010}; \citealt{Cassata2011}; \citealt{Poggianti2013}), selecting galaxies that are already dead and passive at high redshift would imply a systematic selection of the most compact ones. Secondly, we have to be aware of the so called progenitor bias \citep{vanDokkum2001}. If new galaxies are continuously added onto the red sequence as cosmic time evolves (e.g. \citealt{DeLucia2004}; \citealt{Sanchez-Blazquez2009}; \citealt{Newman2012}; \citealt{Carollo2013}; \citealt{Cassata2013}) this implies that the red sequence population at high redshift does not contain all the progenitors of nearby red sequence population. Nonetheless, our results are consistent with a picture in which the massive red-sequence galaxies studied here do not suffer further evolution on their stellar population properties, being compatible with a passive evolution scenario. Therefore, whether if they evolve or not in size is not relevant for our main conclusions.

\section{Summary}
We have analyzed the stellar populations on an individual galaxy basis of a set of 24 ETGs in RX\,J10152.7, a rich cluster at intermediate redshift (z\,=\,0.83). The results were obtained from applying a combination of commonly used index-index diagrams and a full-spectrum-fitting approach. The quality of the data and a detailed treatment for each galaxy has allowed us to study how their global properties relate to galaxy velocity dispersion and to the local environment. We also linked these relations to the different types of star formation histories found.  To explore the possible evolution of the studied ETGs, we we have further compared our results to a sample of local ETGs in Coma, a cluster that closely resembles the properties and substructure of RX\,J0152.7-1357. Our main goal was to see whether there is an evolutionary link between the ETGs in the two clusters. The main conclusions are summarized as follows:
\begin{itemize}
 \item We find that local environment strongly correlates with galaxy velocity dispersion (i.e. galaxy mass) . Galaxies in the center of the N-SubCl show the highest velocity dispersions ($\sim$\,260\,$\rm{km\,s^{-1}}$), followed by the galaxies in the center of the S-SubCl ($\sim$\,230\,$\rm{km\,s^{-1}}$). On the contrary, the outskirts of the cluster are mainly populated by lower-mass galaxies ($\leq$\,180\,$\rm{km\,s^{-1}}$). 
 \item A correlation is also found between local environment and the stellar population properties: galaxies showing the oldest ages are located at the center of the largest substructure, the N-SubCl.
 \item The derived SFHs for the galaxies depend both on galaxy mass and environment: \textit{(i)} galaxies with single-like old burst are located at the center of the N-SubCl and are the most massive ones; \textit{(ii)} galaxies with a single-like burst at intermediate ages mainly populate the S-SubCl and are massive; \textit{(iii)} the outskirts of the clumps are populated by low-mass galaxies with more extended SFHs.
 \item We have related the derived stellar population parameters to the velocity dispersion of the galaxies. As there exist a tight correlation between local environment, SFH and $\sigma$, we have parametrized these relations by the SFH type. We compare the derived quantities of our galaxies with a similarly-selected sample of ETGs in Coma (our reference low-redshift cluster) to investigate the possible evolution of the galaxies in RX\,J0152.7-1357. Our main findings are listed below:
  \begin{enumerate}
          \item \textit{Line-strengths}: In terms of the stellar populations, the most massive galaxies are always compatible with a passive evolution scenario. On the contrary, we find that the galaxies which deviate from the passive evolution predictions are the less massive ones. These deviations can be explained in terms of their more extended SFHs and their later incorporation into the cluster.
          \item \textit{Mean ages and metallicities}: Both these properties are compatible with the passive evolution scenario. A positive mild relation with $\sigma$ is found, such that more massive galaxies are slightly older and more metal rich.
          \item \textit{Abundance patterns}: This parameter shows that both the C and CN abundance patterns of the ETGs in RX\,J10152.7 already resemble those found in galaxies of similar mass in Coma. This suggests that the abundance pattern seen today was already in place at z$\sim$\,0.83. In addition, this points out to a similar formation timescale for the bulk of the stellar populations on the ETGs of both clusters.
\end{enumerate}
\end{itemize}
Our results favor a picture that is compatible with a downsizing scenario. The most massive galaxies, located preferentially at the centers of the subclumps, have evolved passively since the bulk of their stars formed at high redshift. On the contrary, less massive galaxies are still evolving to the present time, experiencing a more extended star formation most likely due to their later incorporation into the cluster. Both the metal content and the abundance patterns seem to be already at place at intermediate redshifts, which is indicative of a formation epoch at higher redshifts than the one studied here. The present study has allowed us to determine the timescales for the formation of the different substructures of RX\,J10152.7, as seen by the SFHs of its individual members. Extending this type of detailed studies to other clusters at intermediate (and even higher) redshifts, will open a window for a better understanding of the physical mechanisms responsible for the transformation of the galaxy populations closer to the epoch of cluster assembly.

\acknowledgments
Most part of this work was supported by the Programa Nacional de Astronom\'ia y Astrof\'isica of the Spanish Ministry of Science and Innovation under grant AYA2010-21322-C03-02 during AFM's PhD thesis and now by the Japan Society for the Promotion of Science (JSPS) Grant-in-Aid for Scientific Research (KAKENHI) Number 23224005. AFM deeply thanks Jes\'us Falc\'on-Barroso, Vicent Quilis, Reynier Peletier and Sune Toft for their useful comments that helped to improve the original draft. PSB is supported by the Ramon y Cajal program of the Ministerio de Ciencia y Competitividad.

\appendix
\section{Line index analysis}
\subsection{Line-strength measurements}
Line-strength values were measured on the LIS8.4\AA\, system of V10 with the {\tt REDUCEME} package \textit{index}. The errors were obtained from the error spectra associated to the each galaxy's final spectra but were also double-checked by calculating them directly from the formulas in \citet{Cardiel2003}. The comparison between some of our newly measured indices and those published in J05 is shown in Figure A1. For this purpose, we transformed their indices into the LIS system using the webpage of MILES (\textit{http://miles.iac.es/pages/webtools/lickids-to-lis.php}). Our C$_{2}$4668 and CN$_{2}$ values are smaller, while a significant spread is found for G4300 and Fe4383 indices. 

\begin{figure}[h!]
\label{figure:A1}
\centering
\includegraphics[scale=0.63]{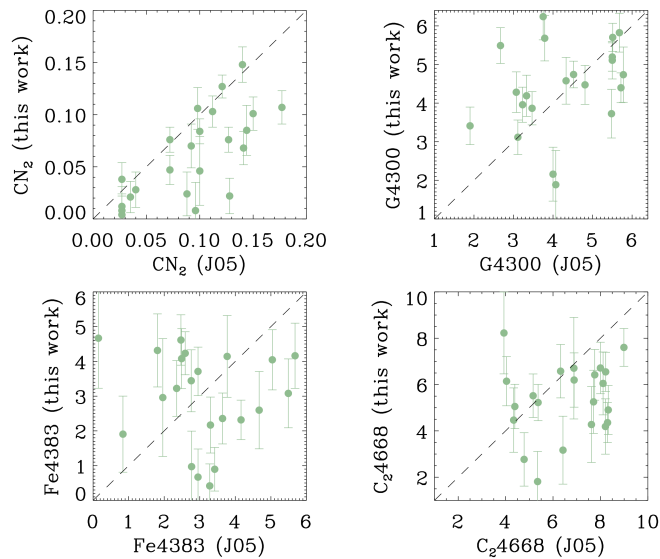}
\caption{Comparison of our index measurements with those of J05. The latter were transformed to the LIS system for a fair comparison.} 
\end{figure} 

\subsection{Ages, metallicities and abundance ratios}
Figure A2 shows the measured indices in the model prediction grids for our best age indicator (H$\gamma$F) \textit{vs} several metallic indices. Galaxies are shown in different colors depending on their velocity dispersion, as in Fig.\,1. Some of the galaxies lie outside the grid, thus their ages and metallicities are difficult to extrapolate with {\tt RMODEL}. We remind the reader that we do not consider extrapolated metallicities above 0.5\,dex. The derived stellar population parameters might be slightly different depending on the pair of indices in use. For example, we can see in table A1 that the CN$_{2}$ and the C$_{2}$4668 in general tend to give higher metallicities than Fe4383. This happens because these galaxies show a non-solar abundance pattern. The Fe4383 grid also exhibits many galaxies that fall outside it. This occurs because Fe4383 tends to give older ages than the other indices, as seen in Figure A3. This figure also highlights the impact of the overabundance in the age estimates. In the first panel the derived ages are in better agreement because C$_{2}$4668 and CN$_{2}$ present similar abundance patterns. On the contrary, the right panel shows that the ages derived with Fe4383 are systematically larger, although still compatible within the error bars.\\
We have only derived the abundance patterns from the index-index diagrams when all three estimates were reliable (namely age, Z$_{A}$ and Z$_{Fe}$), which occurs only for a few galaxies. Therefore, the combined method is the one employed to derive the abundance patterns values used in Sect.\,6. Its robustness is tested in Figure A4. It shows the metallicities derived from this method compared to those inferred from the index-index diagrams, showing a good agreement, in particular for C$_{2}$4668. 

\begin{figure*}
\centering
\includegraphics[scale=0.75]{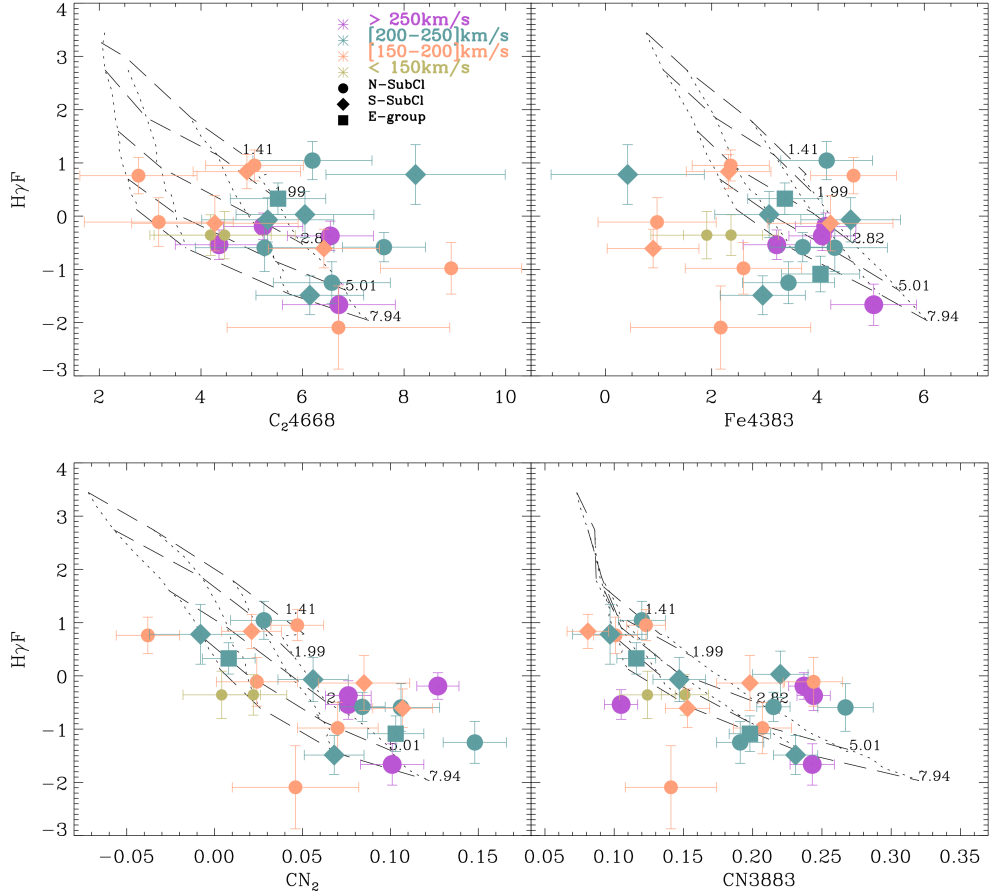}
\label{figure:A2}
\caption{The age-sensitive indicator H$\gamma$F is plotted \textit{vs} several metallic-sensitive indices (C$_{2}$4668, Fe4383, CN$_{2}$ and CN3883), all measured in the LIS-8.4\AA\, flux-calibrated system. The SSP model grids of V10 are plotted, with age (in Gyr) increasing from top to bottom as indicated in the labels, and metallicity from left to right ([Z/H]= -0.71, -0.40, 0.00, +0.22). Galaxies are color- and sized- coded by their velocity dispersion, as indicated in the figure. Membership to the different substructures is shown by the different symbols: circles for the N-SubCl, diamonds for the S-SubCl and squares for the E-group.}
\end{figure*} 

\begin{table*}
\label{table:A1}     
\centering   
\caption{Ages and metallicities from the different index-index pairs}                   
\begin{tabular}{r|r r|r r|r r|r r}   
\hline\hline      
ID & age(Gyr) & [M/H] & age(Gyr) & [M/H]& age(Gyr) & [M/H]& [C/Fe] & [CN/Fe] \\    
 &\multicolumn{2}{c}{C$_{2}$4668-H$\gamma$F} & \multicolumn{2}{|c}{Fe4383-H$\gamma$F}&\multicolumn{2}{|c|}{CN$_{2}$-H$\gamma$F} & from C$_{2}$4668 & from CN$_{2}$\\   
\hline                                                                                                                                                                                                  
338   & 4.73$^{+7.23}_{-2.01}$ &-0.203$^{+0.41}_{-0.61}$& -                      & -                       &10.07$^{+10.15}_{-6.81}$& -0.716$^{+1.12}_{-2.25}$ &   -                     &    -                     \\   
346   & -                      &-                       & -                      & -                       & 1.87$^{+0.54}_{-0.63}$ &  0.006$^{+0.53}_{-0.84}$ &   -                     &    -                     \\   
422   & 5.21$^{+5.84}_{-1.98}$ &-0.230$^{+0.23}_{-0.54}$& -                      & -                       & 7.10$^{+6.58}_{-4.01}$ & -0.416$^{+0.55}_{-1.23}$ &   -                     &    -                     \\   
523   & 2.94$^{+1.10}_{-0.59}$ & 0.005$^{+0.12}_{-0.25}$& 3.01$^{+2.65}_{-1.01}$ & 0.003$^{+0.33}_{-0.45}$ & -                      &   -                      & 0.002$^{+0.35}_{-0.51}$ &    -                     \\   
627   & 6.85$^{+3.25}_{-2.61}$ &-0.579$^{+0.65}_{-1.95}$& -                      & -                       & 5.16$^{+4.41}_{-4.03}$ & -0.351$^{+0.85}_{-1.12}$ &   -                     &    -                     \\   
766   & 7.04$^{+3.06}_{-2.27}$ & 0.144$^{+0.16}_{-0.32}$&11.45$^{+11.03}_{-5.87}$& -0.14$^{+0.47}_{-0.53}$ & 6.99$^{+2.57}_{-1.69}$ &  0.148$^{+0.09}_{-0.11}$ & 0.284$^{+0.50}_{-0.62}$ &  0.288$^{+0.48}_{-0.54}$ \\   
776   & 1.56$^{+1.55}_{-0.81}$ & 0.243$^{+0.15}_{-0.34}$& 2.92$^{+1.95}_{-1.01}$ & -0.45$^{+0.56}_{-1.54}$ & 1.86$^{+1.62}_{-0.66}$ &  0.300$^{+0.39}_{-0.36}$ & 0.693$^{+0.58}_{-1.56}$ &  0.750$^{+0.68}_{-1.58}$ \\   
813   & -                      &-                       & 6.79$^{+6.15}_{-3.06}$ & -0.31$^{+0.34}_{-0.92}$ & 2.45$^{+0.96}_{-0.22}$ &  0.351$^{+0.21}_{-0.24}$ &   -                     &  0.661$^{+0.40}_{-0.95}$ \\   
908   & 5.34$^{+3.88}_{-1.80}$ &-0.200$^{+0.22}_{-0.68}$& 8.83$^{+8.48}_{-4.19}$ & -0.51$^{+0.74}_{-1.31}$ & 2.51$^{+1.31}_{-0.25}$ &  0.257$^{+0.21}_{-0.28}$ & 0.310$^{+0.77}_{-1.45}$ &  0.767$^{+0.77}_{-1.34}$ \\   
1027  & 3.80$^{+6.62}_{-1.08}$ &-0.005$^{+0.33}_{-0.43}$& 4.02$^{+3.94}_{-1.52}$ &-0.007$^{+0.59}_{-0.71}$ & -                      &  -                       & 0.002$^{+0.68}_{-0.83}$ &    -                     \\   
1085  & 2.36$^{+0.67}_{-0.12}$ & 0.469$^{+0.26}_{-0.29}$& 3.72$^{+3.47}_{-1.21}$ & -0.11$^{+0.44}_{-0.51}$ & 3.07$^{+1.52}_{-0.56}$ &  0.478$^{+0.03}_{-0.52}$ & 0.579$^{+0.51}_{-0.59}$ &  0.588$^{+0.44}_{-0.73}$ \\   
1110  & 4.67$^{+5.52}_{-3.65}$ & 0.172$^{+0.36}_{-0.52}$& -                      & -                       & -                      &  -                       &   -                     &    -                     \\   
1210  & -                      &-                       & -                      &   -                     & 4.89$^{+3.47}_{-3.22}$ &  0.006$^{+0.09}_{-0.23}$ &  -                      &    -                     \\   
1458  & 3.85$^{+6.72}_{-2.60}$ &-0.181$^{+0.35}_{-1.01}$& 2.75$^{+2.33}_{-2.09}$ & 0.110$^{+0.30}_{-1.41}$ & -                      &  -                       &-0.291$^{+0.46}_{-1.76}$ &     -                    \\   
1507  & -                      & -                      &  -                     &                         & -                      &  -                       &   -                     &     -                    \\   
1567  & 1.92$^{+1.54}_{-1.13}$ & 0.347$^{+0.31}_{-0.52}$& 5.07$^{+3.94}_{-3.82}$ &-0.422$^{+0.41}_{-1.32}$ & -                      &  -                       & 0.767$^{+0.51}_{-1.42}$ &     -                    \\   
1590  & 2.72$^{+1.41}_{-0.43}$ & 0.309$^{+0.28}_{-0.35}$& -                      & -                       & -                      &  -                       &   -                     &     -                    \\   
1614  & -                      & -                      & 9.68$^{+6.87}_{-3.38}$ &-0.310$^{+0.32}_{-1.02}$ & 3.43$^{+1.65}_{-0.64}$ &  0.387$^{+0.23}_{-0.35}$ &   -                     &  0.697$^{+0.39}_{-1.08}$ \\   
1682  & 1.98$^{+0.89}_{-0.71}$ & 0.240$^{+0.21}_{-0.36}$& 2.62$^{+2.06}_{-0.65}$ &-0.200$^{+0.15}_{-0.67}$ & 4.80$^{+5.06}_{-3.47}$ & -0.530$^{+0.78}_{-1.43}$ & 0.440$^{+0.26}_{-0.76}$ & -0.330$^{+0.79}_{-1.58}$ \\   
1811  & -                      &-                       & -                      &  -                      & -                      &  -                       &   -                     &     -                    \\   
1935  & 6.84$^{+5.45}_{-1.02}$ & 0.073$^{+0.06}_{-0.23}$& -                      &  -                      & 8.40$^{+5.91}_{-3.23}$ & -0.060$^{+0.18}_{-0.36}$ &   -                     &     -                    \\   
\hline                                                                    
\end{tabular}\\
{Ages and metallicities derived from different index-index grids (columns 2-7). Errors were estimated with 1000 Monte-Carlo simulations with {\tt RMODEL} using the errors on the indices and deriving 1$\sigma$ error contours in the age-metallicity space. Columns 8 and 9 are the abundance ratios inferred from the index-index grids.}               
\end{table*}

\begin{figure}
\label{figure:A3}
\centering
\includegraphics[scale=0.8]{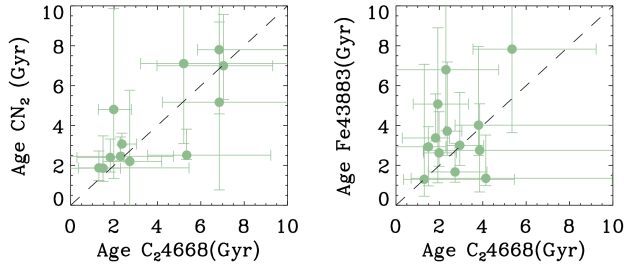}
\caption{Comparison of the derived ages from different pairs of indices, as stated on each label, to emphasize the effect of the overabundance of some elements. } 
\end{figure} 

\begin{figure}
\label{figure:A4}
\centering
\includegraphics[scale=0.55]{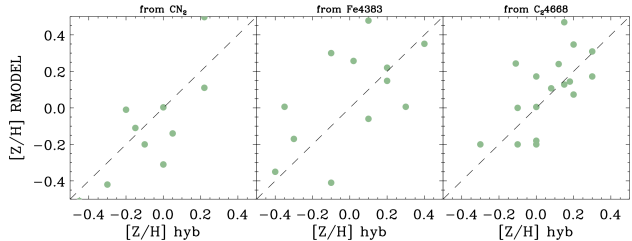}
\caption{Comparison of the derived metallicities from various index-index diagrams and from the combined approach. They are in good agreement, confirming the robustness of the combined method.} 
\end{figure} 

\section{Star Formation Histories}
\subsection{At intermediate redshifts}
Figures B1 and B2 show the 24 ETGs spectra from RX\,J0152.7-1357 (with the new reduction performed here) together with the mixture of SSPs from the V10 models that best fitted the spectra using the full-spectrum-fitting approach (left panels). The adopted IMF slope, which depends on the velocity dispersion of the galaxy, is indicated for each galaxy. The right panels show the derived SFH for each galaxy. For internal checking purposes, we plot in Figure B3 the luminosity-weighted ages derived from the full-spectrum-fitting approach with {\tt STARLIGHT} compared to those inferred from the index-index diagrams. These techniques give slightly different ages. For the old stellar populations, index-index diagrams tend to provide older ages than those obtained from the full-spectrum-fitting (\citealt{Vazdekis2001a}; \citealt{Mendel2007}). On the contrary, for populations with a strong contribution of a young component, the SSP-equivalent age is in better agreement (e.g. \citealt{Serra2007}). 

\begin{figure*}
\label{figure:B1}
\centering                          
        \includegraphics[scale=0.4]{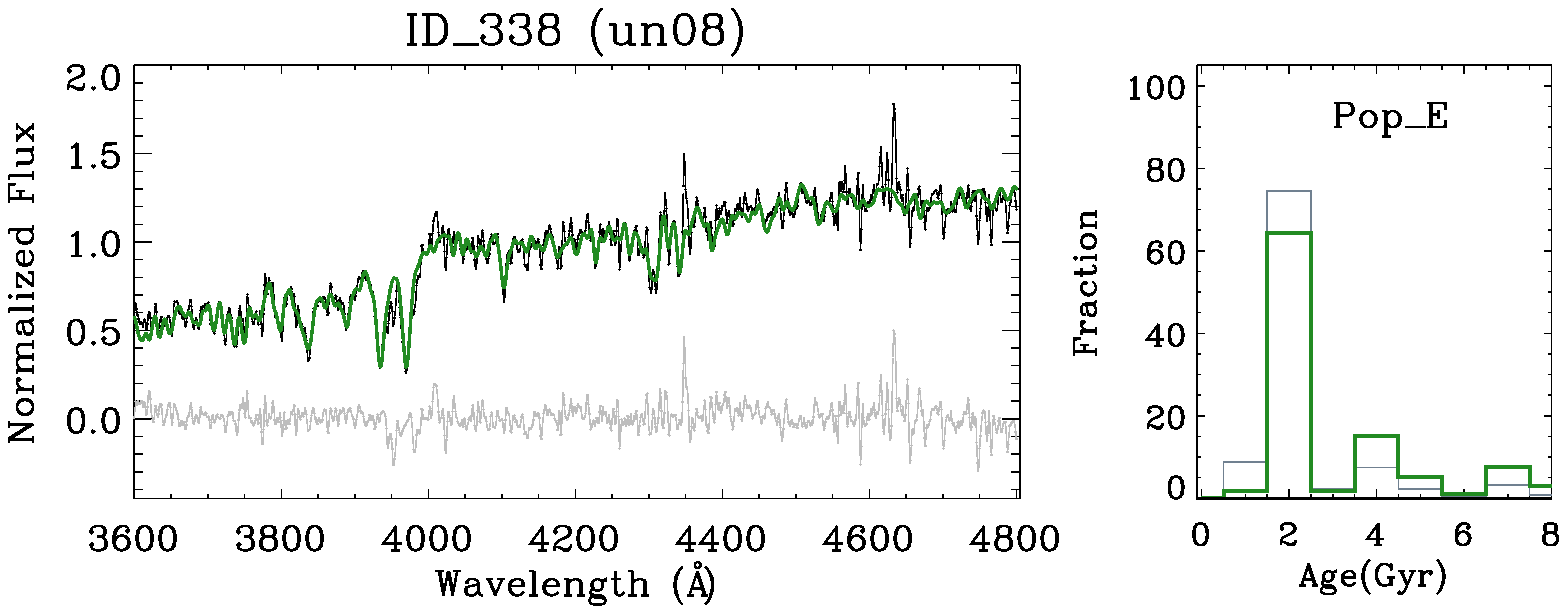}  
        \includegraphics[scale=0.4]{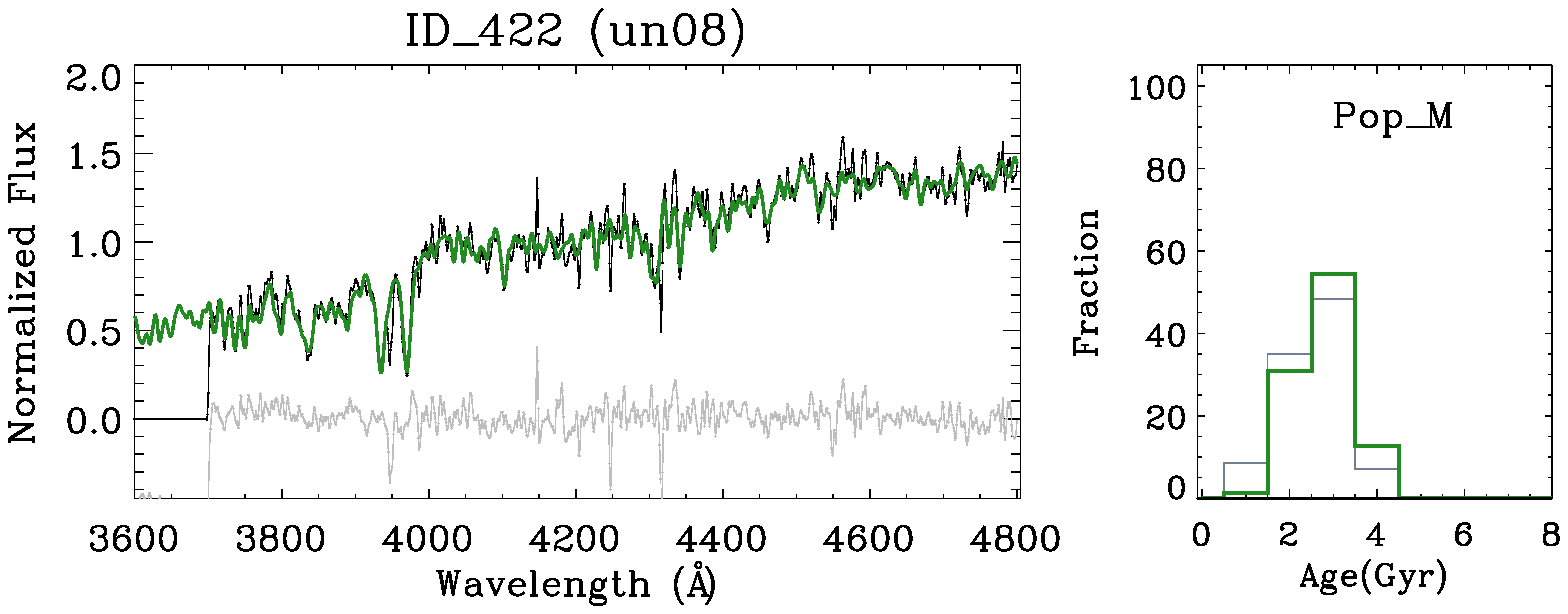}  
        \includegraphics[scale=0.4]{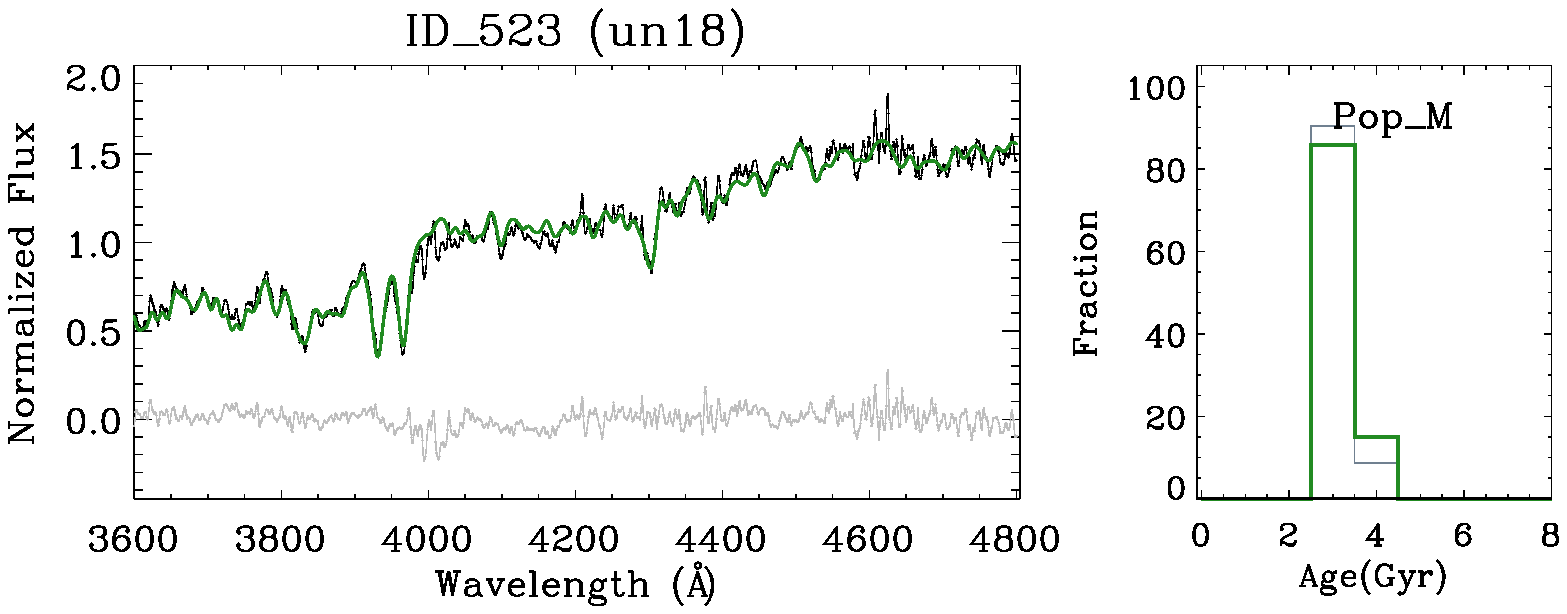}       
        \includegraphics[scale=0.4]{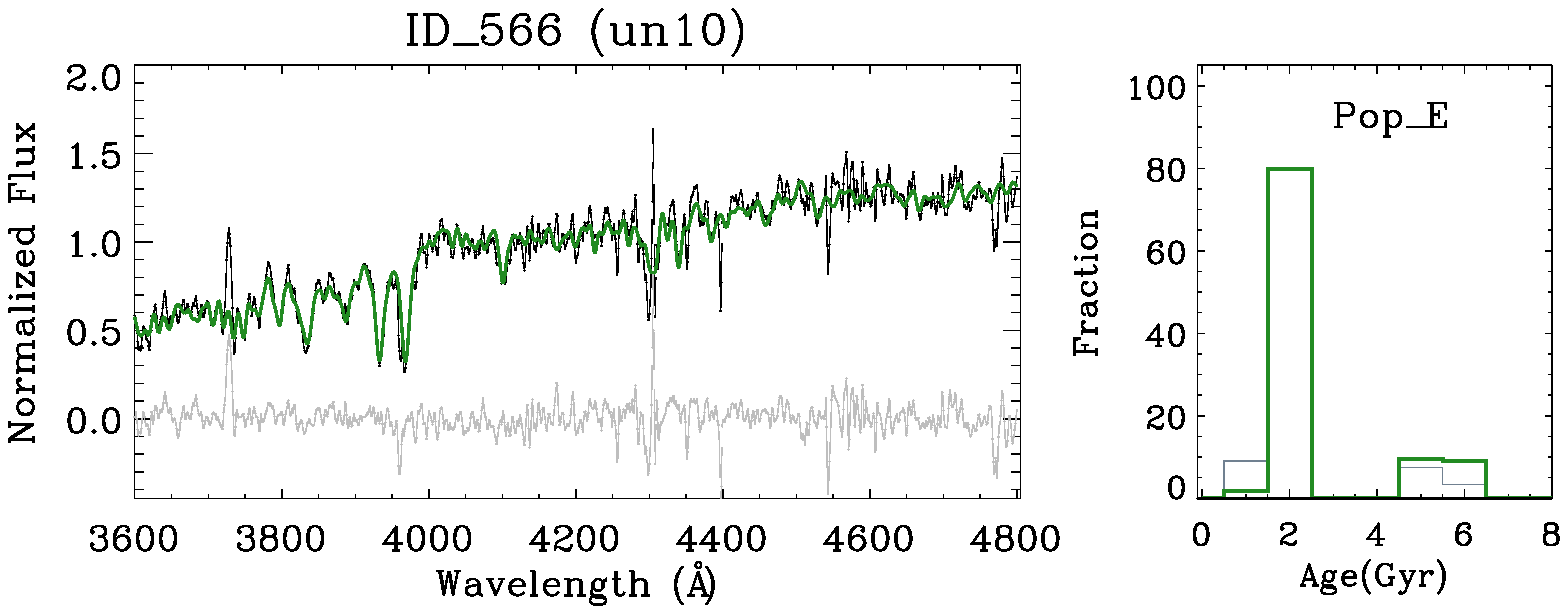} 
        \includegraphics[scale=0.4]{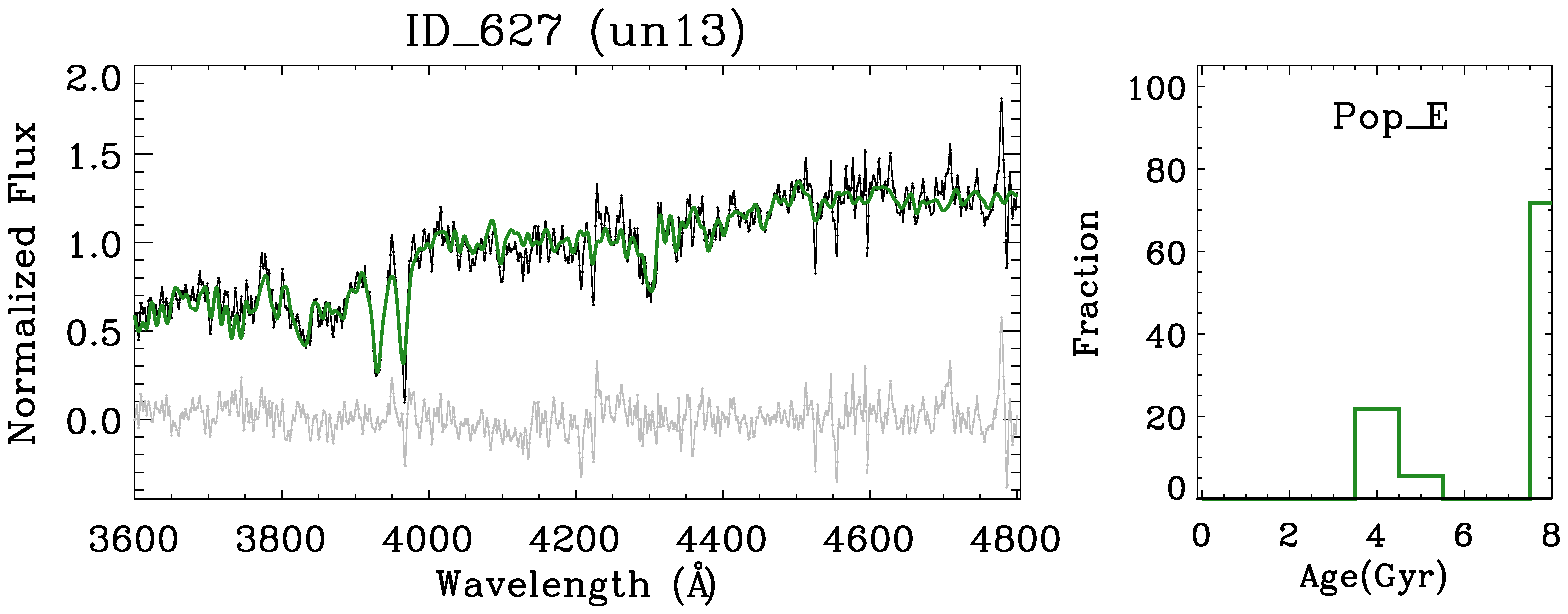}  
        \includegraphics[scale=0.4]{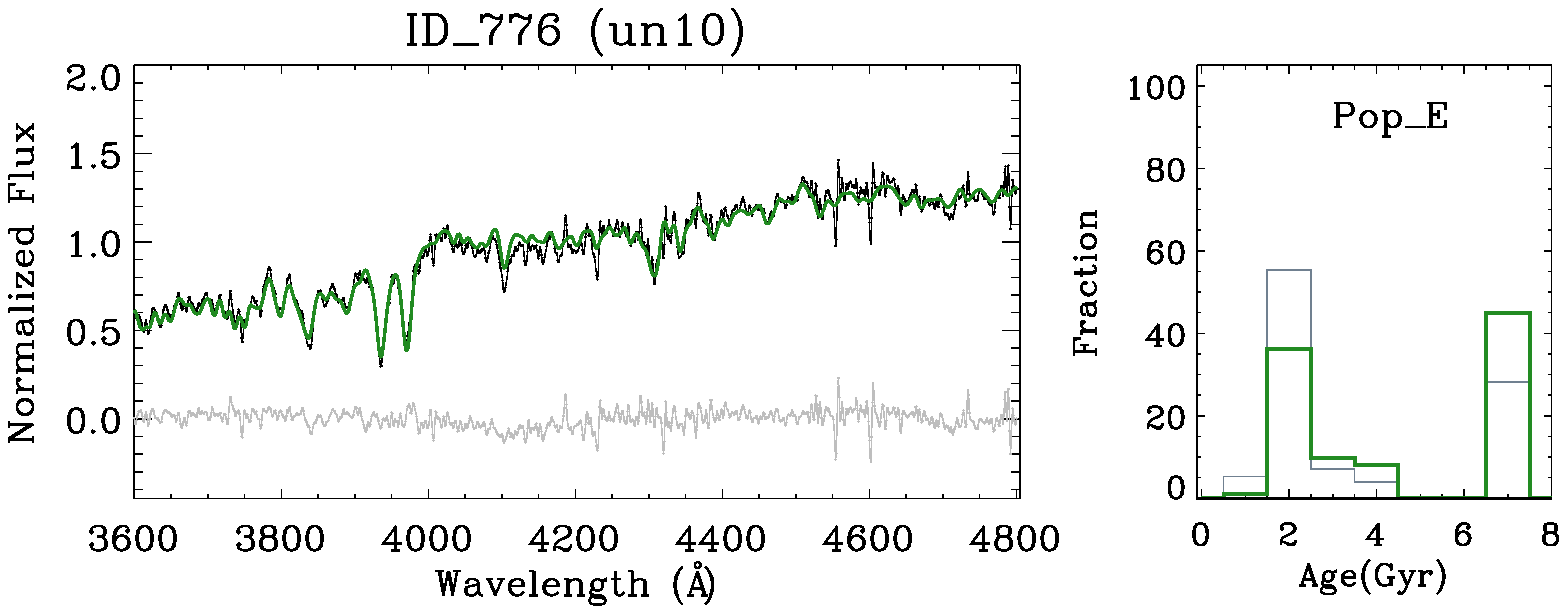} 
        \includegraphics[scale=0.4]{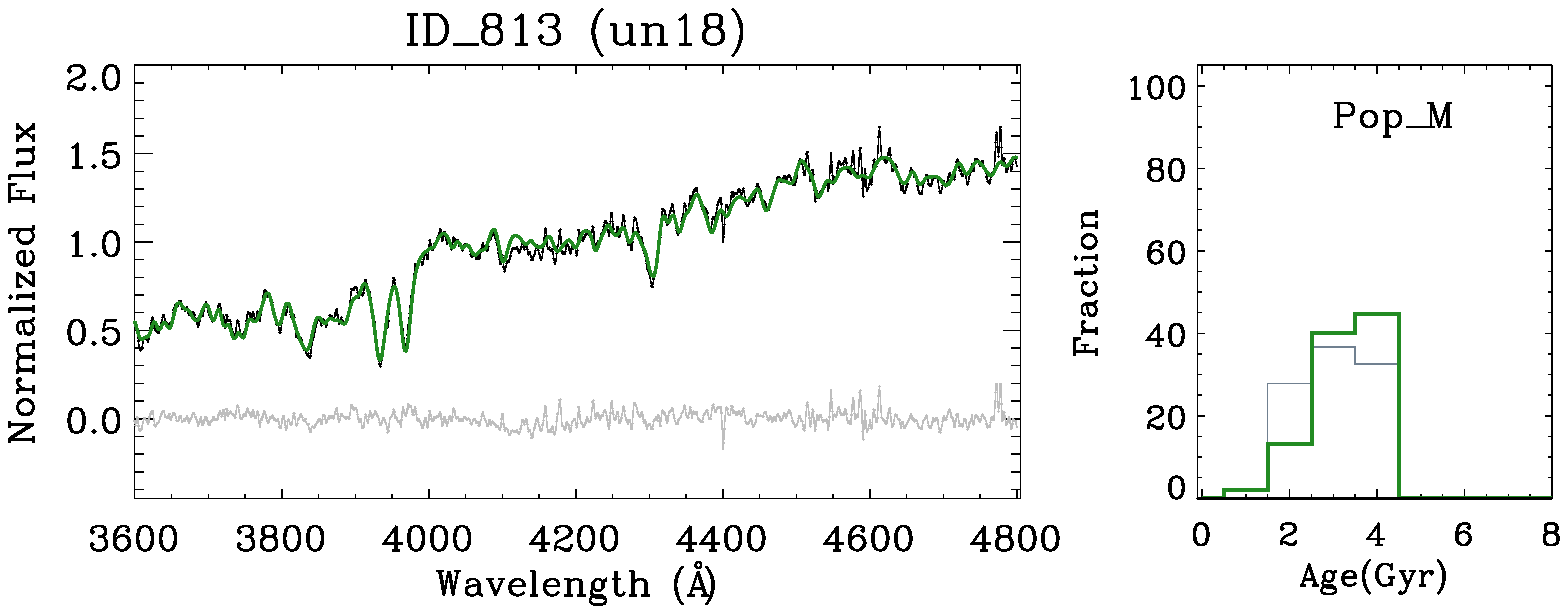} 
        \includegraphics[scale=0.4]{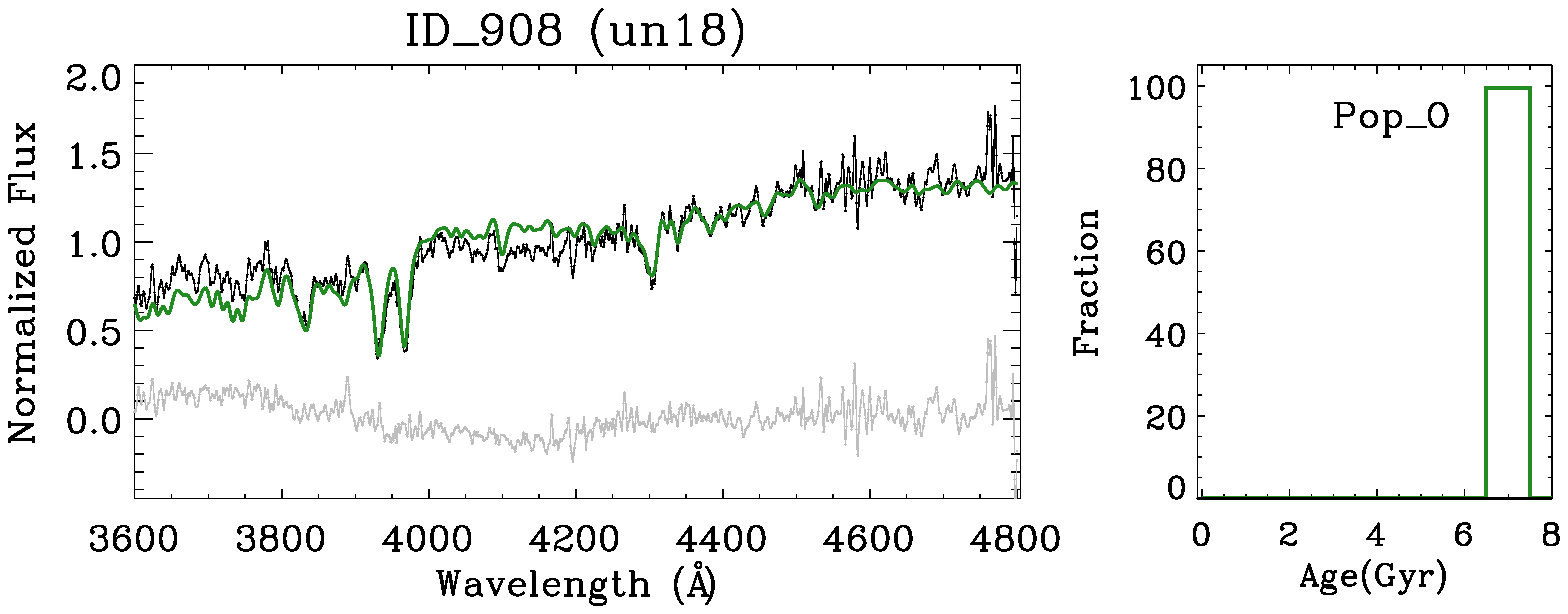} 
        \includegraphics[scale=0.4]{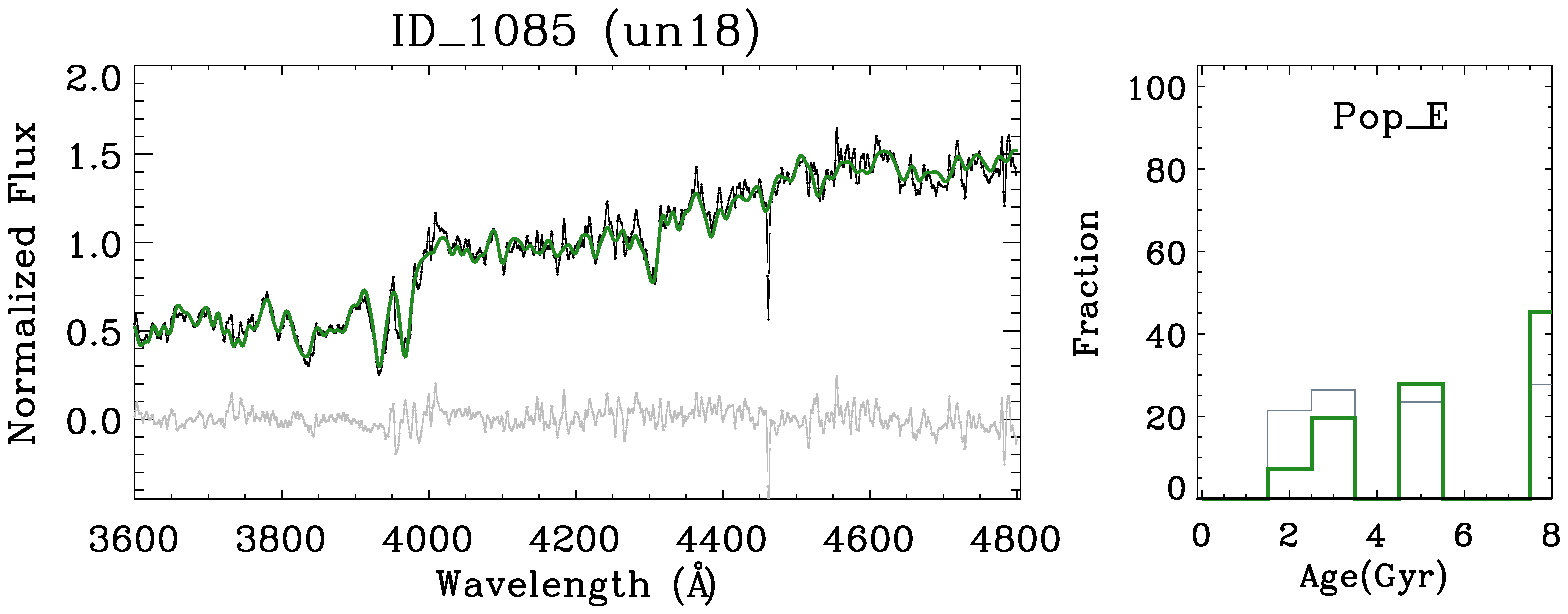}
        \includegraphics[scale=0.4]{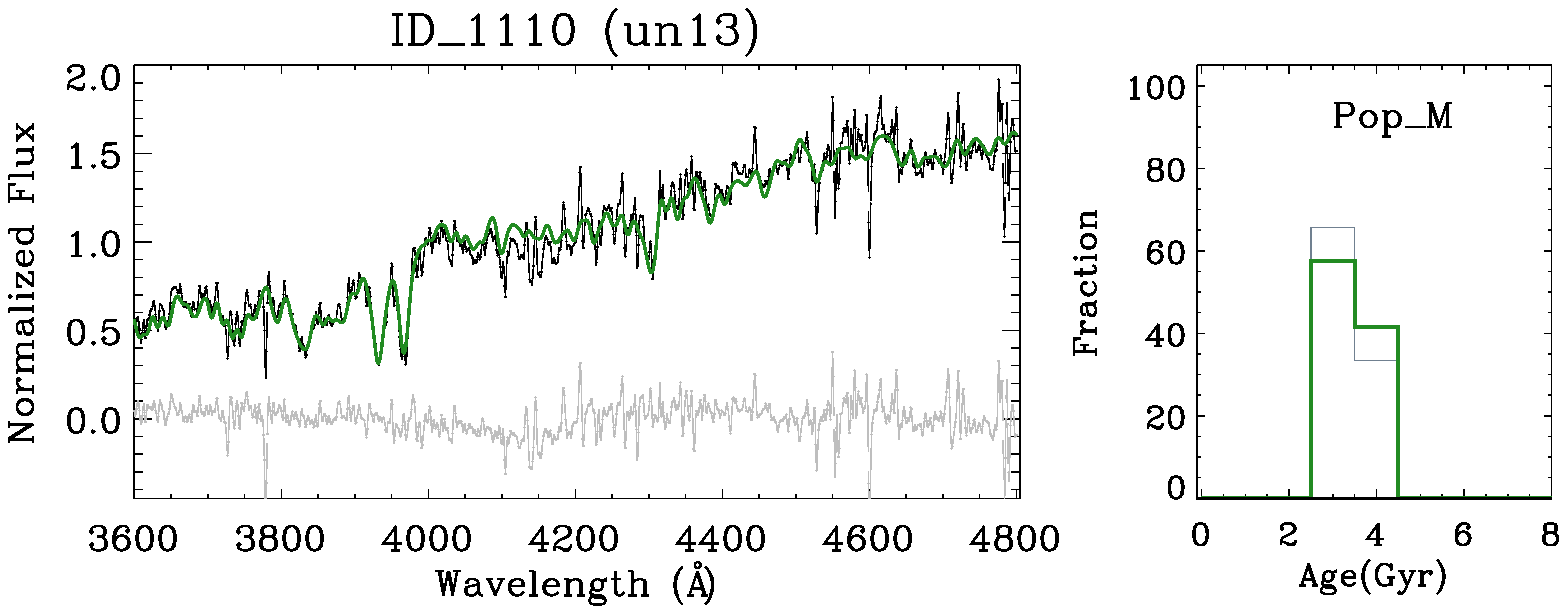} 
        \includegraphics[scale=0.4]{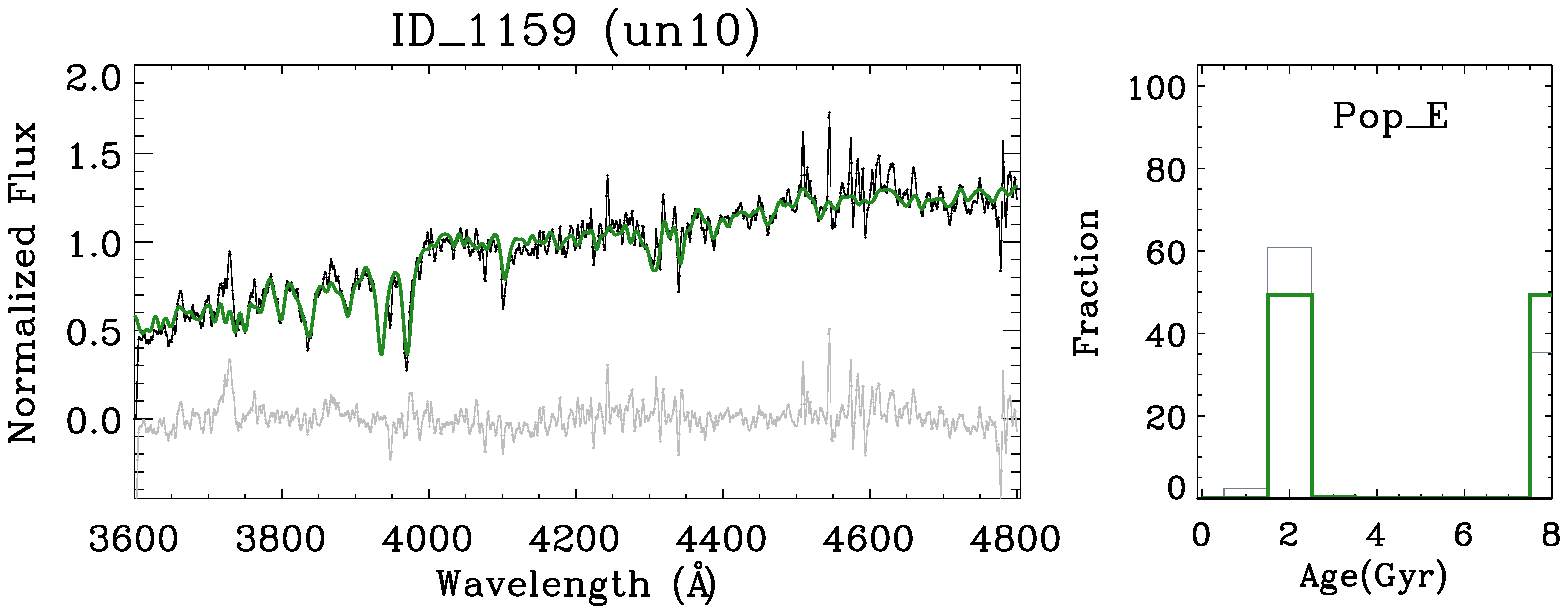}
        \includegraphics[scale=0.4]{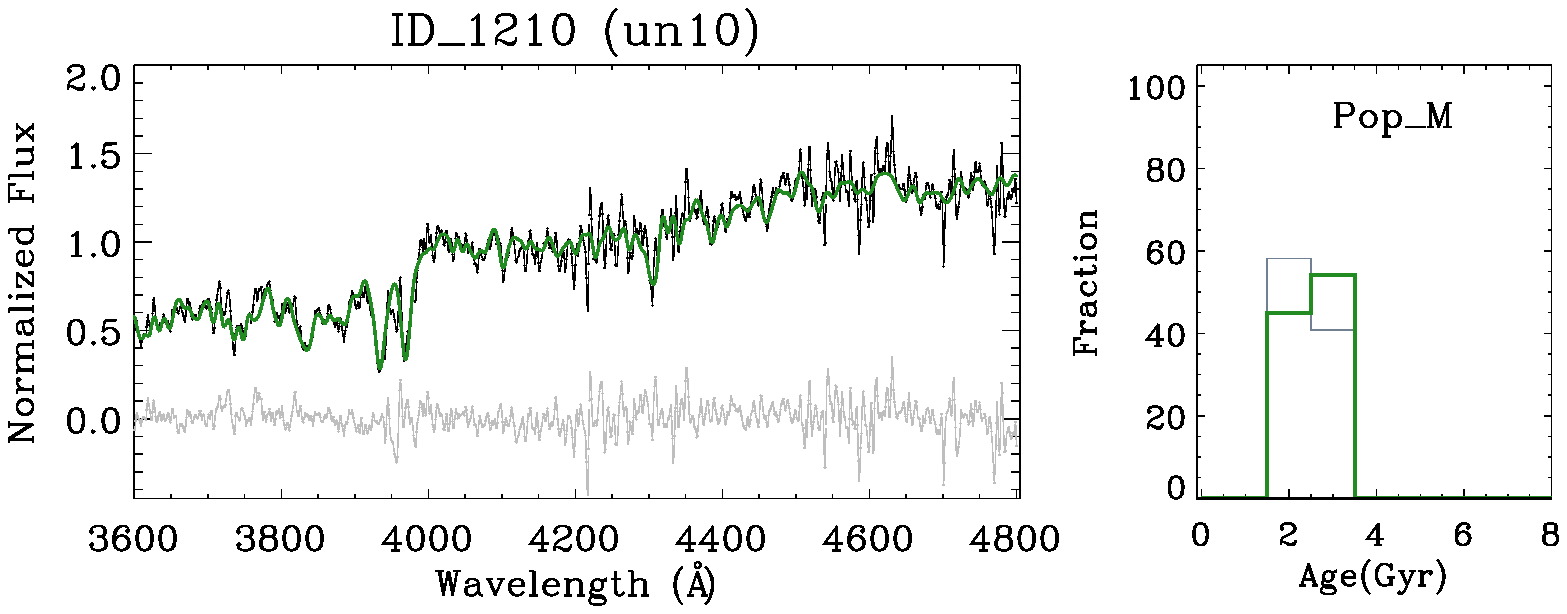}
        \includegraphics[scale=0.4]{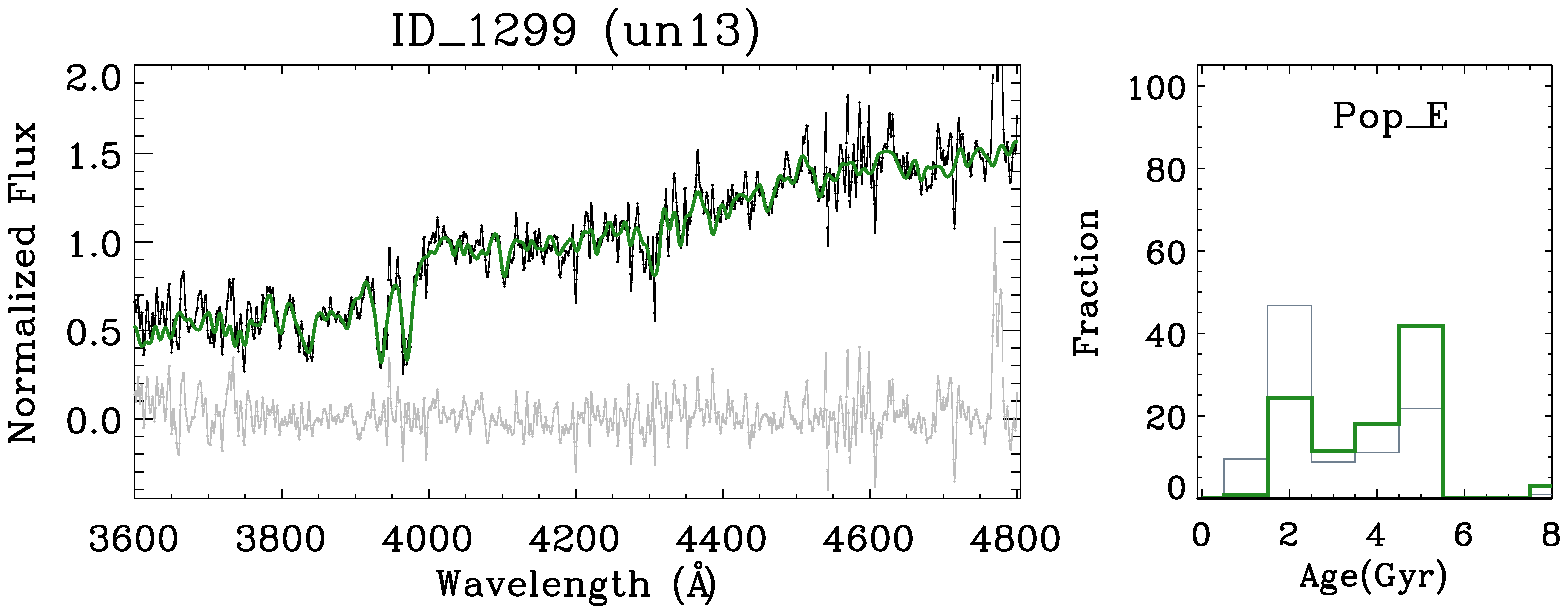}  
        \includegraphics[scale=0.4]{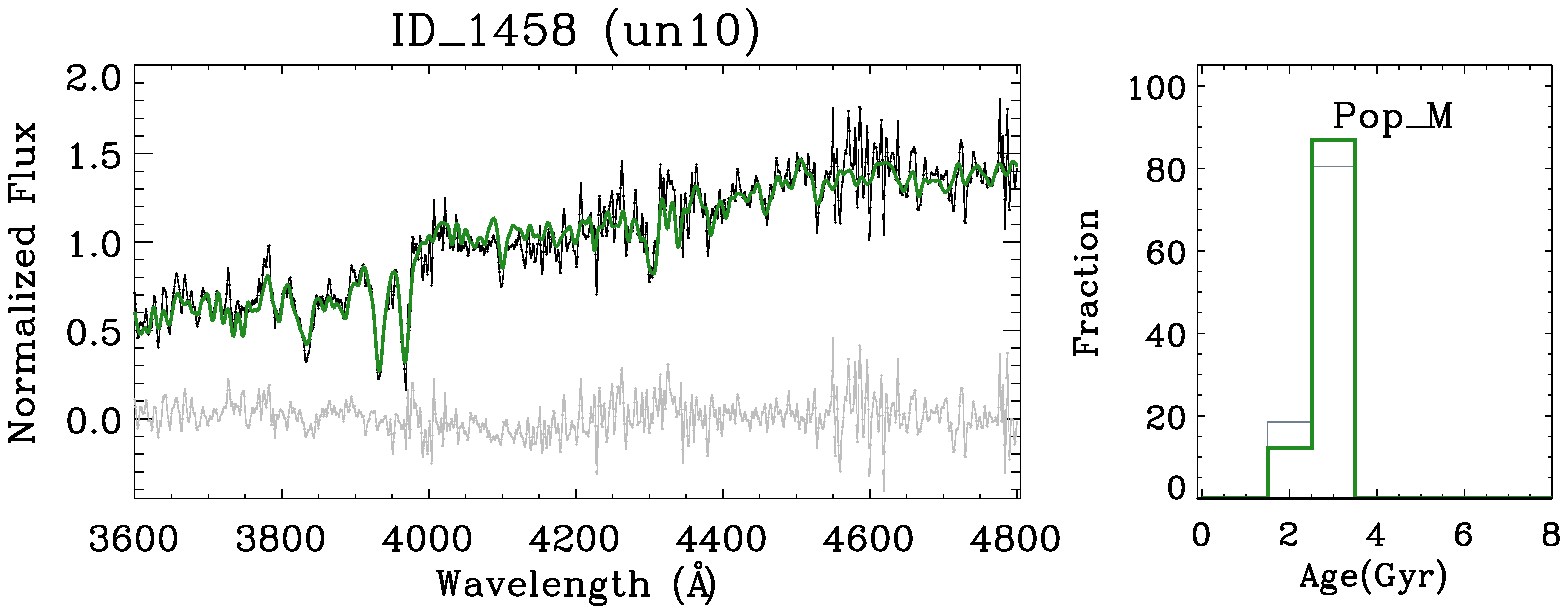} 
        \includegraphics[scale=0.4]{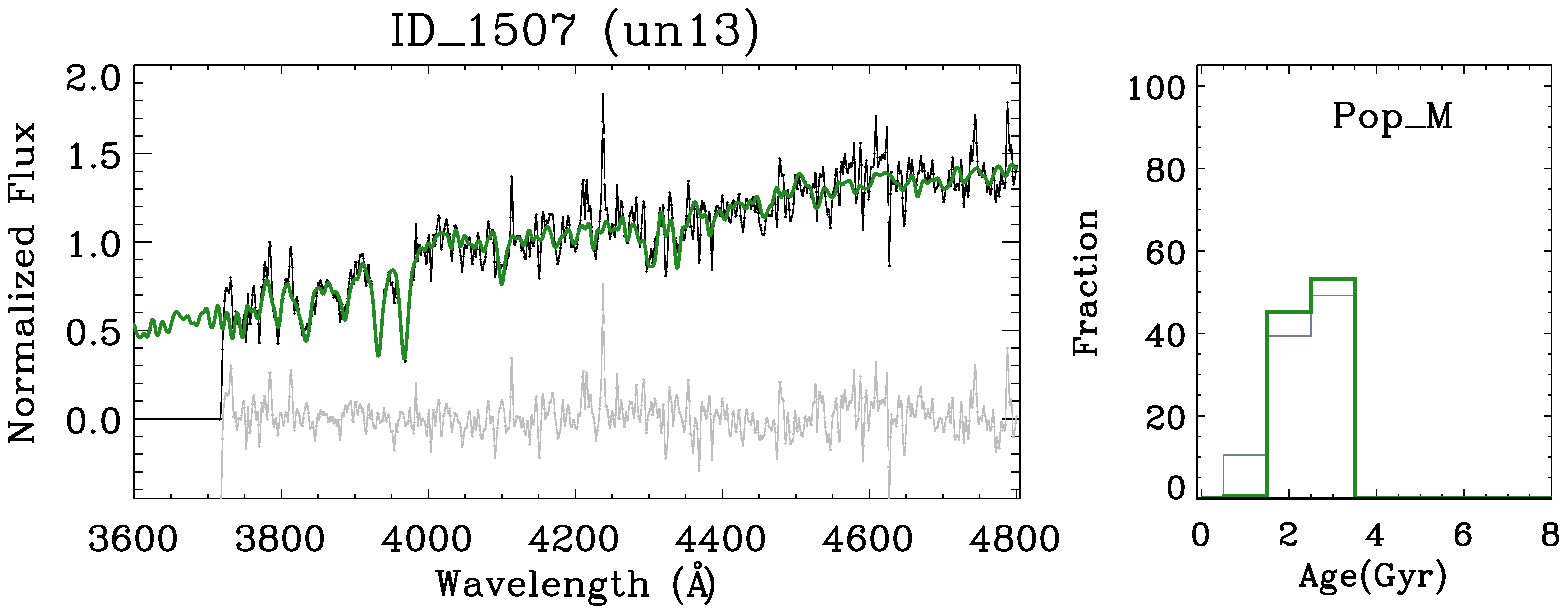} 
        \includegraphics[scale=0.4]{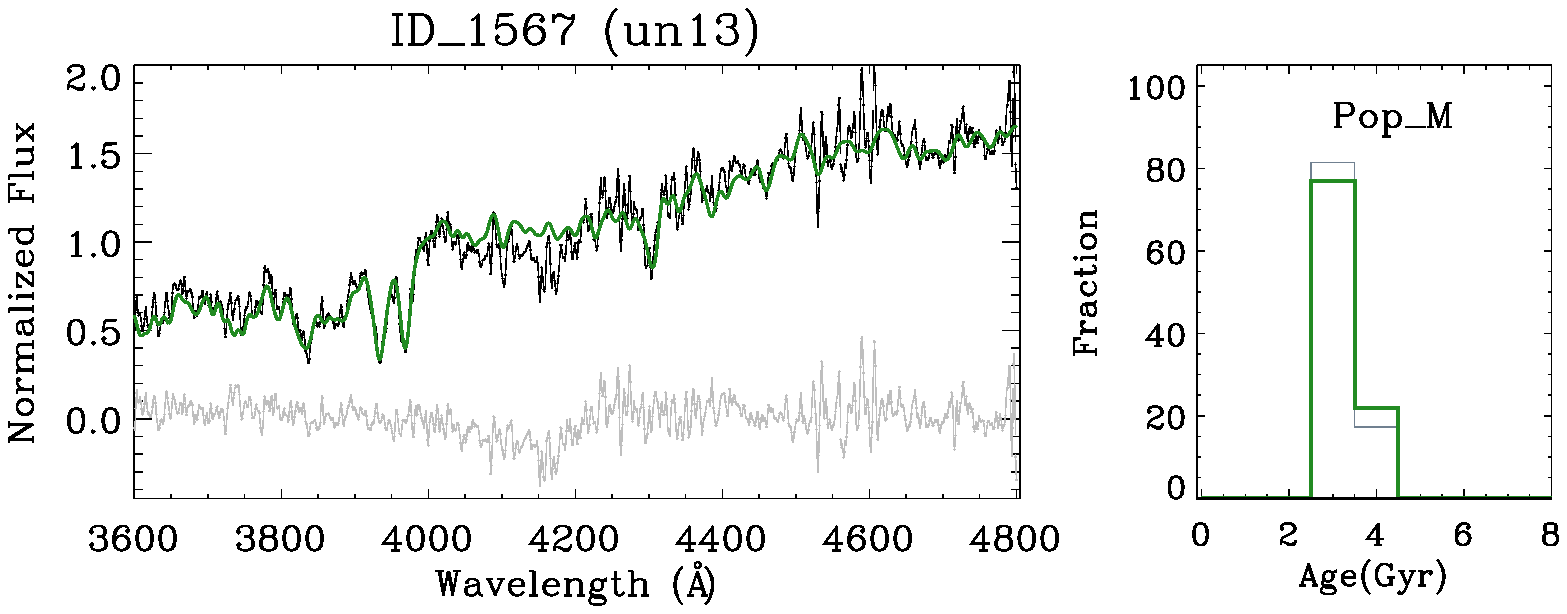}  
\caption{The rest of RX\,J0152.7-1357 cluster galaxy spectra are plotted here, together with the best fit from {\tt STARLIGHT} and their residuals (left panels). The IMF slope adopted is stated on top of each panel. We plot in green the histograms corresponding to the mass-weighted SFHs whereas in gray we plot the luminosity-weighted SFHs (right panels). The  different types of SFHs, classified according to the age of the dominant burst, are stated on each panel.} 
\end{figure*} 

\begin{figure*}
\label{figure:B2}
\centering             
        \includegraphics[scale=0.4]{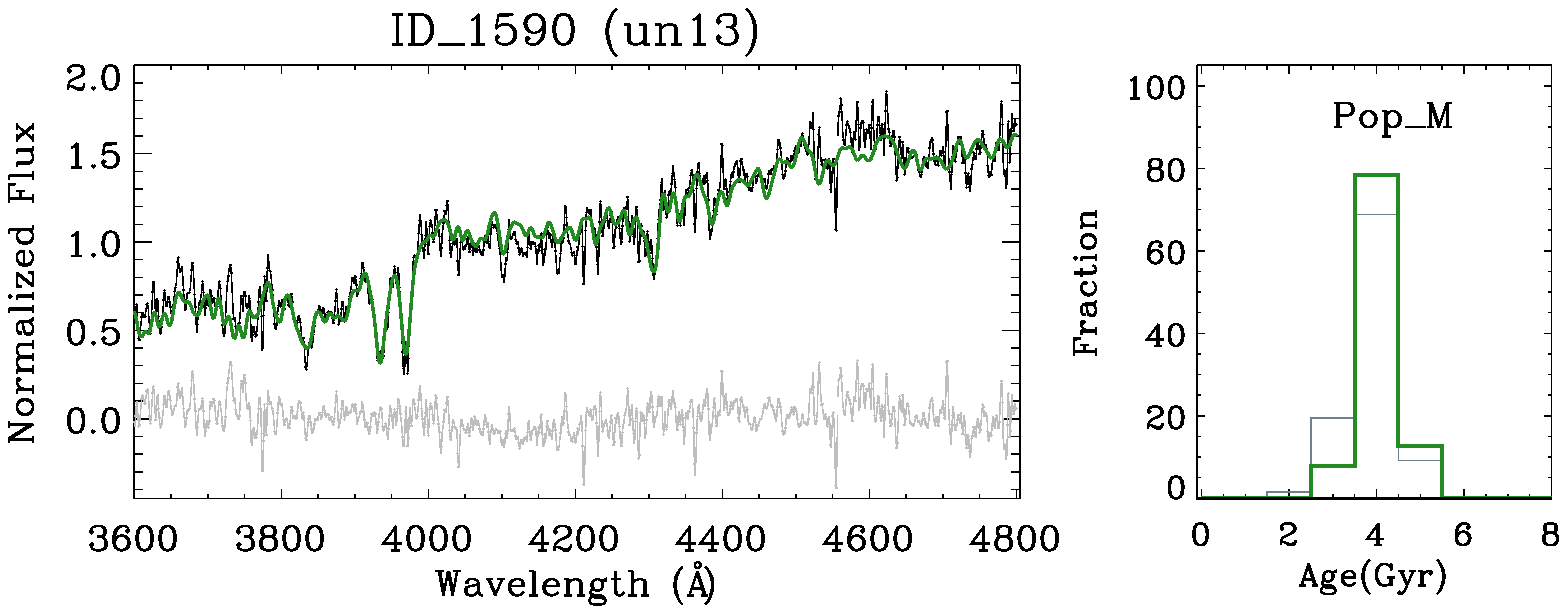}        
        \includegraphics[scale=0.4]{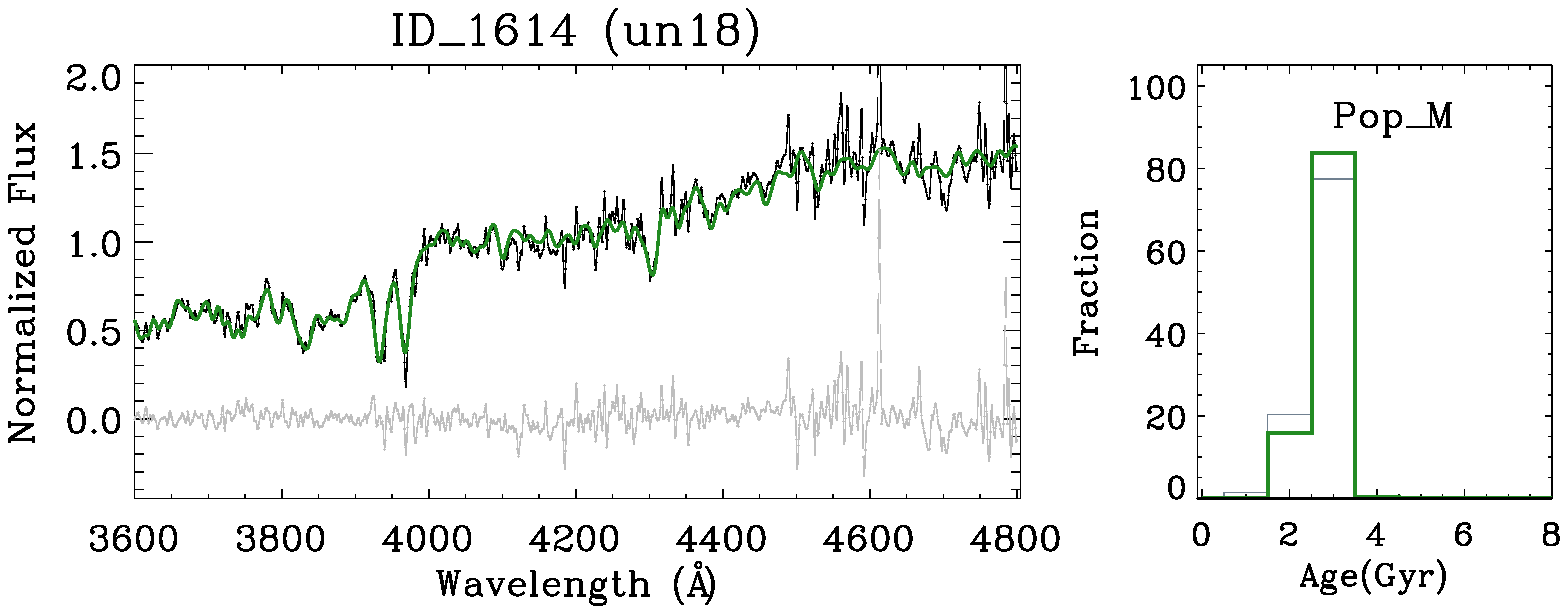}  
        \includegraphics[scale=0.4]{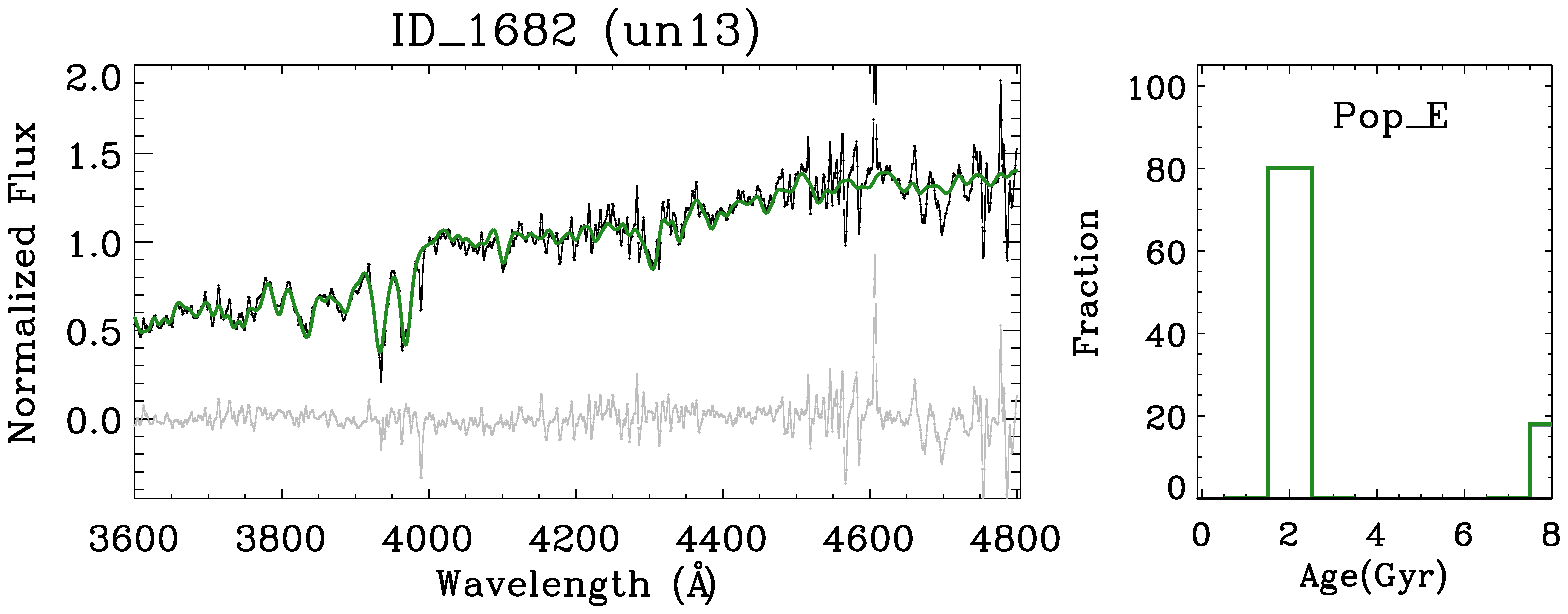} 
        \includegraphics[scale=0.4]{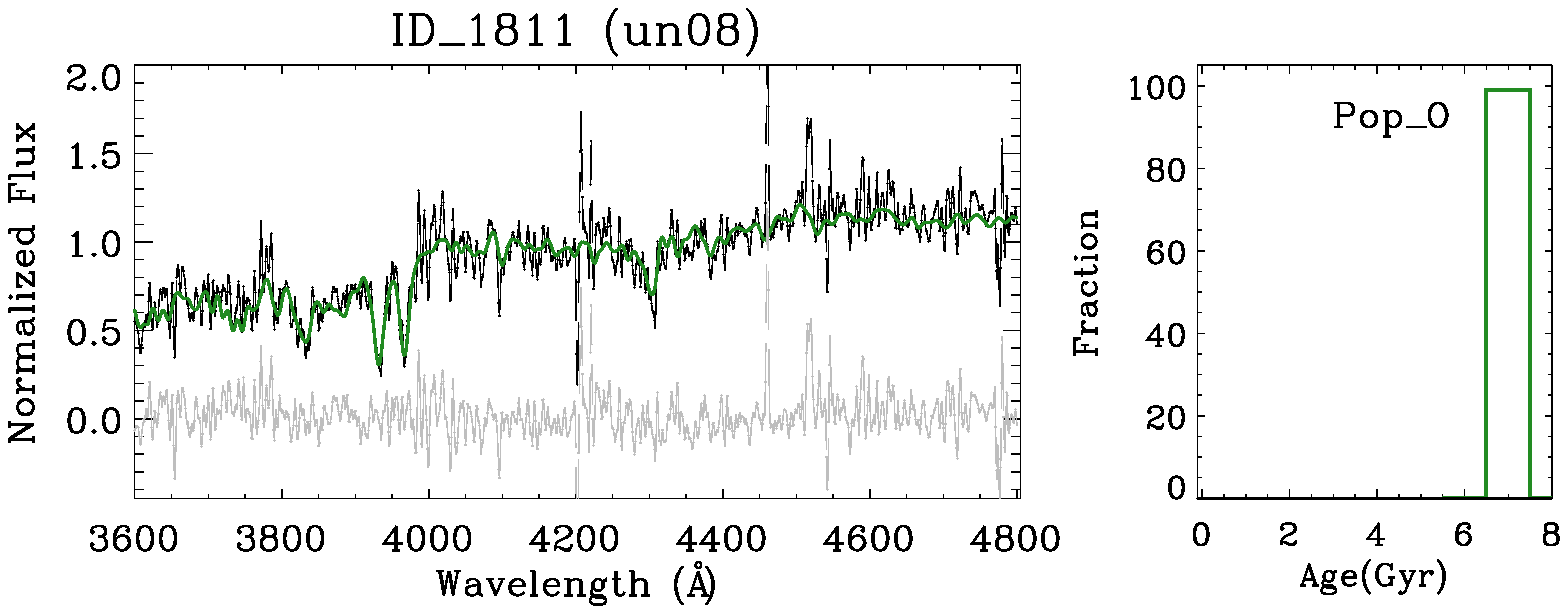} 
        \includegraphics[scale=0.4]{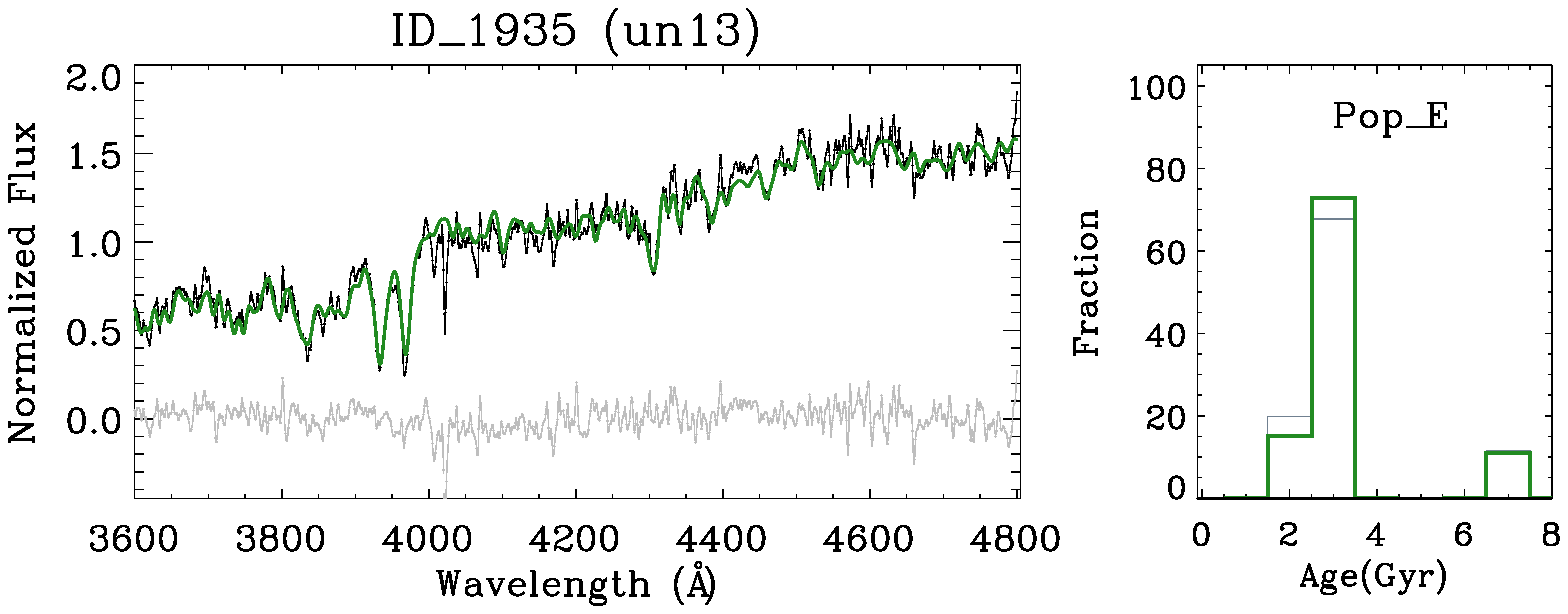}
\caption{Continuation of Figure B1} 
\end{figure*} 

\begin{figure}
\label{figure:B3}
\centering
\includegraphics[scale=0.7]{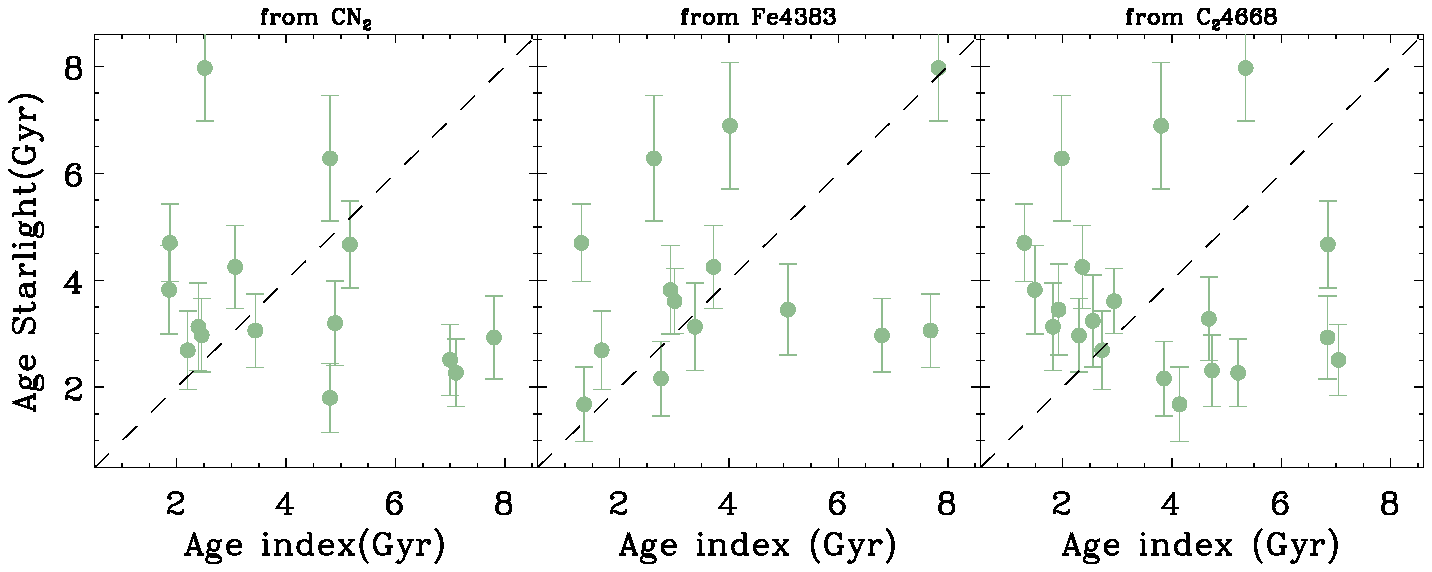}
\caption{Derived ages from different pairs of indices compared to the mean luminosity-weighted ages derived from {\tt STARLIGHT}.} 
\end{figure} 

\subsection{In the local Universe}
Figures B4 and B5 present the spectra and the SFHs for each galaxy considered in our control Coma sample. In this case, we have limited our models to ages below 14.12\,Gyr, to be consistent with the methodology followed for the intermediate-z cluster. We have tested the impact of this choice by comparing the derived ages using the whole set of models (up to 17\,Gyr) with the ones limited to 14\,Gyr. We see in Figure B6 that only the oldest ones are affected, changing from 17\,Gyr to 14\,Gyr as expected. On the contrary, for those with ages $<$12\,Gyr, their estimates remain practically unchanged. However, the derived metallicity is independent of the ages assumed. \\
Finally, we have also compared our results to the ones originally derived in S\'anchez-Bl\'azquez et al. (2006b) using a full-spectrum-fitting approach. Although the codes used were not the same, they are in good agreement, in particular for the metallicity (see Figure B7).

\begin{figure*}
\label{figure:B4}
\centering
        \includegraphics[scale=0.4]{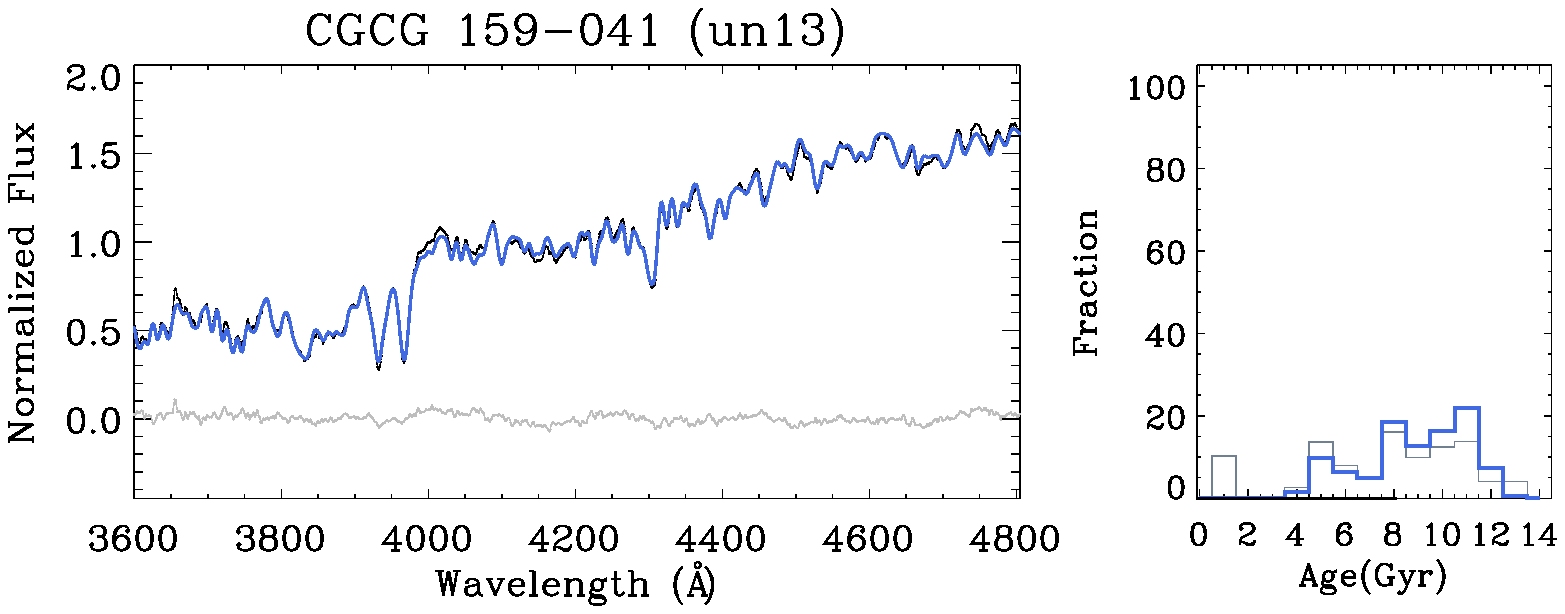}  
        \includegraphics[scale=0.4]{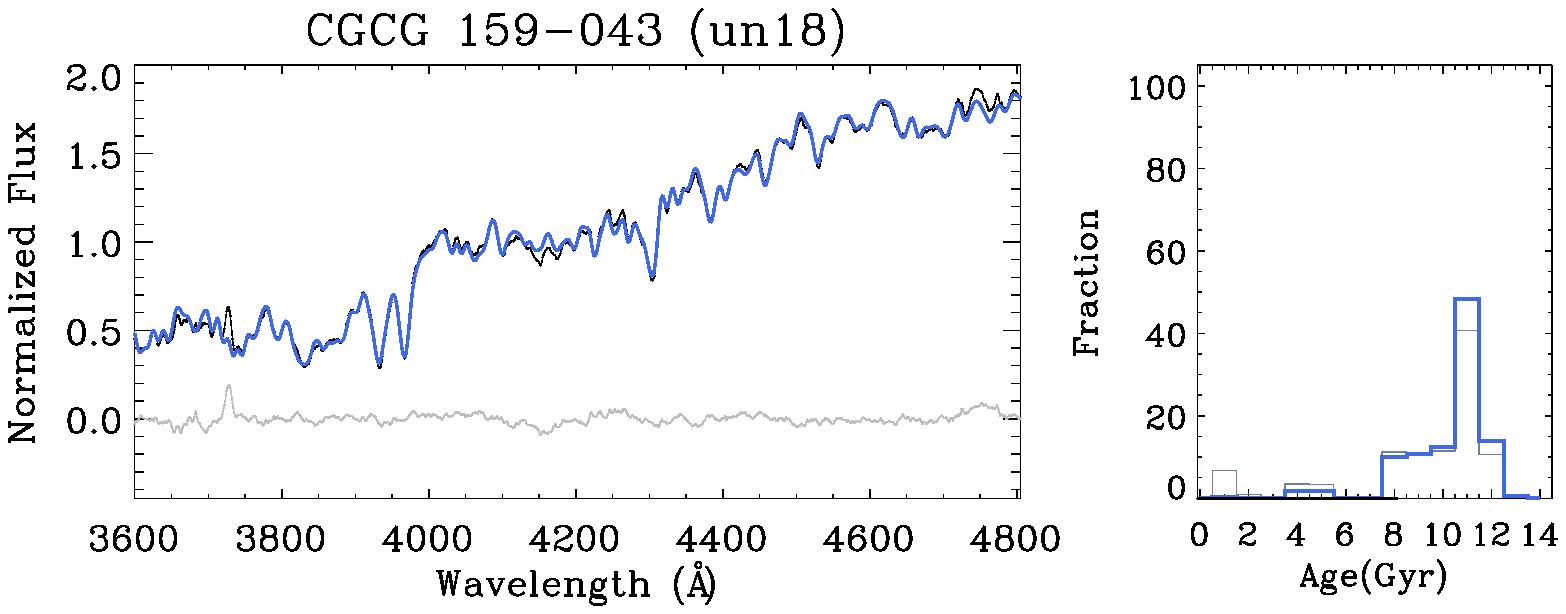} 
        \includegraphics[scale=0.4]{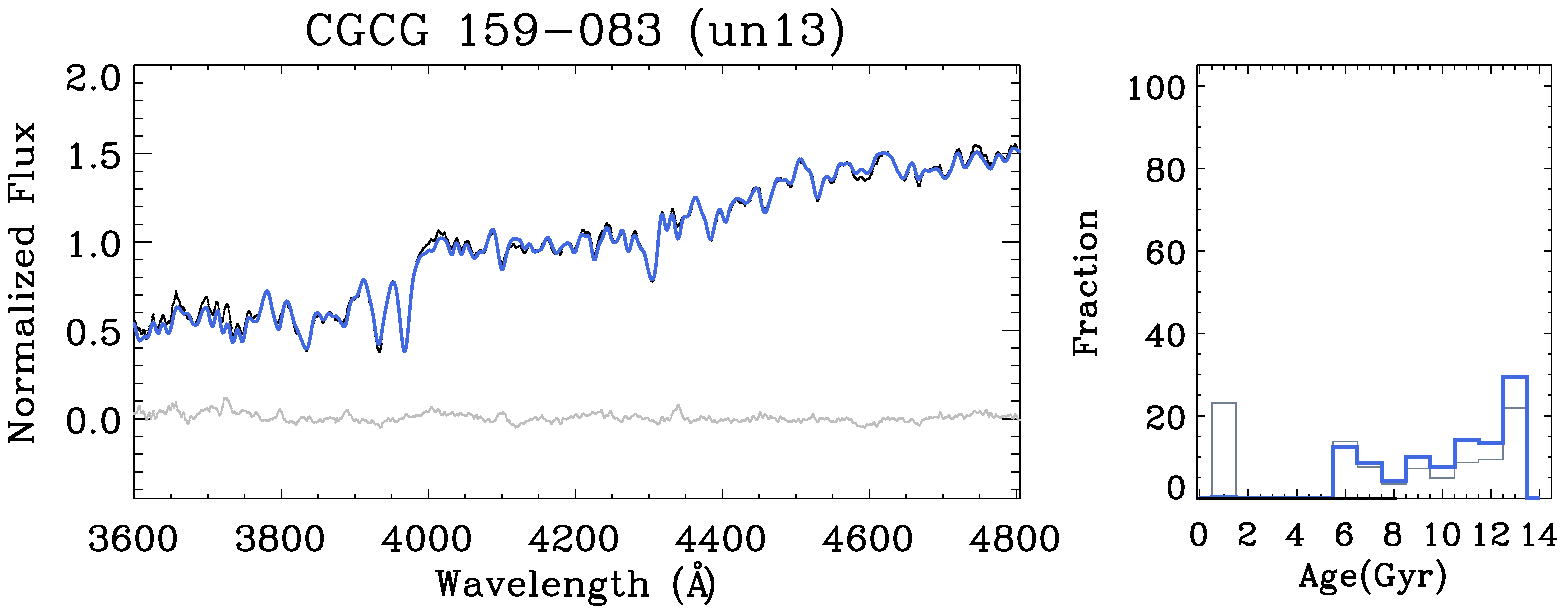}  
        \includegraphics[scale=0.4]{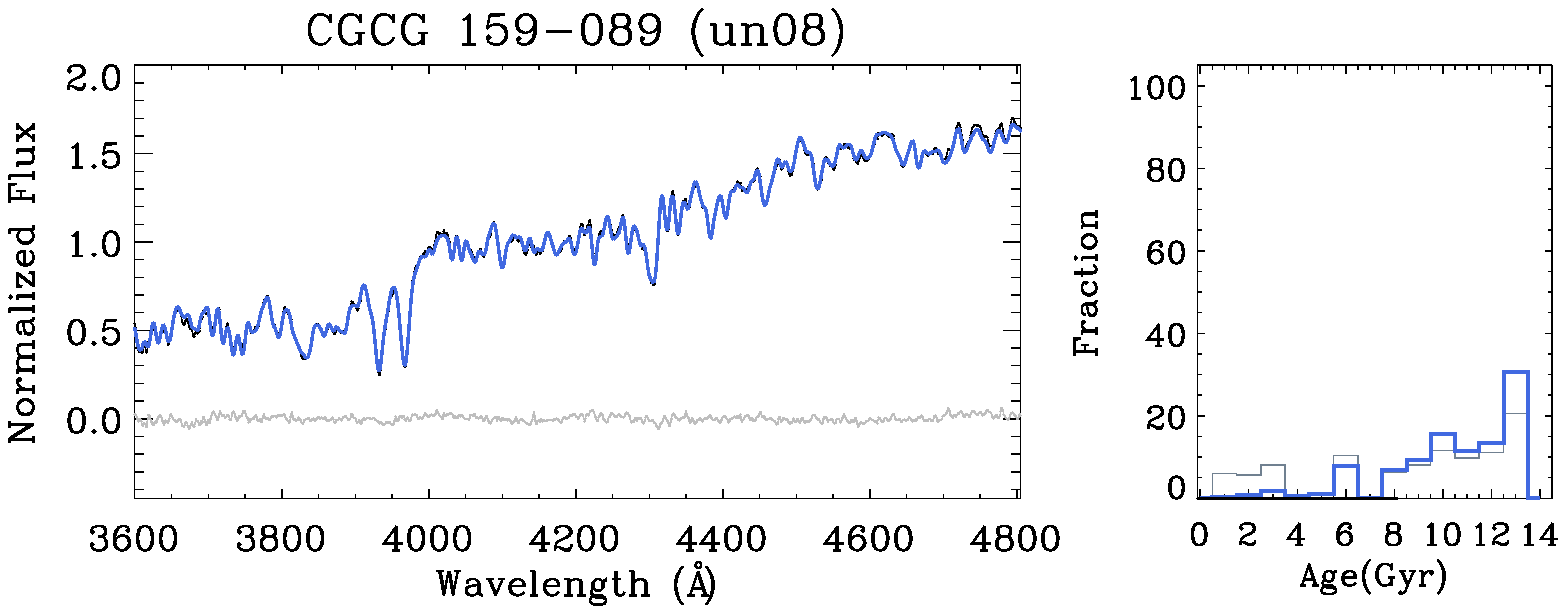}       
        \includegraphics[scale=0.4]{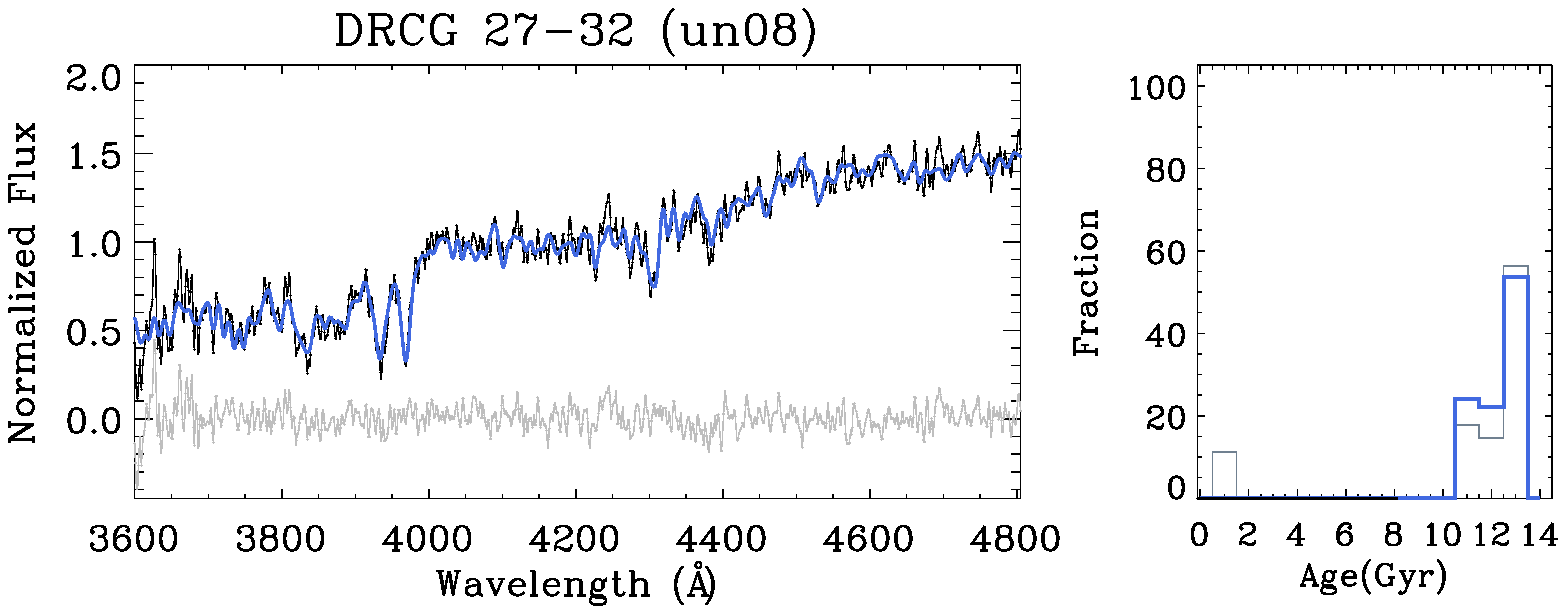} 
        \includegraphics[scale=0.4]{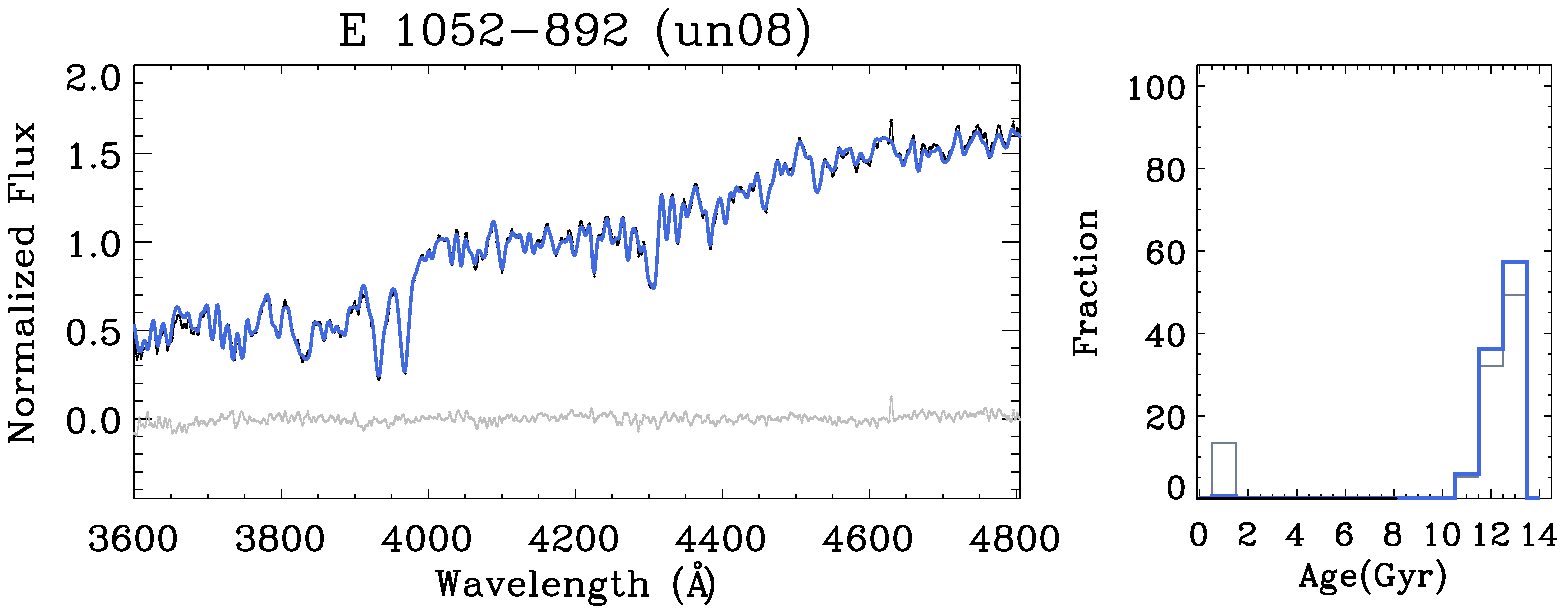}  
        \includegraphics[scale=0.4]{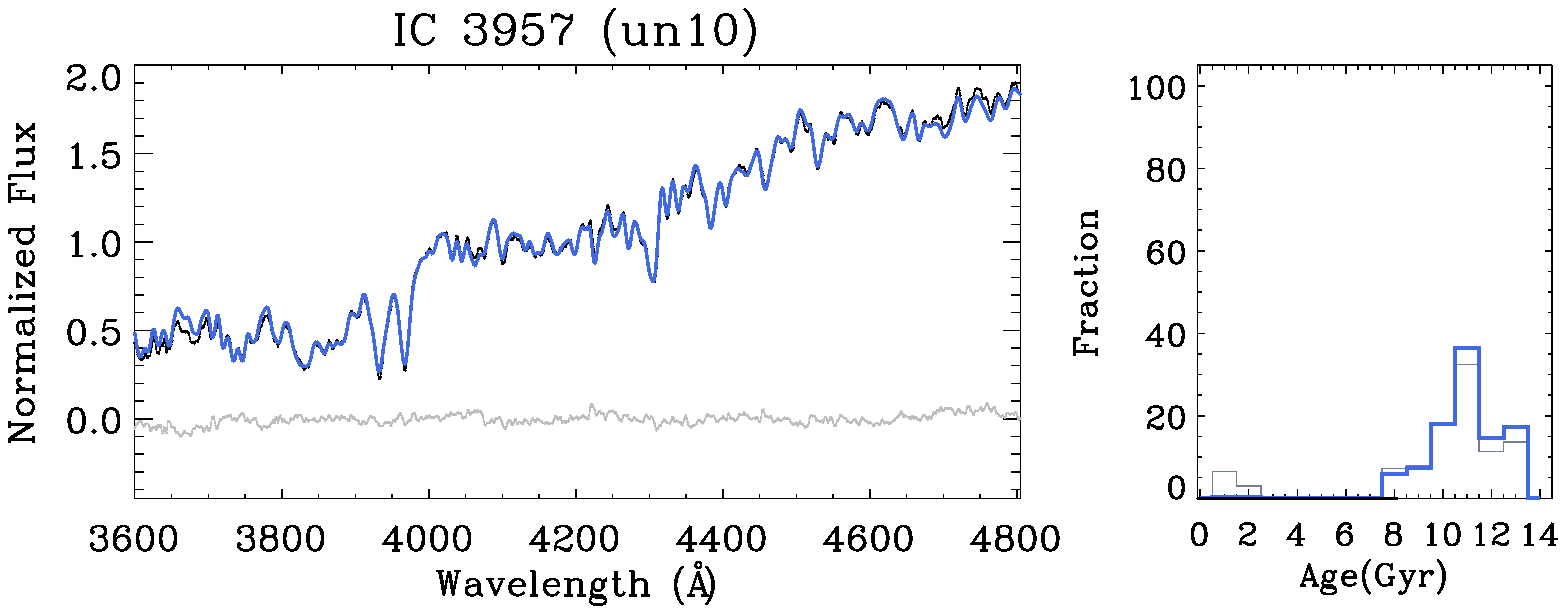} 
        \includegraphics[scale=0.4]{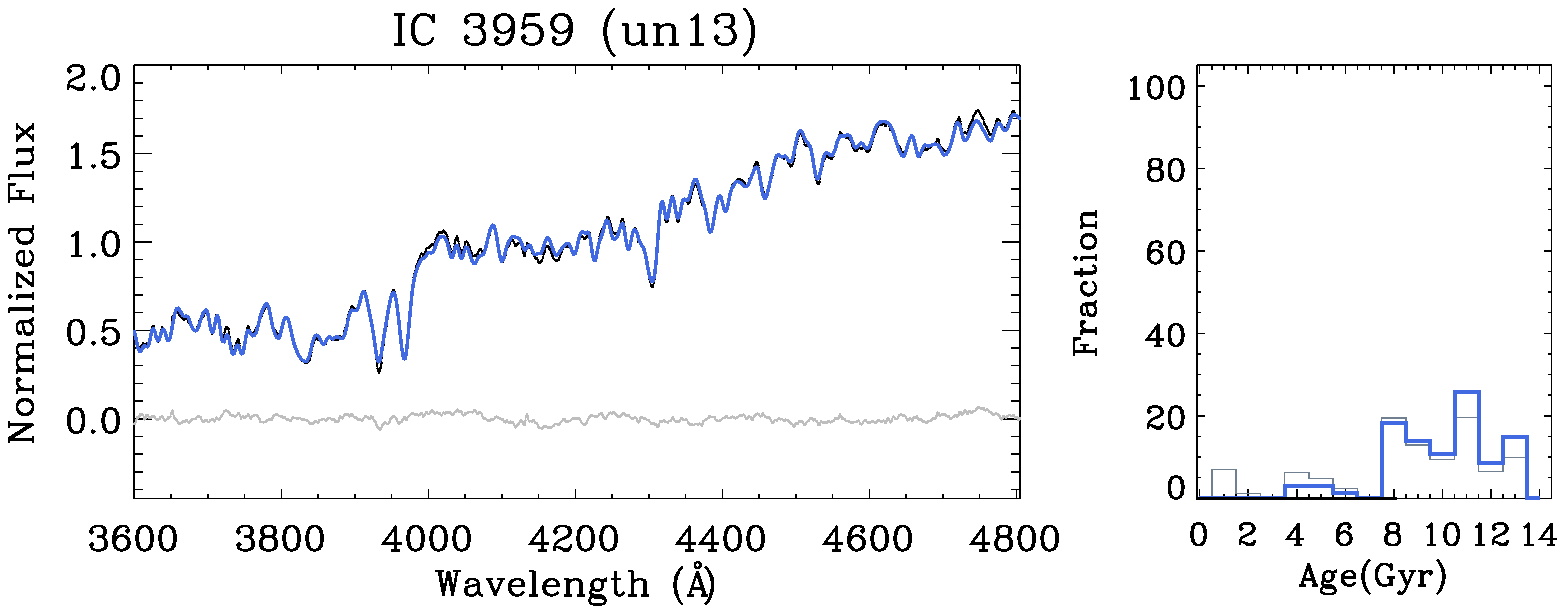} 
        \includegraphics[scale=0.4]{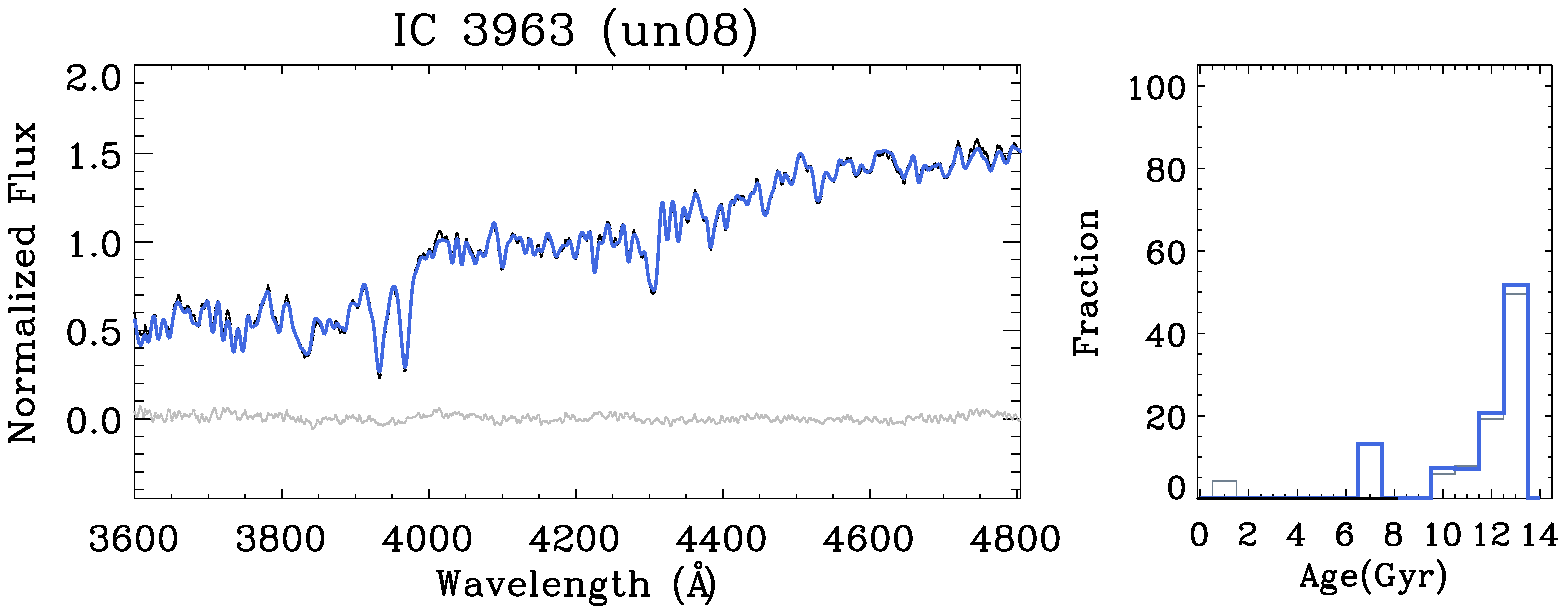} 
        \includegraphics[scale=0.4]{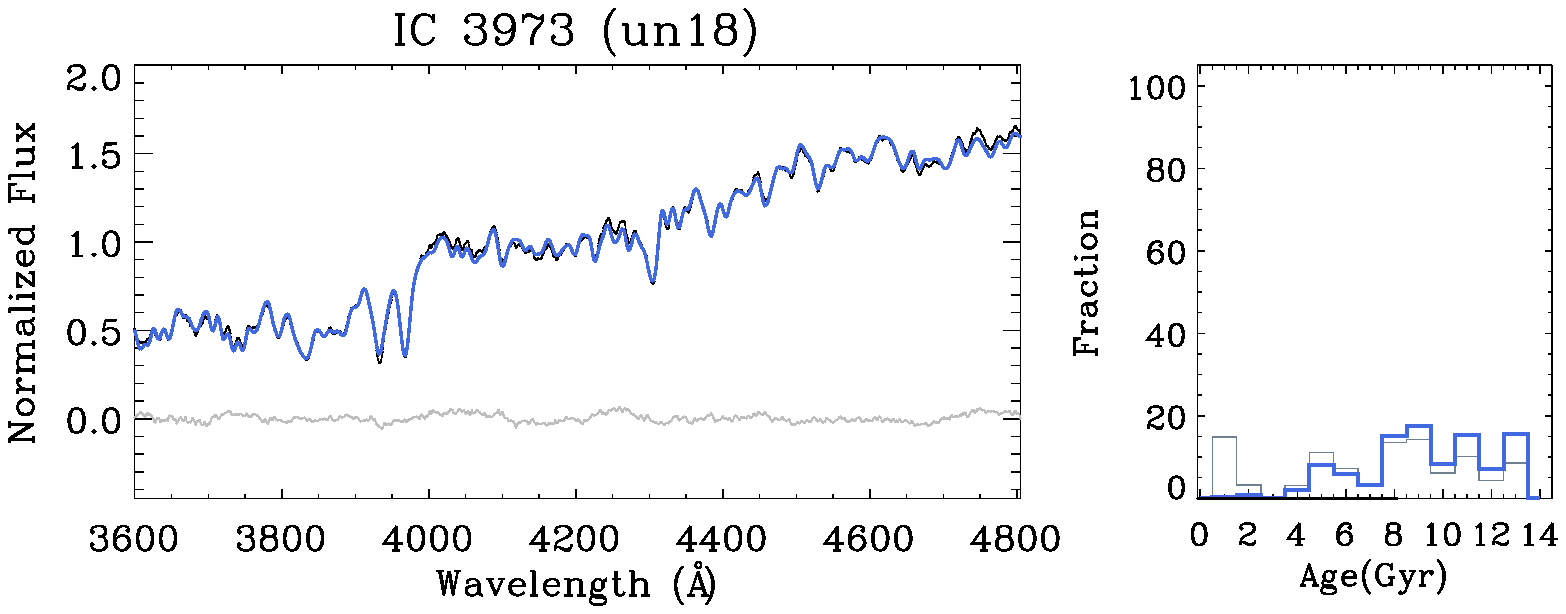} 
        \includegraphics[scale=0.4]{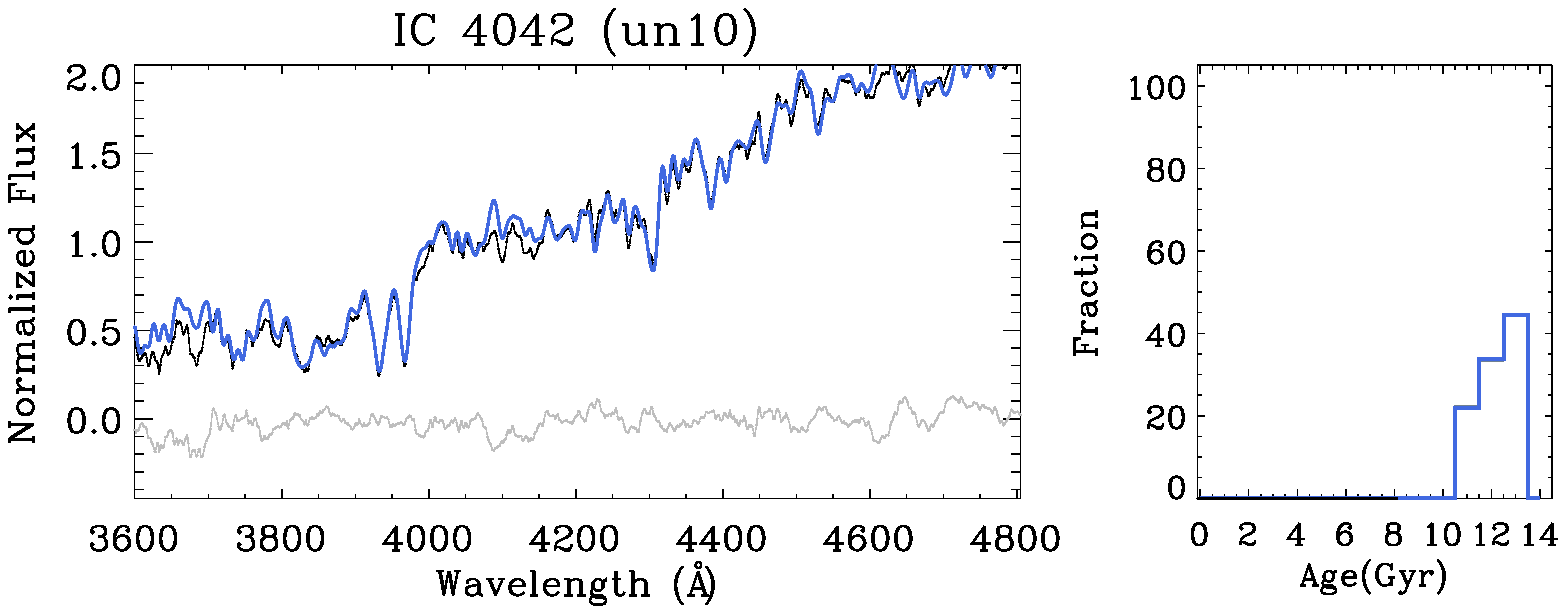}
        \includegraphics[scale=0.4]{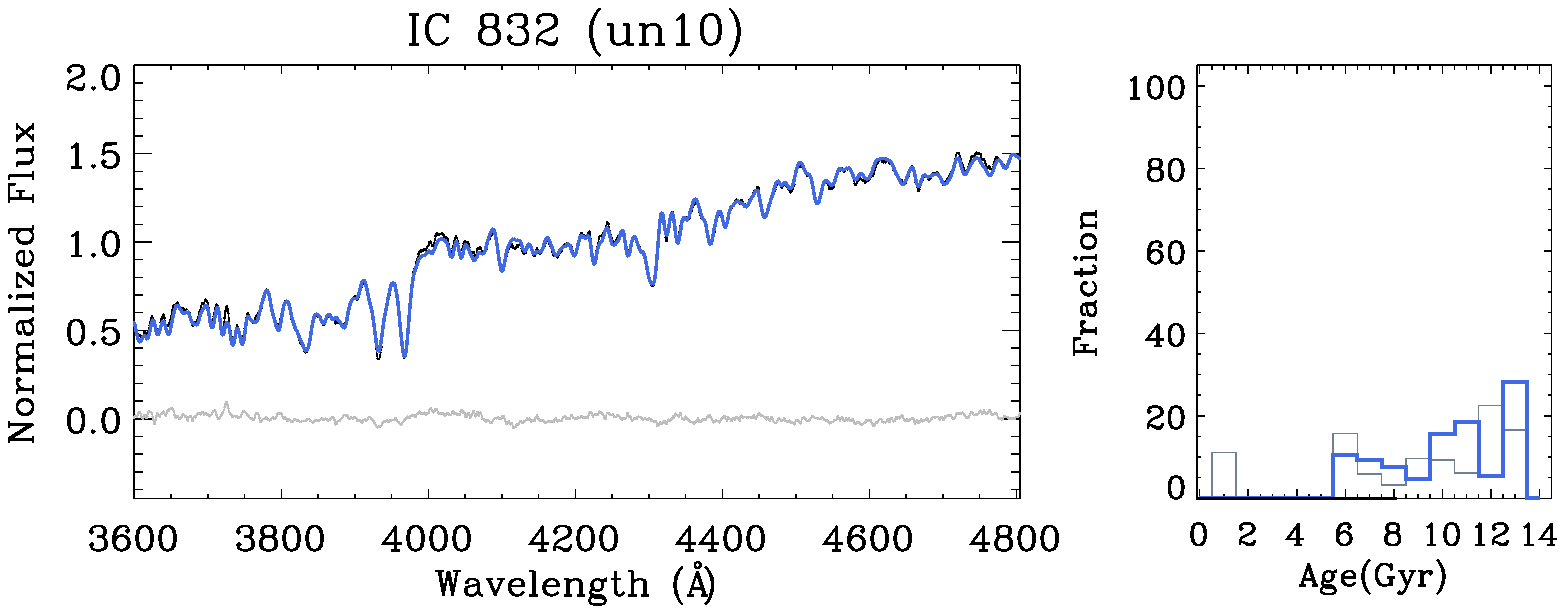}
        \includegraphics[scale=0.4]{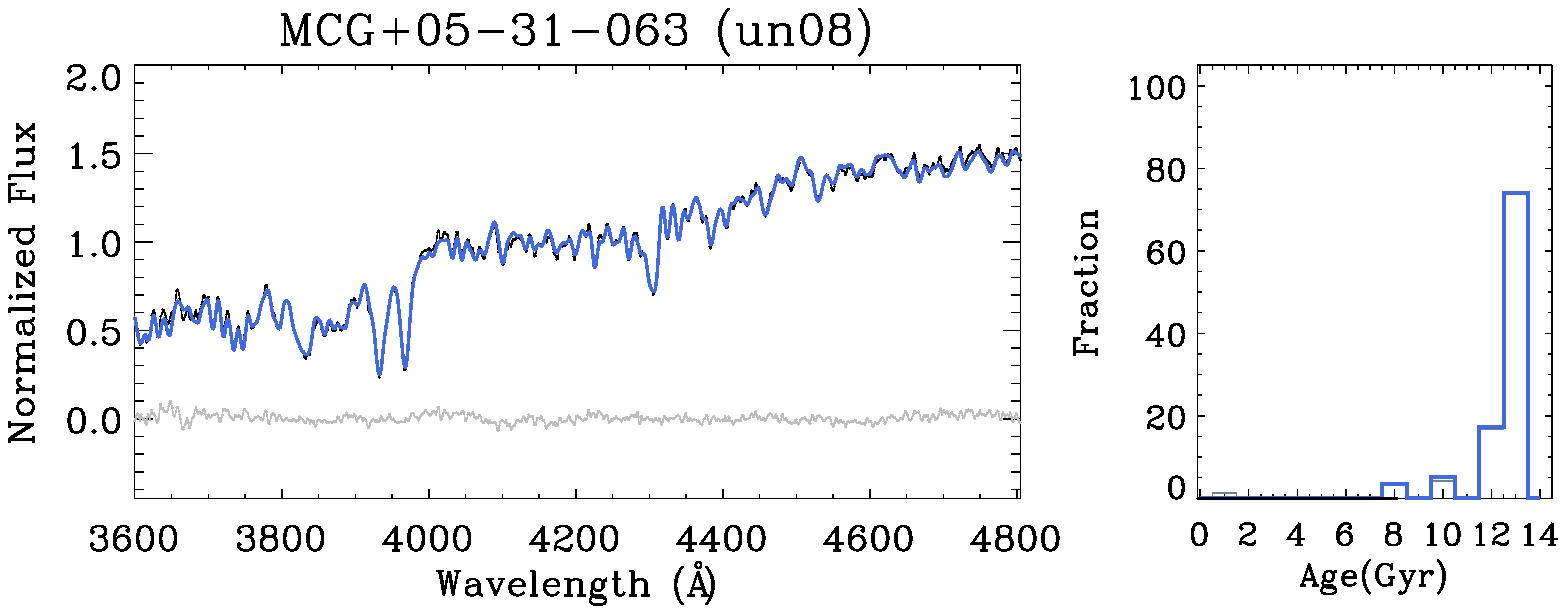}
        \includegraphics[scale=0.4]{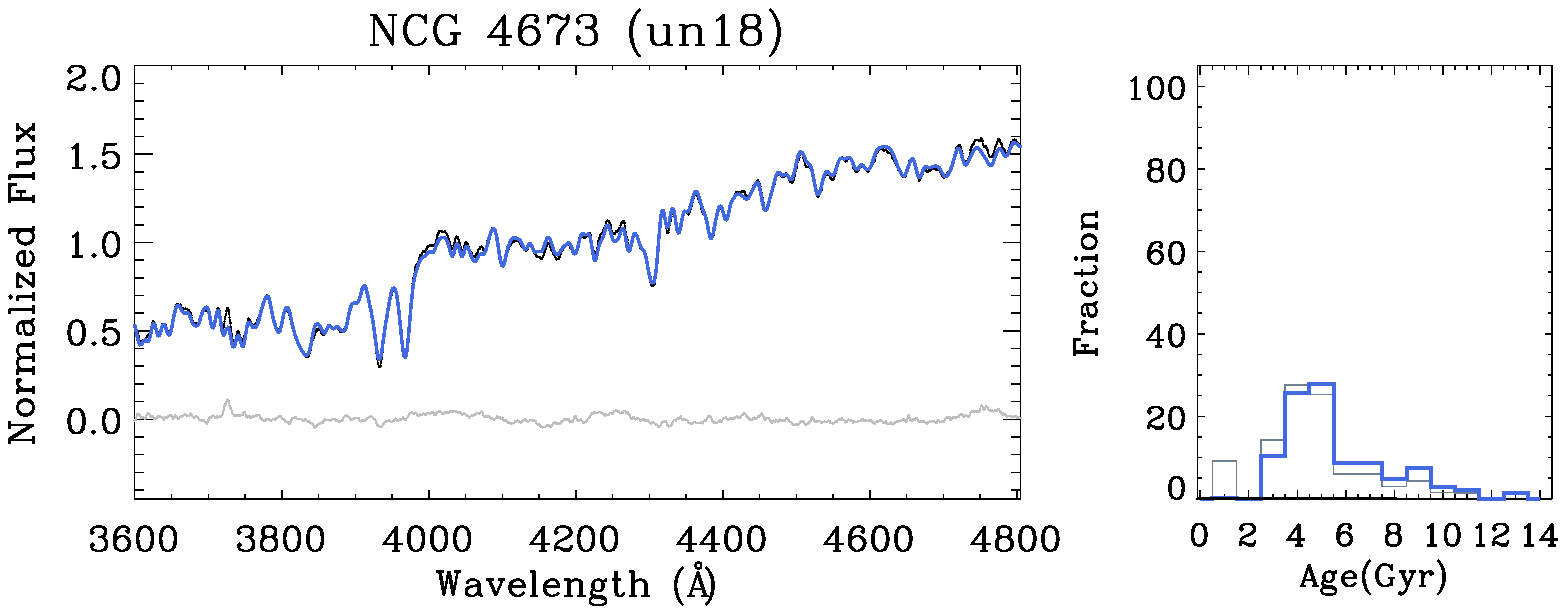}
        \includegraphics[scale=0.4]{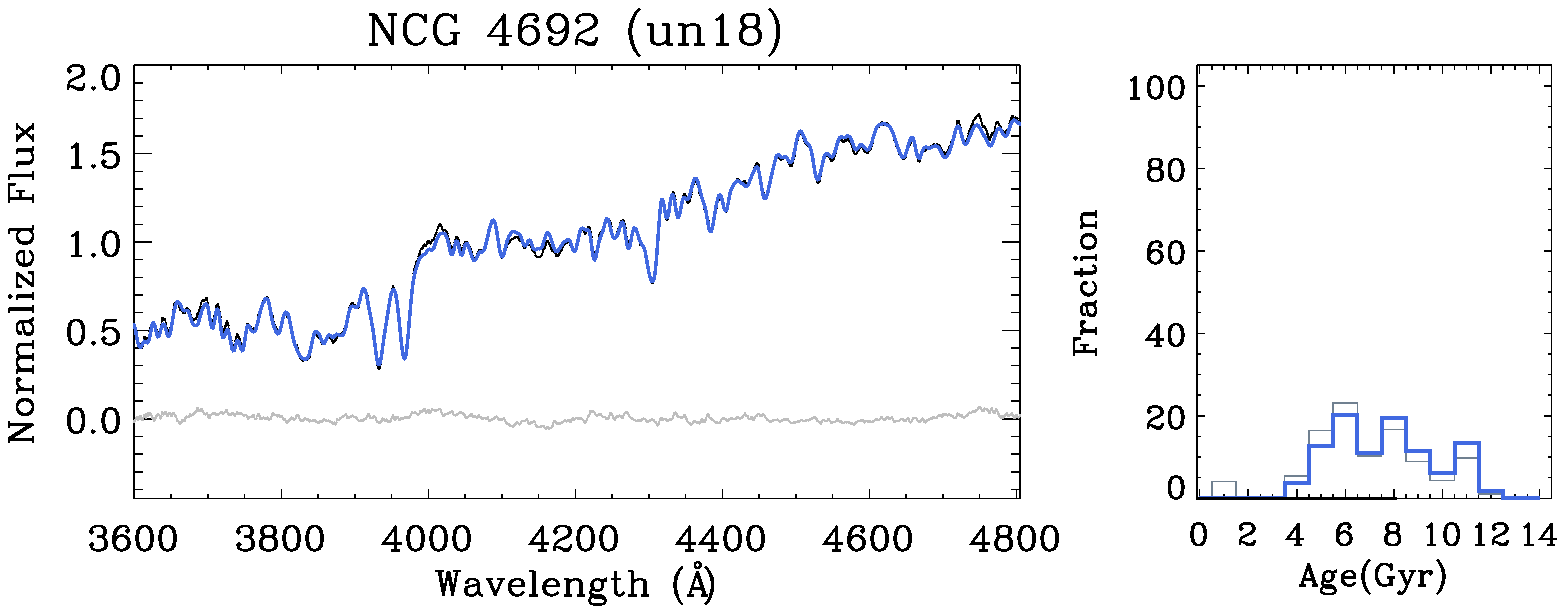}
        \includegraphics[scale=0.4]{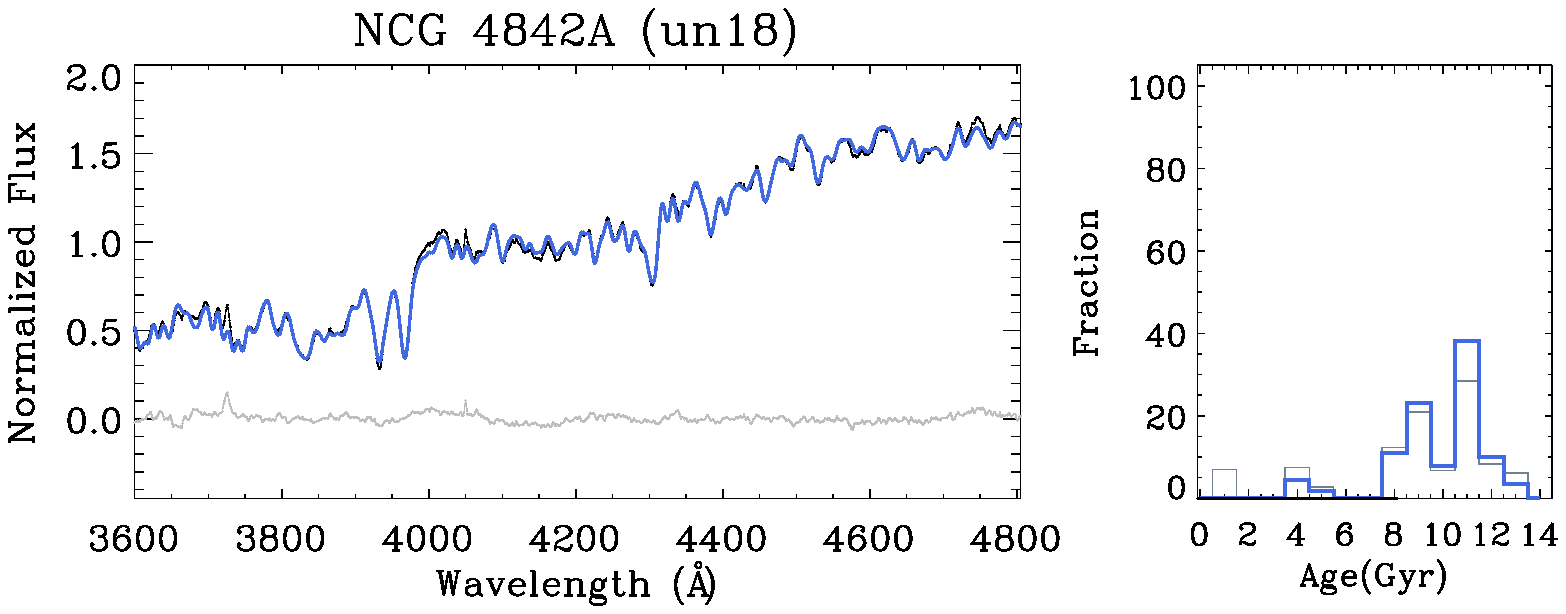}
\caption{Coma cluster galaxy spectra used as the control sample are plotted, together with the best fit from {\tt STARLIGHT} and the residual (left panels). The derived Star Formation Histories are shown on the right panels, where the blue histograms represent the mass-weighted derived SFH and the gray line represents the luminosity-weighted estimate.} 
\end{figure*}

\begin{figure*}
\label{figure:B5}
\centering
       \includegraphics[scale=0.4]{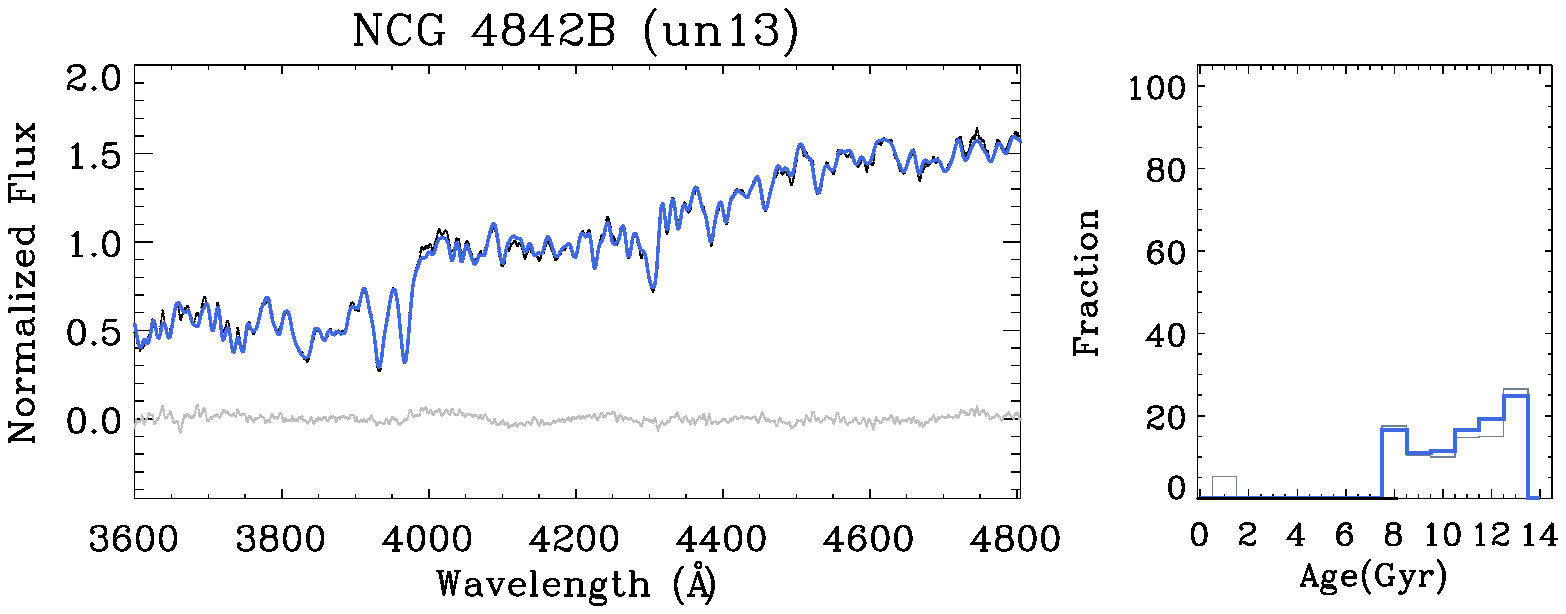}
       \includegraphics[scale=0.4]{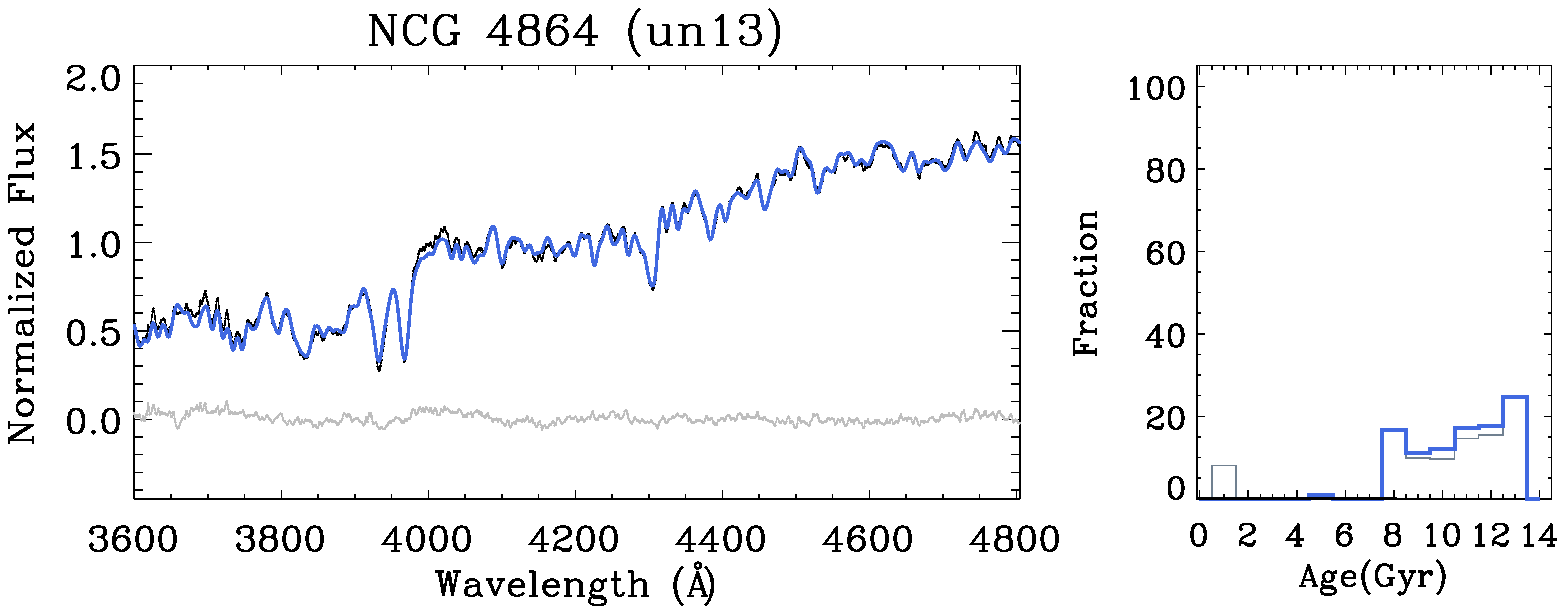}
       \includegraphics[scale=0.4]{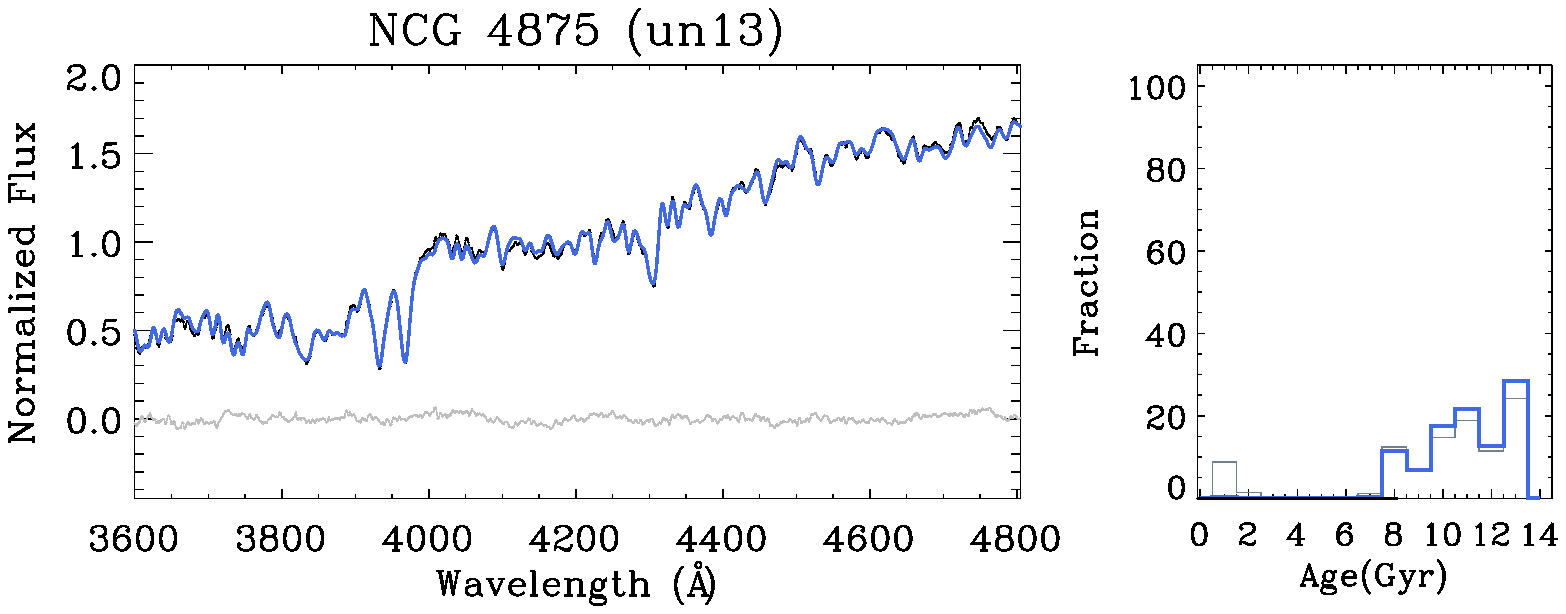}
       \includegraphics[scale=0.4]{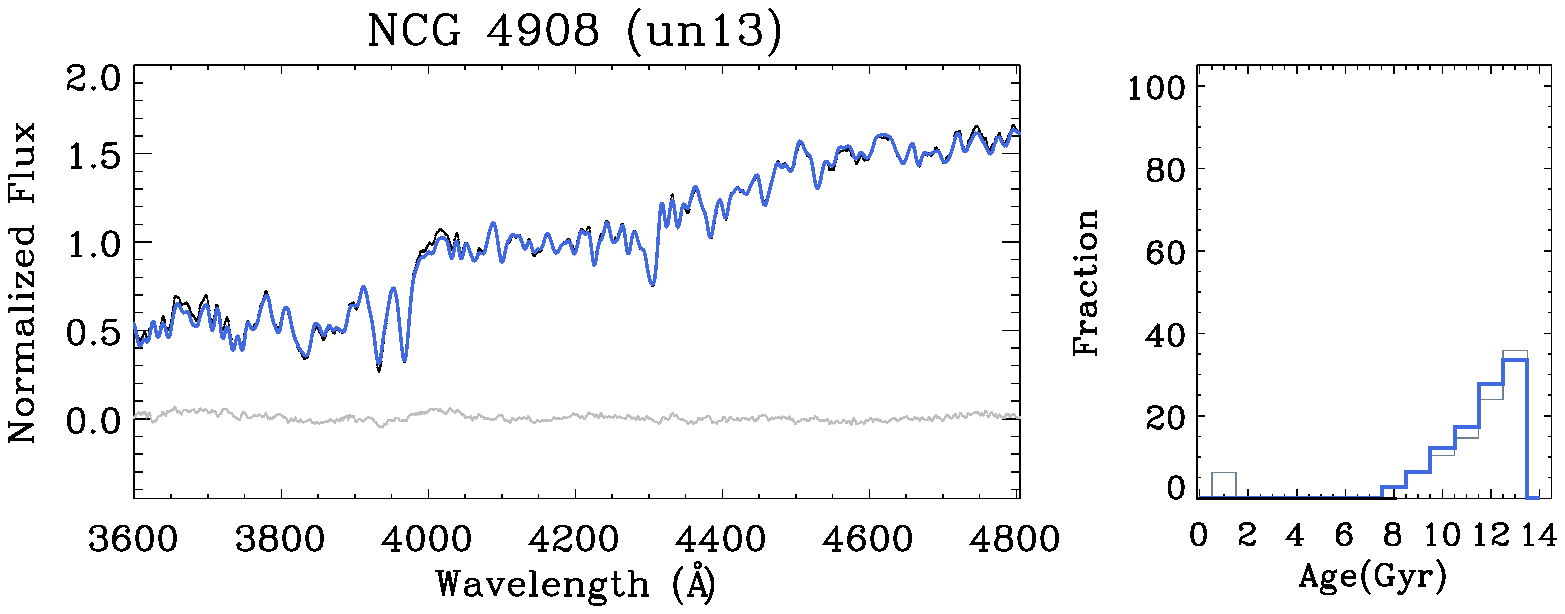}
       \includegraphics[scale=0.4]{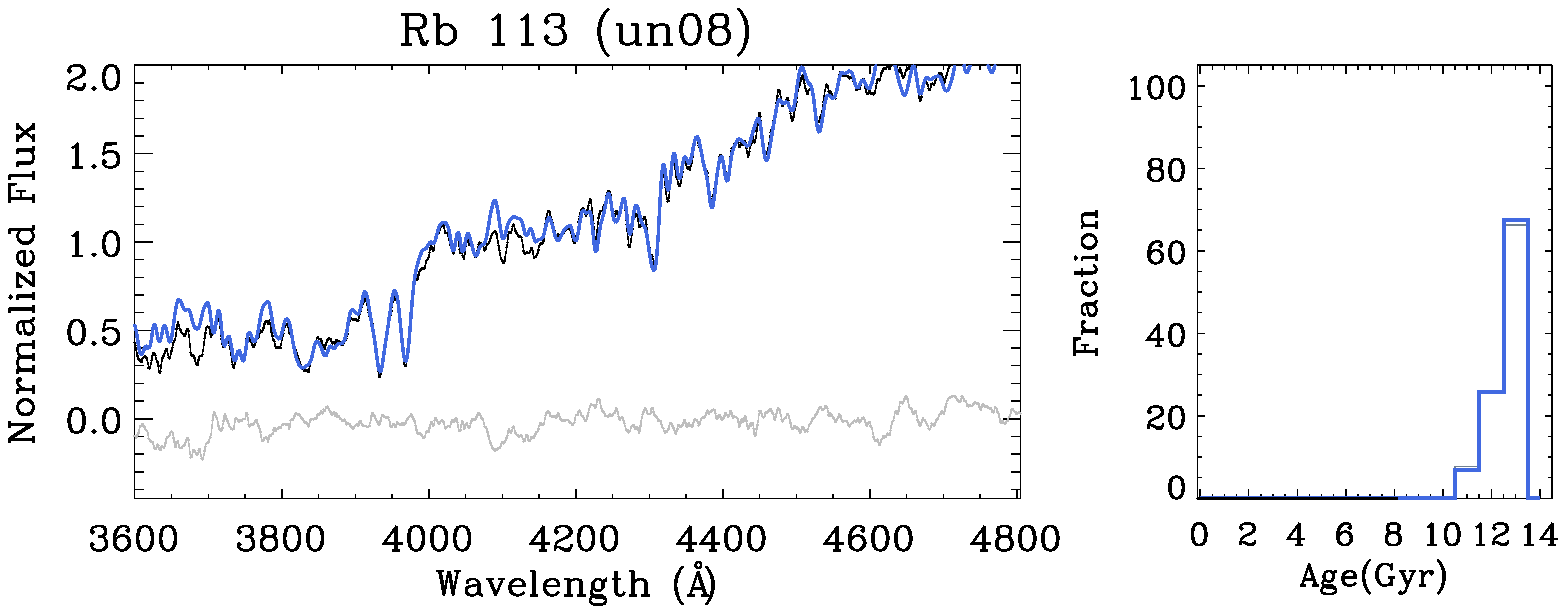}
       \includegraphics[scale=0.4]{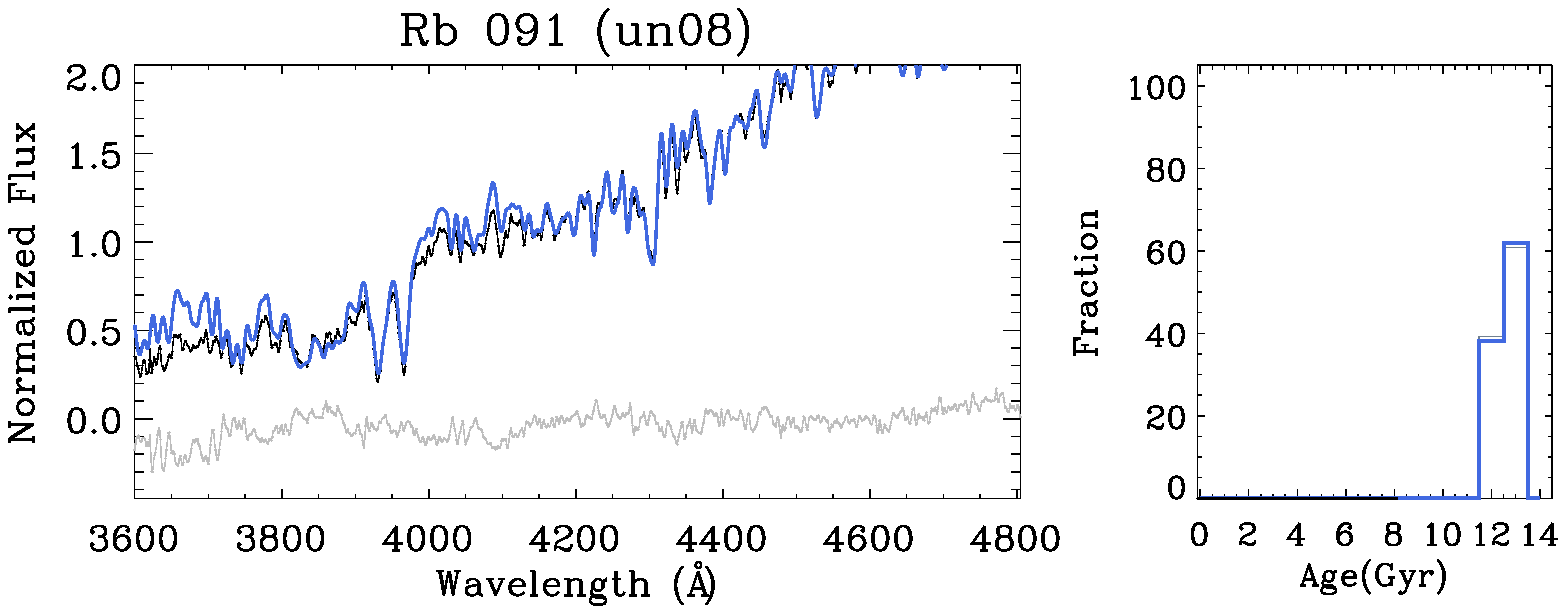}      
\caption{Figure B4, continued.} 
\end{figure*} 

\begin{figure*}
\label{figure:B6}
\centering
\includegraphics[scale=0.9]{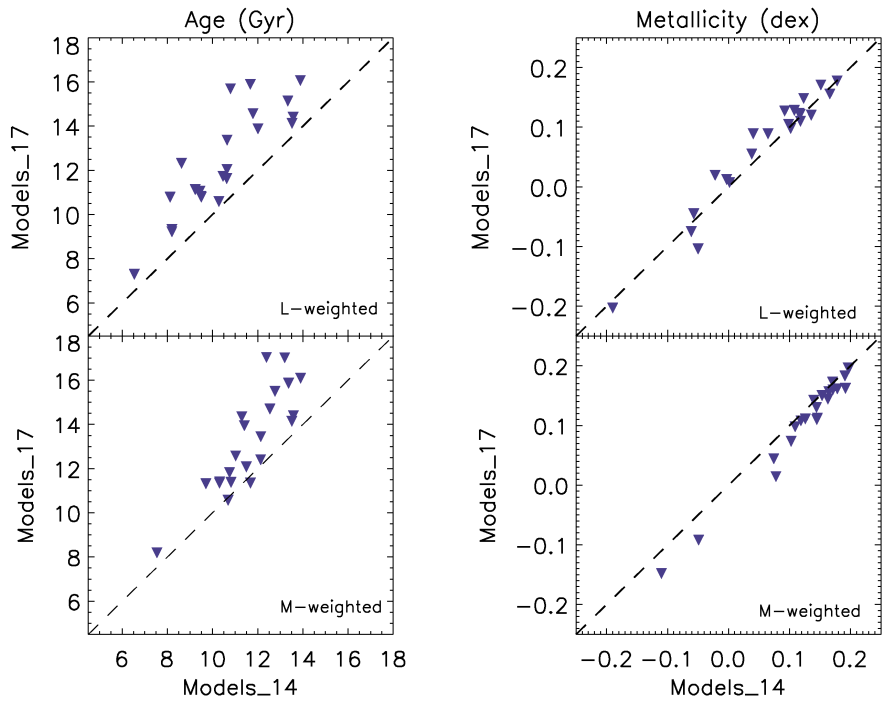}
\caption{Test  of the robustness of selecting the SSP models limited to the age of the Universe at the redshift of the cluster to derive the SFHs. In this case, we test the results with Coma. It is seen that only for the oldest ones there is a variation according to the limit imposed, while the younger ones are in good agreement.} 
\end{figure*} 

\begin{figure*}
\label{figure:B7}
\centering
\includegraphics[scale=0.8]{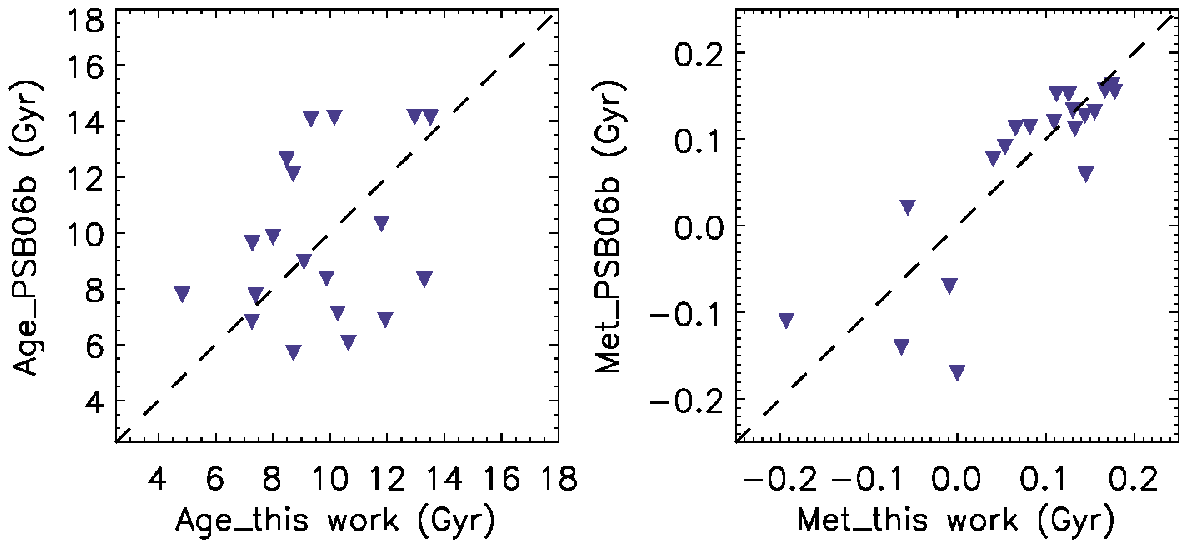}
\caption{Comparison of the ages derived from the full-spectrum-fitting in this work and in S\'anchez-Bl\'azquez et al. (2006b).} 
\end{figure*} 

\subsection{IMF variations}
Figure B8 shows the differences in the stellar populations derived employing a non-standard \textit{vs} a universal Kroupa IMF. We see that the largest differences are found for the oldest galaxies in the Coma set. This occurs because the IMF effect is particularly relevant for the old populations while it does not affect the youngest ones as much \citep{Ferre-Mateu2013}. Because we have removed the models corresponding approximately half the age of the Universe in the high redshift cluster, this dependence is not seen, while it is more relevant for the galaxies in Coma. 

\begin{figure*}
\label{figure:B8}
\centering
\includegraphics[scale=0.9]{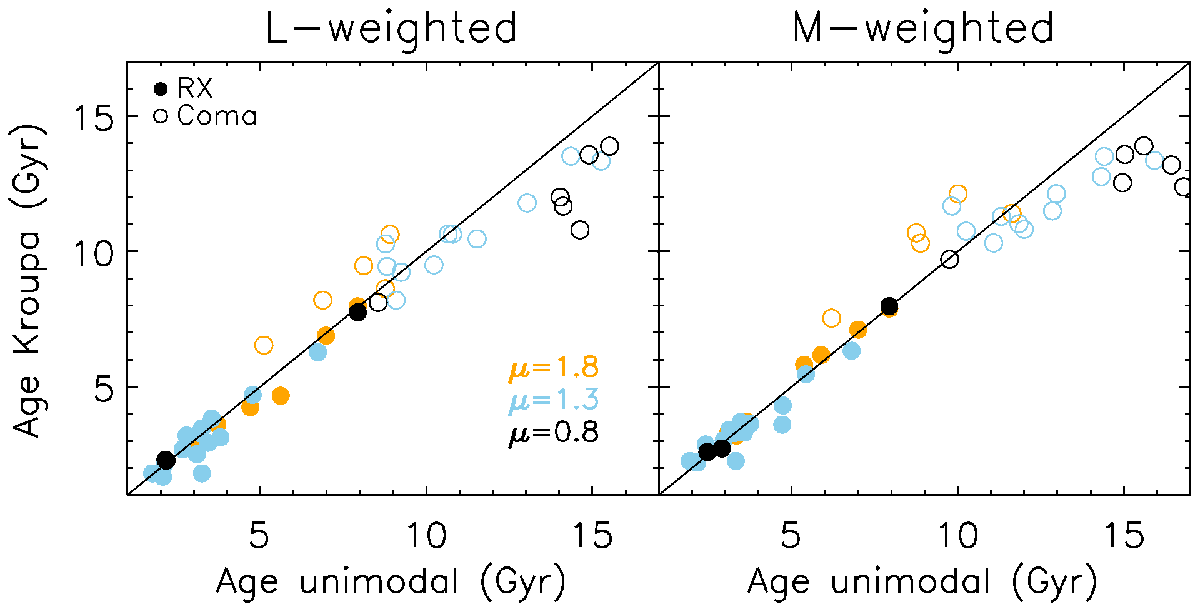}
\caption{Ages derived assuming an standard Kroupa-like IMF slope \textit{versus} the ages assuming different IMF slopes according to the velocity dispersion of the galaxy. Filled dots represent the intermediate-z cluster, while open dots represent the galaxies in Coma. Three different IMF slopes are shown in different colors, both in the luminosity and the mass weighted panels.} 
\end{figure*} 

\bibliography{biblio_clusters}
\bibliographystyle{mn2e}

\end{document}